\documentclass[aip,jcp,reprint]{revtex4-2}

\usepackage{graphicx}
\usepackage{dcolumn}
\usepackage{bm}
\usepackage[version=4]{mhchem}
\usepackage{chemscheme}
\usepackage{hyperref}
\usepackage{amsmath,amssymb}
\usepackage{floatrow}
\usepackage{chemfig}
\usepackage{chemformula}
\usepackage{soul}
\usepackage{xr}
\usepackage{subfiles}
\begin{document}

\preprint{AIP/propargyl}

\title[Propargyl radical]{\texorpdfstring{The X-ray absorption spectrum of the propargyl radical, \ch{C3H3^{.}.}       } {}}

 \author{Dorothee Schaffner}
 \affiliation{Institute of Physical and Theoretical Chemistry, University of 
 W{\"u}rzburg, D-97074 W{\"u}rzburg, Germany.
 }
 \author{Theo Juncker von Buchwald}%
\affiliation{DTU Chemistry, Technical University of Denmark, DK-2800 Kgs. Lyngby, Denmark.}

 \author{Jacob Pedersen}%
\affiliation{DTU Chemistry, Technical University of Denmark, DK-2800 Kgs. Lyngby, Denmark.}
\affiliation{Department of Chemistry, Norwegian University of Science and Technology, N-7491 Trondheim, Norway.}

 \author{Andreas Rasp}%
\affiliation{Institute of Physical and Theoretical Chemistry, University of 
 W{\"u}rzburg, D-97074 W{\"u}rzburg, Germany.}
 
 \author{Emil Karaev}%
\affiliation{Institute of Physical and Theoretical Chemistry, University of 
 W{\"u}rzburg, D-97074 W{\"u}rzburg, Germany.}
 
 \author{Valentin von Laffert}%
\affiliation{Institute of Physical and Theoretical Chemistry, University of 
 W{\"u}rzburg, D-97074 W{\"u}rzburg, Germany.}

  \author{Alessio Bruno}%
\affiliation{Dipartimento di Chimica e Tecnologia dei Farmaci, Universit{\`a} degli Studi di Roma  ``La Sapienza'', piazzale Aldo Moro 5, I-00185 Rome, Italy.}

\author{Michele Alagia}
\affiliation{CNR - Istituto Officina dei Materiali (IOM), Laboratorio TASC, 34149 Trieste, Italy.}

\author{Stefano Stranges}
\affiliation{CNR - Istituto Officina dei Materiali (IOM), Laboratorio TASC, 34149 Trieste, Italy.}
\affiliation{Dipartimento di Chimica e Tecnologia dei Farmaci, Universit{\`a} degli Studi di Roma  ``La Sapienza'', piazzale Aldo Moro 5, I-00185 Rome, Italy.}

 \author{Ingo Fischer}
  \homepage{E-mail: ingo.fischer@uni-wuerzburg.de (experiment)}
\affiliation{Institute of Physical and Theoretical Chemistry, University of 
 W{\"u}rzburg, D-97074 W{\"u}rzburg, Germany.}
 
\author{Sonia Coriani}
\homepage{E-mail: soco@kemi.dtu.dk (theory)}
\affiliation{DTU Chemistry, Technical University of Denmark, DK-2800 Kgs. Lyngby, Denmark.}

\date{\today}

\begin{abstract}
We report a combined experimental and computational study of the near-edge X-ray absorption fine structure (NEXAFS) spectrum of the propargyl radical, \ch{C3H3^{.}}. As a central intermediate in the formation of polycyclic aromatic hydrocarbons, the propargyl radical is a species of considerable relevance in combustion and astrochemistry and was here generated by pyrolysis from propargyl bromide. The NEXAFS spectrum shows a pronounced band at 282.2\,eV corresponding to transitions from carbon 1s orbitals to singly occupied molecular orbitals. \textit{Ab initio} calculations show that 
two transitions to the lowest lying states 1~$^2\!$A$_1$ and 2~$^2\!$A$_1$, which take place from the C1s orbital of the two terminal carbon atoms, contribute to this band.
In addition, a 420\,meV spacing of the first band is visible and is assigned to a vibrational progression in the symmetric CH$_2$ stretch. Transitions at higher energies are also described reasonably well by theory. The fragmentation pattern was investigated at the different resonant transitions and 
shows the cleavage of one as well as both C--C bonds.
\end{abstract}

\maketitle

\section{\label{sec:introduction} Introduction}
The mechanism of the formation of polycyclic aromatic hydrocarbons (PAHs) is a major research topic in astrochemistry as well as in combustion science.\cite{Richter2000, Kress2010, Levey2022,Byrne_2023} In interstellar space, PAHs and their cations are investigated as possible carriers of the diffuse interstellar bands (DIB) as well as the unidentified infrared bands (UIB).\cite{tielens2013molecular} They have been postulated to comprise a significant part of the carbon budget in space.\cite{tielens2008interstellar} However, the formation mechanism of PAHs in the low pressure and low temperature environment of space is an open question.\cite{Levey2022} Barrierless or low-barrier reactions, such as ion-molecule and radical-radical reactions, offer a possible explanation. In combustion, the dimerization of propargyl, \ch{C3H3^{.}}, was found to be an efficient reaction that proceeds in a sequence of barrierless steps to benzene and phenyl radical.\cite{Alkemade1989, Stein_1991} 
Propargyl, shown on the right-hand side of Scheme~\ref{scheme:pyrolysis}, is a resonantly stabilized radical of C$_{2\text{v}}$ symmetry with a $\tilde X$\textsuperscript{ 2}B\textsubscript{1} ground state.\cite{botschwina2010calculated}
For the nominal resonance structures ethynyl methyl (\ch{H2C^{.}-C+CH}) and allenyl (\ch{H2C=C=C^{.}H}), a ratio of 61\% to 39\% was inferred from recent microwave measurements analyzing the spatial distribution of the unpaired electron.\cite{Changala_2024}
Propargyl is a well-known intermediate in flames and plasmas and has recently been detected in the cold dark Taurus Molecular Cloud (TMC-1)\cite{Agundez2021, Agundez2022} where the corresponding cationic species has been observed as well.\cite{Silva_2023} The detection was based on prior laboratory microwave spectra.\cite{Tanaka1997} It was suggested that propargyl is ``one of the most abundant radicals detected in TMC-1, and it is probably the most abundant organic radical with a certain chemical complexity ever found in a cold dark cloud.''\cite{Agundez2021} Most likely, it is also involved in the atmospheric chemistry of the Saturn moon Titan.\cite{Hebrard2013}  Given its abundance, the propargyl radical may be involved in the synthesis of aromatic molecules in space as well. For example, the ion-molecule reaction \ch{H2CCCH^{.}}~+~\ce{C3H3+}~$\rightarrow$~\ce{C6H5+}~+~\ch{H^{.}} has been suggested to be an efficient source for formation of the phenyl cation.\cite{herbst1989gas} 
Due to the importance of the propargyl radical in astrochemistry and in combustion chemistry, its spectroscopy and chemistry have been broadly investigated. Specifically, infrared,\cite{Jacox_1974,Jochnowitz_2005} UV/Vis\cite{Ramsay_1966,Fahr_2005} and photoelectron spectra,\cite{Hemberger_2011,Merkt_2013,Garcia_2018} including a study that followed the steps to benzene formation,\cite{Savee2022} 
have been reported. The photodissociation has been studied by H-atom photofragment spectroscopy,\cite{Deyerl1999} high-n Rydberg atom time-of-flight (HRTOF) spectroscopy,\cite{Zheng2009} and in a Laval nozzle expansion.\cite{Broderick2018} 
Hydrogenation reactions of \ce{C3H3+} and related cations important for carbon chemistry in space have been studied in ion traps.\cite{Savic_2005,Gerlich_2005}
The bimolecular recombination of propargyl was studied in the gas-phase and branching ratios of the isomeric \ce{C6H6} reaction products were determined.\cite{Kaiser_2021,Hrodmarsson_2024} The further efficient growth of PAHs by a subsequent addition of propargyl units to the first ring has been investigated by IR/UV ion dip spectroscopy.\cite{Constantinidis_2017} Both the structure as well as reactions of the propargyl radical 
have also been investigated 
computationally, with a focus on the propargyl dimerization.\cite{botschwina2010calculated, Klippenstein_2003} 
Possible mechanisms for PAH formation and growth reactions that include propargyl as one reactant have been studied experimentally\cite{Savee_2015,Osborn_2023,Kaiser_2023,Zhang_2025} and computationally.\cite{Lindstedt_2002,Matsugi_2012,Mebel_2020}
Propargyl has also been investigated as precursor for polycyclic aromatic nitrogen containing heterocycles (PANH) formation in space.\cite{Schleier_2025}

In contrast to the investigations in other spectral regions, the response of polyatomic organic radicals to X-ray excitation has hardly been studied. The astrochemical relevance of this topic originates from the X-ray emission in young stellar objects that might have a significant influence on the circumstellar molecular composition.\cite{stauber2005} 
So far, X-ray absorption near the first C1s ionization threshold (XAS, NEXAFS) in hydrocarbon radicals has only been investigated for allyl, \ch{C3H5^{.}},\cite{Alagia_Allyl} methyl, \ch{CH3^{.}},\cite{Alagia_Methyl} and its isotopomer \ch{CD3^{.}},\cite{Ekstrom_CD3_CH3} as well as \textit{tert}-butyl, \ch{C4H9^{.}}.\cite{Schaffner_2024}
A recent study focused on the C1s NEXAFS spectroscopy of different isomeric \ce{C3H3+} cations, including the propargylium ion. There, site-specific resonances and fragmentation channels leading to doubly charged fragments were investigated.\cite{Reinwardt_2024}

In the present work, we address the NEXAFS spectroscopy of neutral propargyl, one of the astrochemically most important radicals, in a combined experimental and computational study. This work will also help to identify the propargyl radical in future studies using time-resolved X-ray spectroscopy.

\section{\label{sec:methods} Methods}
\subsection{\label{sec:methods_experimental} Experimental}
The experiments were carried out at the Gas Phase Photoemission beamline at the Elettra Synchrotron, Trieste/IT. The setup has been detailed previously,\cite{Blyth_1999, Alagia_2003} but a brief description is given in the following.
Synchrotron light at the C1s edge and at 21.2\,eV was provided by an undulator and spectrally dispersed by a variable angle, spherical grating type monochromator. Different gratings were used for different photon energy regions, and the photon energy resolution was controlled by the entrance and exit slit width.

\begin{scheme}[hbpt!]
    \centering
\includegraphics[width=\linewidth]{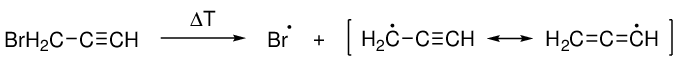}
\caption{Generation of the propargyl radical, \ch{H2CCCH^{.}}, by pyrolysis of propargyl bromide. On the right-hand side the ethynyl methyl and allenyl resonance structures of the propargyl radical are shown.}
\label{scheme:pyrolysis}
\end{scheme}

The precursor propargyl bromide (stabilized with MgO) was purchased from TCI and used after several freeze-pump-thaw cycles. The precursor was cooled to $-$30\,\textdegree C, and the vapor of the precursor diluted in helium (dilution of 0.1\,\%) passed through a resistively heated SiC pyrolysis tube. Scheme~\ref{scheme:pyrolysis} shows the pyrolysis of propargyl bromide that yields the propargyl radical and a bromine atom.\cite{Deyerl1999, Hemberger_2011}
The pyrolysis products were expanded into the vacuum and entered the experimental chamber through a skimmer. The molecular beam was crossed by the synchrotron radiation, and the resulting electrons and ions were extracted in opposite directions. Electrons and ions were detected in coincidence by a single start-multiple stop technique. The electrons were detected by an MCP detector, set close to the ionization region, and used as start signal for the ion time-of-flight (TOF). Ions were detected by a Wiley-McLaren TOF mass spectrometer.

The pyrolytic conversion of the precursor according to Scheme~\ref{scheme:pyrolysis} was investigated by valence photoionization mass spectrometry at 21.2\,eV.
For C1s NEXAFS spectra the total electron yield (TEY) was recorded while the photon energy was scanned. The photon flux was monitored by a photodiode and used for normalization of the spectra. To correct for the helium background in the spectra, a spectrum of pure helium was measured and subtracted. 
In a diagnostic chamber in front of the experimental chamber, methane was measured simultaneously to calibrate the photon energy by its C1s $\to$ 3p-Rydberg transition at 288.00\,eV.\cite{de_Simone_2022} The photon bandwidth around the C1s edge was 50\,meV.
The C1s NEXAFS spectra in the region 280.3--297.2\,eV were measured with a step size of 20\,meV and an acquisition time of 45\,s (pyrolysis at 35\,W heating power) and 35\,s (pyrolysis off) per data point. High-resolution spectra in the range 281.3--283.3\,eV were measured with a step size of 10\,meV and an acquisition time of 75\,s. Valence photoionization mass spectra at 21.2\,eV were recorded with acquisition times between 300\,s and 1800\,s, while mass spectra at the C1s edge were acquired for 6000\,s.

\subsection{Computational\label{sec:methods_computational}}


Two \textit{ab initio} electronic-structure approaches were adopted for the computational study of the electronic transitions, namely the frozen-core core-valence-separated equation-of-motion coupled cluster singles and doubles (fc-CVS-EOM-CCSD) method,~\cite{Vidal2019fcCVS} and the extended core-valence-separated second-order algebraic diagrammatic construction (CVS-ADC(2)-x) method,~\cite{dreuw2015algebraic} as implemented in \texttt{Q-Chem}.~\cite{Qchem541}
A preliminary basis set study was performed at the fc-CVS-EOM-CCSD level using both restricted-open (ROHF) and unrestricted (UHF) Hartree-Fock reference orbitals, and the 6-311++G**,~\cite{clark1983a,krishnan1980a,marchetti2008accurate} cc-pVTZ,~\cite{dunning1989a} 
and aug-cc-pVTZ~\cite{kendall1992a} basis sets on all atoms. 
Unrestricted Hartree-Fock reference orbitals and the aug-cc-pVTZ basis set were adopted at the CVS-ADC(2)-x level. Core ionization energies were obtained at the fc-CVS-EOM-IP-CCSD level. 


We used the CCSD(T*)-F12a~\cite{adler2007simple,knizia2009simplified}/cc-pVTZ-F12~\cite{peterson2008a}-optimized molecular geometry (planar C$_{2\text{v}}$ structure) from \citeauthor{botschwina2010calculated}~\cite{botschwina2010calculated} in all calculations of the purely electronic spectra. 
This structure is rather similar to the recently reported semi-experimental equilibrium geometry of \citeauthor{Changala_2024}\cite{Changala_2024}
The bond lengths of both structures are reported in 
Table~S8 
in the SI.
As observed by the authors of Ref.~\citenum{Changala_2024}, the deviation of the C--C bond lengths in the ground state of propargyl from pure single, double and triple bonds 
are 
``[$\cdots$] evidence of $\pi$-delocalization of the unpaired electron along the carbon backbone''.

Natural transition orbitals (NTOs)~\cite{Vidal2019fcCVS,krylov2020orbitals} were computed at the corresponding fc-CVS-EOM-CCSD and CVS-ADC(2)-x levels of theory and used for the assignment of the spectral peaks. 
The spectra were constructed from the computed excitation energies and oscillator strengths using a Lorentzian convolution with a full width at half maximum (FWHM) broadening of 200\,meV.
All calculations were carried out on DTU HPC resources.~\cite{DTU-DCC}

Three different strategies were adopted to investigate vibrational effects in the spectrum. In the first strategy, the nuclear ensemble approach (NEA),~\cite{crespo2012spectrum} we sampled 200 vibrational structures of the ground state using a Wigner sampling at a temperature of 300\,K. The ground state frequencies used for sampling the vibrational structures are calculated from a B3LYP~\cite{stephens1994ab}/aug-cc-pVTZ optimized geometry. For all structures the NEXAFS spectrum was then calculated using both fc-CVS-EOM-CCSD and CVS-ADC(2)-x with the aug-cc-pVTZ basis set. 
In the second and third strategy, both a time-independent (TI)
and a time-dependent (TD) quantum approach were utilized for the calculation of the vibrational structure of the selected electronic transitions in the NEXAFS spectrum.~\cite{avilaferrer2012verticaladiabatic,cerezo2023fcclasses3}
The (eigenstate-free)
TD approach delivers directly the vibrationally resolved spectral profile, whereas the  TI approach is based on the individual stick transitions of all possible vibronic states, which are then broadened.
The potential energy surfaces (PES) associated with the electronic states involved in the transition that are required by both methods were obtained 
from the harmonic Vertical Gradient (VG)~\cite{macak2000simulations,biczysko2012time,avilaferrer2012verticaladiabatic} and Adiabatic Shift (AS)~\cite{biczysko2012time,avilaferrer2012verticaladiabatic} approximations, respectively. \texttt{FCclasses3}~\cite{cerezo2023fcclasses3} was used for these calculations. 
The VG and AS approximations both utilize the ground state energy, gradient, and Hessian as well as the excitation energy and a 1-norm dipole moment along the same direction (\textit{x}, \textit{y}, and \textit{z}) as the transition dipole moment. The two methods differ in the way they treat the excited state geometry and gradient when reconstructing the PES of the final electronic state.~\cite{avilaferrer2012verticaladiabatic}  
For the VG approximation (also known as linear coupling model~\cite{macak2000simulations}),
the ground state geometry is used also for the excited state, and the excited state gradient is calculated at this geometry. 
In the AS approximation, the excited state geometry is optimized, and the excited state gradient is
practically zero within the chosen optimization threshold.
The ground and excited state(s) share the same normal modes and frequencies, as the same Hessian is used for both.~\cite{avilaferrer2012verticaladiabatic}
The spectra are scaled afterwards using the transition dipole strength. We used a UMP2/aug-cc-pVDZ optimized geometry from Gaussian~\cite{frisch2016gaussian} 
followed by excited state geometry optimizations at the ADC(2)-x level for the first three bright states using \texttt{PySCF}~\cite{pyscf,pyscf_2} for the Hartree-Fock calculation, \texttt{ADCC}~\cite{herbst2020adcc} for the excitation energy calculation, \texttt{cvs-adc-grad}~\cite{cvs-adc-grad} for the excited state gradient, and \texttt{geomeTRIC}~\cite{wang2016geometry} for the excited state geometry optimization.
The vibrationally resolved spectral bands were broadened by a Voigt function with a Gaussian FWHM of 50\,meV and a Lorentzian FWHM of 100\,meV, corresponding to the experimental photon bandwidth and the approximate lifetime width of the C1s core hole, respectively. We computed both TI and TD spectra at 0\,K, as well as TD ones at 300\,K.
We refer to 
Section~S2.1 
in the Supplementary Information (SI) for additional computational details.


\section{\label{sec:results} Results and Discussion}
\subsection{\label{sec:NEXAFS_exp}Experimental NEXAFS spectrum}
\begin{figure}[hbpt!]
    \centering
\includegraphics[width=\linewidth]{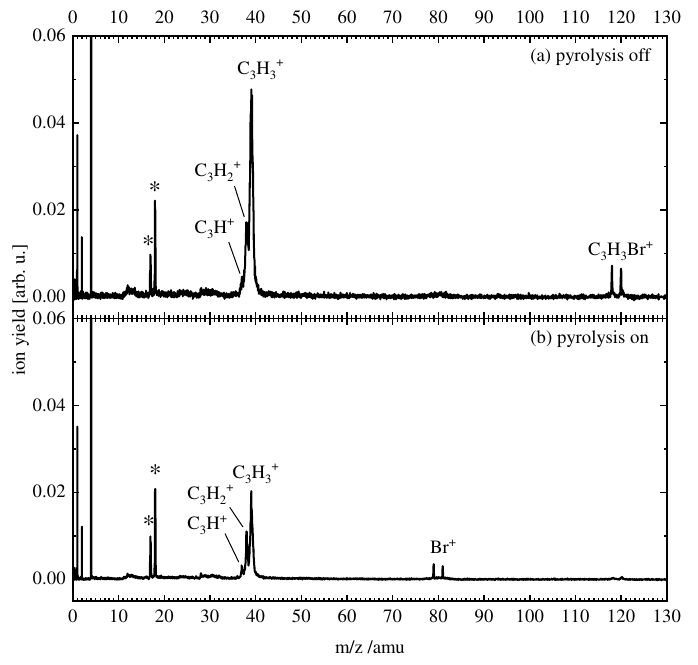}
\caption{TOF mass spectra at 21.2\,eV with (a) pyrolysis off and (b) pyrolysis on. The spectra were normalized to the He\textsuperscript{+} signal at \textit{m/z}~4. Asterisks indicate water as impurity.}
    \label{fig:TOF_valence}
\end{figure}

Figure~\ref{fig:TOF_valence} shows the comparison of the valence photoionization mass spectrum without pyrolysis (upper trace, a) and with optimized pyrolysis conditions (lower trace, b). The spectra were recorded at 21.2\,eV to confirm the pyrolytic conversion of the precursor according to scheme~\ref{scheme:pyrolysis}.
Both spectra were normalized to the mass signal of the carrier gas helium at \textit{m/z}~4.
Without pyrolysis the parent ion of the precursor, \ce{C3H3Br+}, is observed at \textit{m/z}~118 and 120 with a double peak structure originating from the two isotopes \textsuperscript{79}Br and \textsuperscript{81}Br. Dominant signals at \textit{m/z}~39, 38, and a weaker one at \textit{m/z}~37 result from the dissociative photoionisation (DPI) of the precursor to \ce{C3H3+}, \ce{C3H2+} and \ce{C3H+}. For \ce{C3H3+} an appearance energy of 10.88\,eV was determined previously,\cite{Holmes_1979} and also \ce{C3H+} was formerly identified as a DPI product of the precursor in spectra taken at 15.77\,eV.\cite{Schuessler_2005}
Measurements with pyrolysis were performed at a power of 35\,W, corresponding to a SiC capillary temperature of approximately 1200\,\textdegree C. The radical, which has a rich vibrational manifold of states with many low frequency vibrational modes, is assumed to be in the interaction region vibrationally cold, due to the efficient adiabatic expansion in the jet source.
The corresponding mass spectrum in Figure~\ref{fig:TOF_valence}b) shows new bromine signals from the pyrolysis and only a negligibly small \ce{C3H3Br+} mass signal of the unpyrolysed precursor. The peak at \textit{m/z}~39 is lower in intensity and narrower compared to the spectrum of the pure precursor and is likely to belong both to the pyrolytically generated propargyl radical \ch{C3H3^{.}} and to the DPI product of the unconverted precursor. The latter shows a broad pedestal below this peak, which results from the kinetic energy release in the DPI. Compared to the pure precursor spectrum the relative intensity of both \ce{C3H2+} and \ce{C3H+} to \ce{C3H3+} is increased. \ce{C3H2+} and \ce{C3H+} are known DPI products of the propargyl radical that are associated with appearance energies of 12.5--13\,eV and $>$15.5\,eV.\cite{Schuessler_2005}

\begin{figure}[hbpt!]
 \centering
\includegraphics[width=\linewidth]{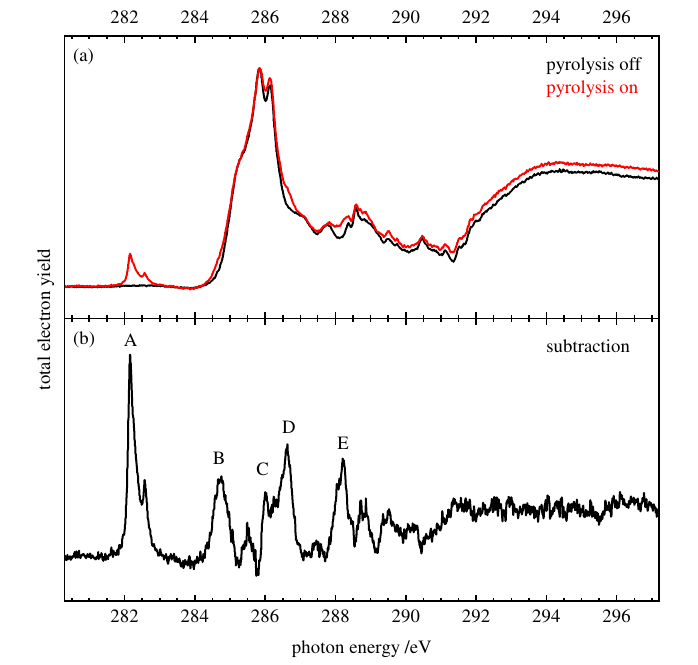}
    \caption{(a) C1s NEXAFS spectra of the precursor propargyl bromide without and with pyrolysis. (b) C1s NEXAFS spectrum of the propargyl radical obtained from subtraction.}
    \label{fig:NEXAFS_gapscan}
\end{figure}
NEXAFS spectra at the C1s edge were recorded as TEY with and without pyrolysis and are shown in the upper trace of Figure~\ref{fig:NEXAFS_gapscan}.
The spectrum of the precursor (black line) sets in above 284\,eV and shows an intense signal around 286\,eV with a double peak structure with a spacing of 300\,meV. At higher photon energies several less intense signals are observed, and the TEY increases above 291\,eV, when the C1s ionization thresholds are approached.
The absorption spectrum with pyrolysis (red line) corresponds to the spectrum of the precursor and the propargyl radical. A new band is found at 282.2\,eV and its shift to much lower photon energies compared to the pure precursor is a fingerprint of an open shell species. It is attributed to a transition from a C1s orbital into a singly occupied molecular orbital (SOMO) of the unpaired electron 
of the propargyl radical (section~\ref{sec:NEXAFS_theo}). 
The spectrum of the pure radical was obtained by subtraction of the spectra with and without pyrolysis where the onset and the maximum of the band at 286\,eV were used to scale the spectrum of the pure precursor. The resulting spectrum of the propargyl radical is shown in Figure~\ref{fig:NEXAFS_gapscan}(b). Band A that corresponds to the transition into the SOMO shows a vibrational progression of 420\,meV. 
A broad structureless feature is found at 284.8\,eV (B), which is red-shifted with respect to the main signal of the precursor. In contrast, signals at 286.0\,eV (C), 286.6\,eV (D) and 288.2\,eV (E) correspond to new signals at the maximum or the high energy flank of the precursor's main signal and are visible from Figure~\ref{fig:NEXAFS_gapscan}(a). 


\begin{figure}[hbpt!]
\centering
\includegraphics[width=\linewidth]{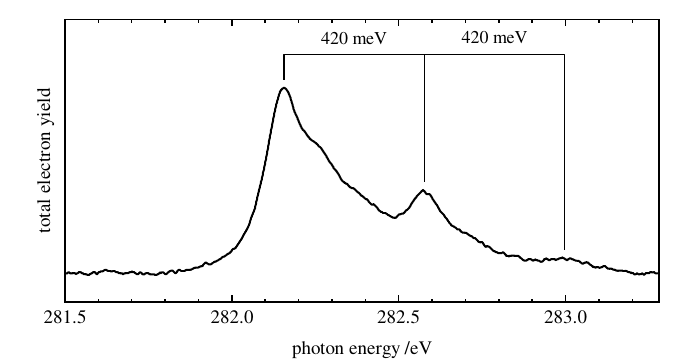}
    \caption{High resolution NEXAFS spectrum of the C1s to SOMO transition (band A).}
    \label{fig:NEXAFS_r0_scan}
\end{figure}

To investigate the vibrational structure of the lowest band in the propargyl spectrum, band A was scanned with a smaller step size and is displayed in Figure~\ref{fig:NEXAFS_r0_scan}. A progression with three components was identified. 
The spacing of 420\,meV suggests the excitation of 
a C--H stretching vibration upon the C1s to SOMO excitation. 
The first component at 282.16\,eV (and possibly also the second one at 282.58\,eV) also reveals a high-energy shoulder that may belong to a simultaneously excited bending vibration. 
To understand how electronic and/or vibrational excitations influence the appearance of this first band (A), a theoretical assignment of the spectrum is indispensable.

\subsection{Computational results and spectral assignment}
\label{sec:NEXAFS_theo}

\begin{figure}
\centering
\includegraphics[width=\linewidth]{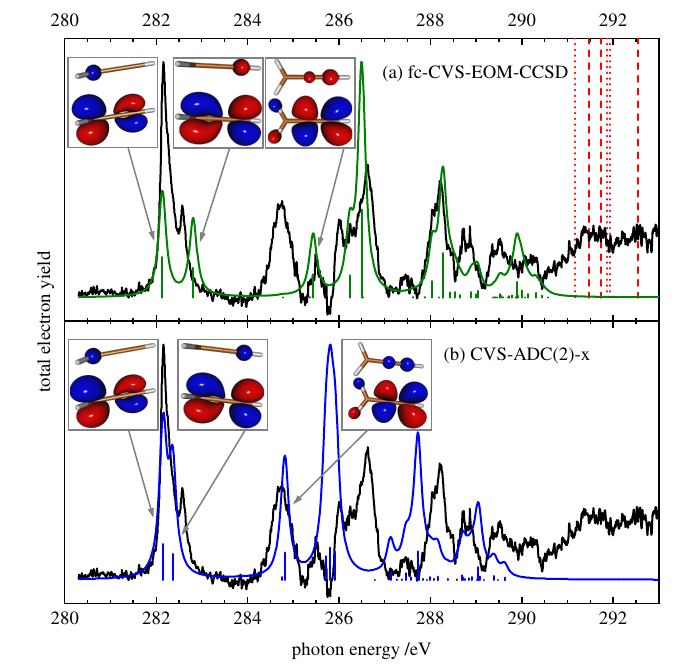}
\caption{Comparison between experimental NEXAFS spectrum (black line) and computed (a) fc-CVS-EOM-CCSD/aug-cc-pVTZ and (b) CVS-ADC(2)-x/aug-cc-pVTZ NEXAFS spectra of the propargyl radical using UHF reference orbitals. The sticks were convoluted with a Lorentzian lineshape function with a FWHM of 200\,meV. The vertical red dashed lines are the fc-CVS-EOM-IP-CCSD ionization potentials to singlet ionic states and the dotted ones to triplet ionic states.}
    \label{fig:calc:spectra}
\end{figure}

For the computational discussion, we denote the CH$_2$ carbon of the propargyl radical as C1 and the CH as C3, with the carbon in between as C2. The bands in Figure~\ref{fig:NEXAFS_gapscan} are assigned based on \textit{ab initio} calculations at the fc-CVS-EOM-CCSD and CVS-ADC(2)-x levels of theory using the aug-cc-pVTZ basis set. All computed excitation energies and their associated oscillator strengths are summarized 
in Table~S3
of the SI.
Additional fc-CVS-EOM-CCSD results obtained with other basis sets are collected in
Table~S4.
NTOs are given in 
Tables~S5 and~S6, 
for fc-CVS-EOM-CCSD and CVS-ADC(2)-x, respectively. The corresponding simulated spectra are shown in 
Figures~S2 and~S3. 
A comparison of the simulated spectra in 
Figure~S2 
shows that a change in basis set leads to a shift in the spectrum when compared to the experimental results and the splitting of the first two transitions narrows when changing from an open-shell restricted to an unrestricted Hartree-Fock reference. Overall, the fc-CVS-EOM-CCSD spectral profiles are rather similar. 

Comparing the unrestricted fc-CVS-EOM-CCSD/aug-cc-pVTZ spectrum to the CVS-ADC(2)-x/aug-cc-pVTZ one in Figure~\ref{fig:calc:spectra}, we note that the energy separation of the first two transitions narrows further at the CVS-ADC(2)-x level, and the third transition shifts in position and increases in intensity, overlapping with the experimental band B at 284.8\,eV. 


Based on the computed spectral data and NTOs, band A at 282.2\,eV is assigned to a combination of the 
transitions to the 
1~$^2$A$_1$ state \mbox{(C1 1s~$\to$~SOMO)}
and to the 
2~$^2$A$_1$ state 
\mbox{(C3 1s~$\to$~SOMO)}.
These are located at 282.15\,eV and 282.36\,eV, respectively, by CVS-ADC(2)-x, and at 282.13\,eV and 282.81\,eV by fc-CVS-EOM-CCSD. 
We note that the SOMO particle NTO is of (ethynyl methyl) p(C1)-$\pi$(C2-C3)  character for the 
transition to the 1 $^2$A$_1$ state, and of (allenyl) $\pi$(C1-C2)-p(C3) character for the transition to the 2 $^2$A$_1$ state.
This is 
suggestive of 
the core excitation taking place from either an ethynyl methyl  or an allenyl resonance structure (compare Scheme~\ref{scheme:pyrolysis}).

The experimental band B at 284.8\,eV overlaps with the CVS-ADC(2)-x 
transition to the 2 $^2$A$_2$ state at 284.82\,eV, which consists of a core excitation from the 1s orbitals of C2 and C3 to an in-plane $\pi^*$ orbital (suggesting the presence of a triple bond between C2 and C3, as expected in the nominal ethynyl methyl resonance structure). 
The third transition in the fc-CVS-EOM-CCSD calculation is less intense and located at higher energy than the experimental band B (overlapping with the less intense peak at its right) but has the same character as the CVS-ADC(2)-x one.

Moving towards the region of bands C--D of the experimental spectrum (286.0--286.6\,eV), we note that CVS-ADC(2)-x yields three dipole allowed transitions with sizable intensity, whereas fc-CVS-EOM-CCSD gives only two. 
In both cases, our computed transitions are slightly too closely spaced, yielding a narrower and more intense band when convoluted. 
Above 285.5\,eV, the CVS-ADC(2)-x transitions are also more red-shifted than the fc-CVS-EOM-CCSD ones, and do not overlap well with the experimental bands.
Nonetheless, the C--D region of the experimental spectrum is dominated, according to our calculations with both methods, by transitions from the C2 1s orbital (with some mixing of the C3 one) to either the in-plane or the out-of-plane $\pi$* 
particle NTOs 
(shown in Table~S5 and S6).

As for the band E just above 288\,eV in the experimental spectrum, several transitions, some with rather low intensity, contribute. At fc-CVS-EOM-CCSD level, two transitions are most intense, with the strongest one at 288.28\,eV due to a C2 1s $\to$ (out-of-plane) $\pi^*$.
Similarly, several transitions contribute to the band E at CVS-ADC(2)-x level, with the most intense one at 287.73\,eV (again red-shifted compared to experiment), and the same character as the intense E transition from fc-CVS-EOM-CCSD. 

We also computed core ionization energies for all three carbon atoms by the fc-CVS-EOM-IP-CCSD approach, summarized in 
Table~S2 
and given as red vertical lines in Figure~\ref{fig:calc:spectra}.
For the formation of a singlet cation, the lowest ionization energy of 291.5\,eV was found for the acetylenic carbon (C3), and the value of the central carbon is only slightly higher by 0.1 to 0.2\,eV.
For the C1 carbon, the highest value of 292.6\,eV was found. The ionization energy value for the formation of a triplet cation is found to be higher than the one of the singlet state for C2, while the reverse is observed for the terminal carbon atoms.




\begin{figure}
\centering
\includegraphics[width=\linewidth,trim=0cm 0.1cm 0cm 0.2cm, clip]
{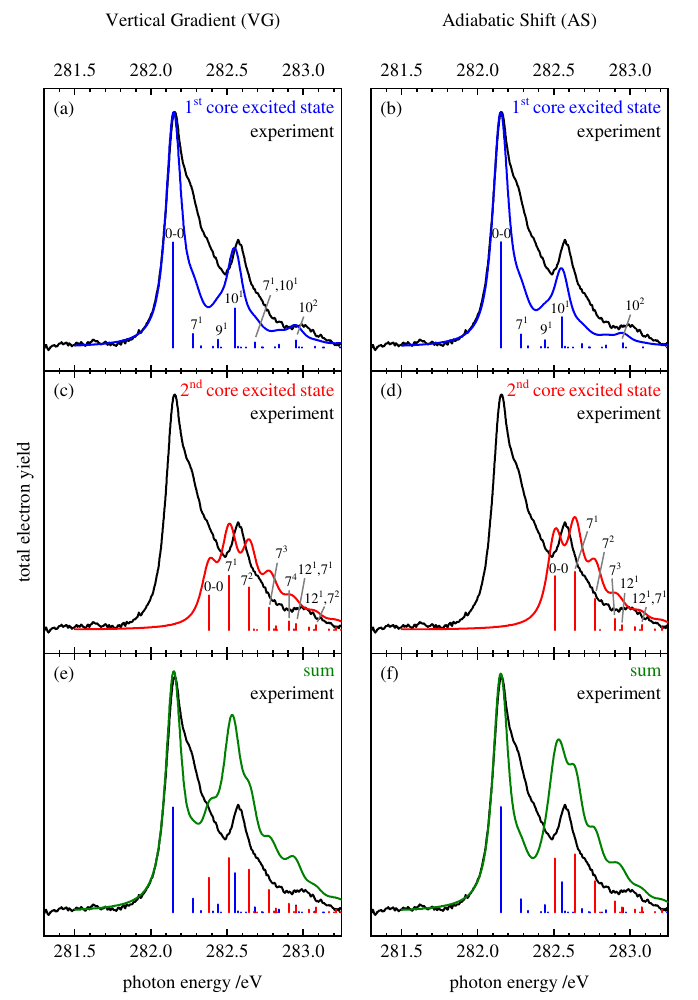}
\caption{Comparison between the high-resolution experimental NEXAFS spectrum (black line) and vibrational structure of the (a, b) first, (c, d) second and (e,f) sum of both first and second CVS-ADC(2)-x electronic transitions 
    at 0\,K, with PES obtained from the Vertical Gradient (left) and Adiabatic Shift (right) approximations. Vertical lines show the vibronic stick transitions from the vibrational ground state in the initial electronic state obtained using the time-independent formalism.~\cite{avilaferrer2012verticaladiabatic} 
    The labels correspond to the vibrational state in the final electronic states, represented as $X^v$, where $X$ is the normal mode index and $v$ is the number of quanta.
    All VG results have been shifted by +175\,meV and all AS ones by +318\,meV to align with the first experimental peak.
    The spectra have been scaled such that the intensity of the 0-0 band of the first electronic transition overlaps with experiment.}

\label{fig:vibrational}
\end{figure}

To conclude this section, we return to the first electronic transitions associated to band A, and discuss the results of our calculations of its vibrational structure. The Wigner samplings shown in 
Figure~S5 
indicate a vibrational splitting of the A band. 
This splitting becomes even more evident in the vibrationally resolved spectra that were obtained from the 
TI calculations based on the VG and AS potential energy surfaces and that are shown in Figure~\ref{fig:vibrational} for CVS-ADC(2)-x. 
The TD calculations for both fc-CVS-EOM-CCSD and CVS-ADC(2)-x are presented in 
Figure~S4.


Figure~\ref{fig:vibrational}(a) and (b) show that the first electronic transition has a vibrational splitting (three bands) consistent with the experimental results. Assignment of which vibronic transitions contribute to each vibrationally resolved band can be done based on the stick transitions from the TI results (with labels $X^v$, where $X$ indicates the mode involved and $v$ the amount of quanta in the mode). 
The most intense band at 282.16\,eV is dominated by the 0-0 transition, with a smaller contribution from the transition from the $v=0$ level of the electronic ground state to the 7$^{1}$ vibrational state of the core excited state. The second band of the progression at 282.58\,eV is dominated by the 10$^{1}$ vibrational state of the core excited state, with a smaller contribution from 9$^{1}$. The third band is mainly 10$^2$, i.e., a two-quanta vibrational state transition of mode 10.
As summarized in 
Table~S7 
(and Figure~S8), 
normal mode 10 is the symmetric \ce{CH2} stretching (of C1), mode 7 is dominated by the C1--C2 stretching and mode 9 by the C2--C3 stretching. 
Inclusion of temperature effects in the TD treatment does not seem to have any visible effect, 
see Figures~S6-S7, 
which is not surprising giving the level of approximation used for the vibrational calculations.

The vibrationally resolved spectrum of the second electronic transition shown in Figure~\ref{fig:vibrational}(c) and (d) is 
an asymmetric band that overlaps with the 10$^1$ band of the first electronic transition. The band is not peaked at the 0-0 transition, but rather at the 7$^1$ transition, and with its additional quanta $v=2, 3, 4$ and 5 (together with some excitation of mode 12, which is the C3--H stretch) contributing to its long tail. 
The sum of the spectra of both the first and second electronic state in Figure~\ref{fig:vibrational}(e) and (f) indicates that the overall intensity of the band at 282.58\,eV is overestimated.
According to our calculations, the ratio of the oscillator strengths between the first and second excited states is 0.58:0.42. This shows a similar trend for the two nominal resonance structures as the 0.61:0.39 ratio found in the recent study by \citeauthor{Changala_2024}\cite{Changala_2024}
%
The dominant excitation of vibrational mode~10 (symmetric \ce{CH2} stretch of C1) and of mode~7 (C1--C2 stretch) upon transition into the first and second core excited state, respectively, can be related to the change in the molecular geometry upon the C1s excitation. Comparing the structure of the ground state with those of the lowest core excited states 
(given in Table~S8), 
we note that for the first electronic transition the strongest geometry change is the reduction of the C1--H bond length, and of the C1--C2 bond length upon excitation of the second electronic transition. 

From these results, we conclude that the splitting of band A in the experimental spectrum mainly results from the vibrational structure of the first electronic transition (from the ethynyl methyl radical form), specifically the 0-0 transition and the symmetric stretching of the two C1--H bonds, possibly with contributions from the vibrational structure of the second electronic transition 
(allenyl radical form), due to 0-0 and 7$^v$ transitions ($v=1-5$),
with some participation of mode 12, too. 

\subsection{Comparison of the propargyl and allyl radical}
As both propargyl (\ch{C3H3^{.}}) and allyl (\ch{C3H5^{.}}) are resonance-stabilized hydrocarbon radicals, it is worth to briefly compare their NEXAFS spectra. Both radicals are characterized by a terminal C1s $\to$ SOMO transition at around 282\,eV. The presence of a transition shifted to lower photon energies compared to closed-shell molecules is a characteristic of radicals and has also been observed for the \textit{tert}-butyl radical.\cite{Schaffner_2024} In fact, also the shape of band A at 282.2\,eV resembles one of the lowest-energy core excitation of the allyl radical, which was observed at 281.99\,eV,\cite{Alagia_Allyl} and shows a progression that was assigned to a combination of a C--H stretch and an out-of-plane bending mode. However, as the terminal carbon atoms are equivalent in allyl (two equivalent resonance structures), the band is somewhat simpler, as it can be described by a single electronic transition  
(whereas two electronic transitions contribute to band A of propargyl). 
In addition, its NEXAFS spectrum shows fewer bands. However, transitions B--E in Figure~\ref{fig:NEXAFS_gapscan} can be related to the central and terminal C1s to LUMO transitions in allyl that were observed at 285.27\,eV, 285.71\,eV and 287.50\,eV.\cite{Alagia_Allyl}

\subsection{Fragmentation after core excitation}
\begin{figure}[hbpt!]
    \centering
\includegraphics[width=\linewidth]{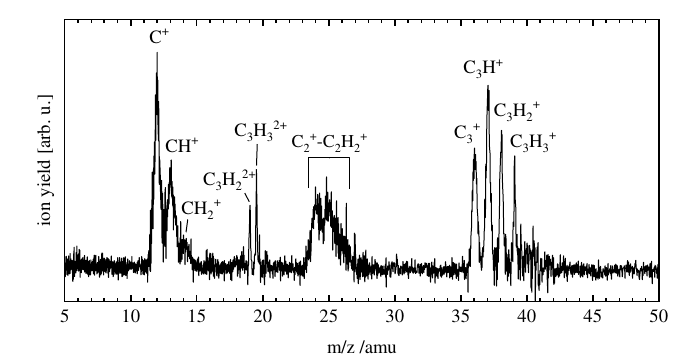}
    \caption{TOF mass spectrum of single ion events upon excitation of the propargyl radical at band A at 282.2\,eV (C1s~$\rightarrow$~SOMO).}
    \label{fig:TOF_Singles_radicalpeak}
\end{figure}

After the resonant core excitation, the core excited states decay via Auger-Meitner processes. The resulting ions were detected by TOF mass spectrometry, and the coincidence setup was used to differentiate between ions from single, double or triple ion events. The fragmentation of the propargyl radical after C1s $\rightarrow$ SOMO excitation was investigated at 282.2\,eV (band A). Figure~\ref{fig:TOF_Singles_radicalpeak} shows the corresponding mass spectrum for single ion events that was corrected for contributions from off-resonant excitation and unpyrolyzed precursor contributions.
The observed fragments and their branching ratios are summarized in Table~\ref{tab:br_singles}.

The parent ion \ce{C3H3+} is observed as a sharp signal at \textit{m/z}~39. Signals at \textit{m/z}~38, 37 and 36 result from a loss of neutral hydrogens.
The signals at \textit{m/z}~19.5 and 19 belong to the dications \ce{C3H3}\textsuperscript{2+} and \ce{C3H2}\textsuperscript{2+} that are produced by double Auger decay.
Most of the fragmentations include the cleavage of one (27\%) or both (34\%) carbon-carbon bonds. Mass signals for \ce{C2H2+} to \ce{C2+} and \ce{CH2+} to \ce{C+} are broader because of the kinetic energy released in the fragmentation.
The absence of \ce{C1} or \ce{C2} fragments carrying three hydrogens in the mass spectrum leads us to conclude that no rearrangement of the hydrogen atoms to the higher-energy isomer 1-propenyl cation, [H\textsubscript{3}C–C–C]\textsuperscript{+},\cite{Reinwardt_2024} 
takes place in the cationic state. For this isomer the detection of \ce{C2H3+} or \ce{CH3+} would have been expected. However, we cannot exclude an isomerisation to the lower-energy cyclopropenyl, \ce{C3H3+}, on the cationic surface based on our mass spectrometric data.

We also investigated excitations into higher core excited states at 286.0\,eV (band C), 286.6\,eV (band D) and 288.1\,eV (band E). The corresponding mass spectra and branching ratios can be found in 
Section~S1 in the SI. Compared to the C1s~$\rightarrow$~SOMO transition these excitations show a stronger fragmentation tendency with less \ce{C3} fragments. This is like our previous observations for the \textit{tert}-butyl radical and was explained by a preference for spectator decay over participator decay for higher excitation energies.\cite{Schaffner_2024}

\begin{table}
\caption{\label{tab:br_singles}Branching ratios of single ion events after excitation of the propargyl radical at 282.2\,eV (band A). Relative intensities were determined from the mass spectrum (Figure~\ref{fig:TOF_Singles_radicalpeak}) by integration. Integrals of overlapping masses were derived from fitting with Gauss functions. The branching ratios were normalized so that all channels sum up to 100\%. Very low branching ratios are not specified and are summarized in ``others''. The error of each branching ratio is estimated to be around $\pm$20\% of the respective branching ratio.}
\begin{ruledtabular}
\begin{tabular}{llc}
\textit{m/z}&species&branching ratio /\%\\
\hline
39&\ce{C3H3+}   &4\\
38&\ce{C3H2+}   &8\\
37&\ce{C3H+}    &13\\
36&\ce{C3+}     &10\\
26&\ce{C2H2+}   &6\\
25&\ce{C2H+}    &12\\
24&\ce{C2+}     &9\\
19.5&\ce{C3H3}\textsuperscript{2+}&2\\
19&\ce{C3H2}\textsuperscript{2+}&--\\
14&\ce{CH2+}    &4\\
13&\ce{CH+}     &13\\
12&\ce{C+}      &17\\
6&C\textsuperscript{2+}&--\\
others& &2\\
\end{tabular}
\end{ruledtabular}
\end{table}


\section{\label{sec:summary} Conclusion}
The NEXAFS spectrum of the propargyl radical at the C1s edge has been recorded using synchrotron radiation. The radical was generated by pyrolysis of propargyl bromide. Although the radical is converted cleanly, conversion of the precursor was not complete and the spectrum of propargyl bromide had to be subtracted. Upon pyrolysis, a new feature emerges at 282.2\,eV that is due to C1s~$\to$~SOMO transitions. 
By comparison with CVS-ADC(2)-x and fc-CVS-EOM-CCSD computations, 
it was found that two core-excited states ($1~^2$A$_1$ and $2~^2$A$_1$) contribute to this band, the first originating from the excitation of a 1s electron from the terminal C1 atom into a SOMO of ethyl allenyl character, and the second from the excitation of a 1s electron from the terminal C3 atom into a SOMO of allenyl character. 
Our results are  consistent with those of a recent study using hyperfine-resolved microwave spectroscopy and accurate \textit{ab initio} computations,~\cite{Changala_2024}
indicating that propargyl possesses a $\pi$-delocalized resonance structure of intermediate ethynyl methyl and allenyl character, with non-equal 0.61:0.39 balance between the relative weights of the two resonance structures.

This first band at 282.2\,eV exhibits a vibrational progression with a spacing of 420\,meV. By means of vertical and adiabatic harmonic calculations, this is mainly assigned to the symmetric \ce{CH2} stretching mode of the first core-excited state (C1 1s~$\to$~SOMO of ethynyl methyl), possibly with contributions from the vibrational structure of the second core excited state (C3 1s~$\to$~SOMO of allenyl).  
In addition, several bands at higher photon energy can be assigned to transitions from C2 and C3 into the $\pi^*$ orbitals. 
%
%
Overall, the appearance of the spectrum resembles the one of the allyl radical, 
\ch{C3H5^{.}}.  

{Core ionization energies between 291.5 eV and 292.6\,eV for ionization into the singlet cation were computed for the different carbon atoms. Compared to the singlet cation, the ionization energy into the triplet cation is lower for the two terminal C atoms, but is slightly higher for the central one.} 

The fragmentation pattern upon \mbox{(terminal) C1s $\rightarrow$ SOMO} excitation shows peaks of the molecular cation and dication, but also a broad distribution of fragments including loss of one to three H-atoms, 
and cleavage of one or both C--C bonds. Interestingly, no CH$_3^+$ is observed, indicating the absence of isomerization to propenyl. 
In comparison, excitation to higher core-excited states is associated with a more pronounced fragmentation dominated by the almost complete disappearance of the molecular cation and an increasing dominance of C$_1^+$ fragment ions. 

\section{\label{sec:SI} Supplementary Material}
See supplementary information for mass spectra at higher photon energies, additional computational details and 
results.

\section{\label{sec:acknowledgement} Acknowledgments}
The work has been financially supported by the German Science Foundation (DFG), contract FI 575/13-2. D.S. acknowledges a fellowship by the Fonds der Chemischen Industrie (FCI).
We acknowledge invaluable discussions with Dr. Fabrizio Santoro and Dr. Iulia Emilia Brumboiu.
J.P. acknowledges financial support from the Technical University
of Denmark within the Ph.D. Alliance Programme. S.C. thanks
Hamburg’s Cluster of Excellence
``CUI: Advanced Imaging of
Matter'' for the 2024 Mildred Dresselhaus Prize.
S.C. and T.J.v.B. acknowledge funding
from the Novo Nordisk Foundation (NNF Grant No. NNFSA220080996).
S.S. and A.B. acknowledge financial support from the University of Rome ``La Sapienza'', project Ateneo 2024 RG1241909C0A0A47, and a PhD fellowship, respectively. 

\section*{Data Availability Statement}
The data are available from the authors upon reasonable request. 

\section{\label{sec:contributions} Author Contributions}
D.S., A.R., E.K., V.v.L., A.B., M.A., S.S., and I.F. designed and conducted the
experiments. T.J.v.B., J.P. and S.C. designed the computational protocols, carried out the 
calculations
and characterized the NEXAFS spectra based on the \textit{ab initio} results.
D.S. analyzed the experimental data and wrote the 
first draft of the paper;  
T.J.v.B. and S.C. wrote
the parts concerning the 
computational results.
All authors discussed the interpretation of the results and contributed to the final version of the paper.

\section*{References}

\bibliography{aipsamp}

@article{stephens1994ab,
  title={Ab initio calculation of vibrational absorption and circular dichroism spectra using density functional force fields},
  author={Stephens, Philip J and Devlin, Frank J and Chabalowski, Cary F and Frisch, Michael J},
  journal={J. Phys. Chem.},
  volume={98},
  number={45},
  pages={11623--11627},
  year={1994},
  publisher={ACS Publications},
  doi={https://doi.org/10.1021/j100096a001}
}

@article{macak2000simulations,
  title={Simulations of vibronic profiles in two-photon absorption},
  author={Macak, Peter and Luo, Yi and {\AA}gren, Hans},
  journal={Chem. Phys. Lett.},
  volume={330},
  number={3-4},
  pages={447--456},
  year={2000},
  publisher={Elsevier},
  doi={10.1016/S0009-2614(00)01096-4},
}

@incollection{biczysko2012time,
  title={Time-independent approaches to simulate electronic spectra lineshapes: From small molecules to macrosystems},
  author={Biczysko, Malgorzata and Bloino, Julien and Santoro, Fabrizio and Barone, Vincenzo},
  booktitle={Computational strategies for spectroscopy},
  pages={361--443},
  year={2012},
  publisher={Wiley Online Library}
}

@Article{avilaferrer2012verticaladiabatic,
author ="Avila Ferrer, Francisco José and Santoro, Fabrizio",
title  ="Comparison of vertical and adiabatic harmonic approaches for the calculation of the vibrational structure of electronic spectra",
journal  ="Phys. Chem. Chem. Phys.",
year  ="2012",
volume  ="14",
issue  ="39",
pages  ="13549-13563",
publisher  ="The Royal Society of Chemistry",
doi  ="10.1039/C2CP41169E",
url  ="http://dx.doi.org/10.1039/C2CP41169E"}

@article{crespo2012spectrum,
  title={Spectrum simulation and decomposition with nuclear ensemble: formal derivation and application to benzene, furan and 2-phenylfuran},
  author={Crespo-Otero, Rachel and Barbatti, Mario},
  journal={Theor. Chem. Acc.},
  pages={1237},
  year={2012},
  volume={131},
  publisher={Springer},
  doi={10.1007/s00214-012-1237-4},
}

@article{dreuw2015algebraic,
  title={The algebraic diagrammatic construction scheme for the polarization propagator for the calculation of excited states},
  author={Dreuw, Andreas and Wormit, Michael},
  journal={WIREs Comput. Mol. Sci.},
  volume={5},
  number={1},
  pages={82--95},
  year={2015},
  publisher={Wiley Online Library},
  doi={10.1002/wcms.1206},
}

@article{wang2016geometry,
  title={Geometry optimization made simple with translation and rotation coordinates},
  author={Wang, Lee-Ping and Song, Chenchen},
  journal={J. Chem. Phys.},
  volume={144},
  number={21},
  year={2016},
  pages={214108},
  doi={10.1063/1.4952956},
  publisher={AIP Publishing}
}

@article{cerezo2023fcclasses3,
  title={\textit{FCclasses3}: Vibrationally-resolved spectra simulated at the edge of the harmonic approximation},
  author={Cerezo, Javier and Santoro, Fabrizio},
  journal={J. Comput. Chem.},
  volume={44},
  number={4},
  pages={626--643},
  year={2023},
  publisher={Wiley Online Library},
doi={10.1002/jcc.27027},
}

@article{pyscf,
  title={\texttt{PySCF}: the {P}ython-based simulations of chemistry framework},
  author={Sun, Qiming and Berkelbach, Timothy C. and Blunt, Nick S. and Booth, George H. and Guo, Sheng and Li, Zhendong and Liu, Junzi and McClain, James D. and Sayfutyarova, Elvira R. and Sharma, Sandeep and Wouters, Sebastian and Chan, Garnet Kin-Lic},
  journal={WIRES Comput. Mol. Sci.},
  volume={8},
  number={1},
  pages={e1340},
  year={2018},
  publisher={Wiley Online Library},
}

@article{pyscf_2,
  title={Recent developments in the \texttt{PySCF} program package},
  author={Sun, Qiming and Zhang, Xing and Banerjee, Samragni and Bao, Peng and Barbry, Marc and Blunt, Nick S. and Bogdanov, Nikolay A. and Booth, George H. and Chen, Jia and Cui, Zhi-Hao and Eriksen, Janus J. and Gao, Yang and Guo, Sheng and Hermann, Jan and Hermes, Matthew R. and Koh, Kevin and Koval, Peter and Lehtola, Susi and Li, Zhendong and Liu, Junzi and Mardirossian, Narbe and McClain, James D. and Motta, Mario and Mussard, Bastien and Pham, Hung Q. and Pulkin, Artem and Purwanto, Wirawan and Robinson, Paul J. and Ronca, Enrico and Sayfutyarova, Elvira R. and Scheurer, Maximilian and Schurkus, Henry F. and Smith, James E. T. and Sun, Chong and Sun, Shi-Ning and Upadhyay, Shiv and Wagner, Lucas K. and Wang, Xiao and White, Alec and Whitfield, James Daniel and Williamson, Mark J. and Wouters, Sebastian and Yang, Jun and Yu, Jason M. and Zhu, Tianyu and Berkelbach, Timothy C. and Sharma, Sandeep and Sokolov, Alexander Yu. and Chan, Garnet Kin-Lic},
  journal={J. Chem. Phys.},
  volume={153},
  number={2},
  year={2020},
  doi={10.1063/5.0006074},
  pages={024109},
  publisher={AIP Publishing}
}

@software{cvs-adc-grad,
  author = {Iulia Emilia Brumboiu},
  title = {\texttt{cvs-adc-grad}},
  version = {0.0.0},
  date = {2025-05-21},
  year = {2025},
  howpublished = {\url{https://gitlab.com/iubr/cvs-adc-grad}},
}

@article{herbst2020adcc,
  title={\texttt{adcc}: {A} versatile toolkit for rapid development of algebraic-diagrammatic construction methods},
  author={Herbst, Michael F and Scheurer, Maximilian and Fransson, Thomas and Rehn, Dirk R and Dreuw, Andreas},
  journal={WIREs Comput. Mol. Sci.},
  volume={10},
  number={6},
  pages={e1462},
  year={2020},
  publisher={Wiley Online Library},
  doi={10.1002/wcms.1462},
}

@misc{frisch2016gaussian,
  title={Gaussian 16},
  author={Frisch, M.J. and Trucks, G. W. and Schlegel, H Bernhard and Scuseria, G. E. and Robb, M. and Cheeseman, J. R. and Scalmani, G and Barone, VPGA and Petersson, G. A. and Nakatsuji, HJRA and others},
  year={2016},
  publisher={Gaussian, Inc. Wallingford, CT}
}

@article{clark1983a,
    author = {Clark, Timothy and Chandrasekhar, Jayaraman and Spitznagel, Günther W. and Schleyer, Paul Von Ragué},
    title = {Efficient diffuse function-augmented basis sets for anion calculations. {III}. {T}he {3-21+G} basis set for first-row elements, {Li-F}},
    journal = {J. Comput. Chem.},
    volume = {4},
    pages = {294-301},
    year = {1983},
    doi = {10.1002/jcc.540040303}
}

@article{krishnan1980a,
    author = {Krishnan, R. and Binkley, J. S. and Seeger, R. and Pople, J. A.},
    title = {Self-consistent molecular orbital methods. {XX}. {A} basis set for correlated wave functions},
    journal = {J. Chem. Phys.},
    volume = {72},
    pages = {650-654},
    year = {1980},
    doi = {10.1063/1.438955}
}

@article{dunning1989a,
    author = {Dunning, Thom H.},
    title = {Gaussian basis sets for use in correlated molecular calculations. {I}. {T}he atoms boron through neon and hydrogen},
    journal = {J. Chem. Phys.},
    volume = {90},
    pages = {1007-1023},
    year = {1989},
    doi = {10.1063/1.456153}
}

@article{kendall1992a,
    author = {Kendall, Rick A. and Dunning, Thom H. and Harrison, Robert J.},
    title = {Electron affinities of the first-row atoms revisited. Systematic basis sets and wave functions},
    journal = {J. Chem. Phys.},
    volume = {96},
    pages = {6796-6806},
    year = {1992},
    doi = {10.1063/1.462569}
}

@article{Qchem541,
author = {Epifanovsky,Evgeny  and Gilbert,Andrew T. B.  and Feng,Xintian  and Lee,Joonho  and Mao,Yuezhi  and Mardirossian,Narbe  and Pokhilko,Pavel  and White,Alec F.  and Coons,Marc P.  and Dempwolff,Adrian L.  and Gan,Zhengting and Hait,Diptarka  and Horn,Paul R.  and Jacobson,Leif D.  and Kaliman,Ilya  and Kussmann,J{\"o}rg  and Lange,Adrian W.  and Lao,Ka Un  and Levine,Daniel S.  and Liu,Jie  and McKenzie,Simon C.  and Morrison,Adrian F.  and Nanda,Kaushik D.  and Plasser,Felix  and Rehn,Dirk R.  and Vidal,Marta L.  and You,Zhi-Qiang  and Zhu,Ying  and Alam,Bushra  and Albrecht,Benjamin J.  and Aldossary,Abdulrahman  and Alguire,Ethan  and Andersen,Josefine H.  and Athavale,Vishikh  and Barton,Dennis  and Begam,Khadiza  and Behn,Andrew  and Bellonzi,Nicole  and Bernard,Yves A.  and Berquist,Eric J.  and Burton,Hugh G. A.  and Carreras,Abel  and Carter-Fenk,Kevin  and Chakraborty,Romit  and Chien,Alan D.  and Closser,Kristina D.  and Cofer-Shabica,Vale  and Dasgupta,Saswata  and de Wergifosse,Marc  and Deng,Jia  and Diedenhofen,Michael  and Do,Hainam  and Ehlert,Sebastian  and Fang,Po-Tung  and Fatehi,Shervin  and Feng,Qingguo  and Friedhoff,Triet  and Gayvert,James  and Ge,Qinghui  and Gidofalvi,Gergely  and Goldey,Matthew  and Gomes,Joe  and Gonz{\'a}lez-Espinoza,Cristina E.  and Gulania,Sahil  and Gunina,Anastasia O.  and Hanson-Heine,Magnus W. D.  and Harbach,Phillip H. P.  and Hauser,Andreas  and Herbst,Michael F.  and Hernández Vera,Mario  and Hodecker,Manuel  and Holden,Zachary C.  and Houck,Shannon  and Huang,Xunkun  and Hui,Kerwin  and Huynh,Bang C.  and Ivanov,Maxim  and J{\'a}sz,\'Ad{\'a}m  and Ji,Hyunjun  and Jiang,Hanjie  and Kaduk,Benjamin  and K{\"a}hler,Sven  and Khistyaev,Kirill  and Kim,Jaehoon  and Kis,Gergely  and Klunzinger,Phil  and Koczor-Benda,Zsuzsanna  and Koh,Joong Hoon  and Kosenkov,Dimitri  and Koulias,Laura  and Kowalczyk,Tim  and Krauter,Caroline M.  and Kue,Karl  and Kunitsa,Alexander  and Kus,Thomas  and Ladj{\'a}nszki,Istv{\'a}n  and Landau,Arie  and Lawler,Keith V.  and Lefrancois,Daniel  and Lehtola,Susi  and Li,Run R.  and Li,Yi-Pei  and Liang,Jiashu  and Liebenthal,Marcus  and Lin,Hung-Hsuan  and Lin,You-Sheng  and Liu,Fenglai  and Liu,Kuan-Yu  and Loipersberger,Matthias  and Luenser,Arne  and Manjanath,Aaditya  and Manohar,Prashant  and Mansoor,Erum  and Manzer,Sam F.  and Mao,Shan-Ping  and Marenich,Aleksandr V.  and Markovich,Thomas  and Mason,Stephen  and Maurer,Simon A.  and McLaughlin,Peter F.  and Menger,Maximilian F. S. J.  and Mewes,Jan-Michael  and Mewes,Stefanie A.  and Morgante,Pierpaolo  and Mullinax,J. Wayne  and Oosterbaan,Katherine J.  and Paran,Garrette  and Paul,Alexander C.  and Paul,Suranjan K.  and Pavo{\v s}evi{\'c},Fabijan  and Pei,Zheng  and Prager,Stefan  and Proynov,Emil I.  and R{\'a}k,{\'A}d{\'a}m  and Ramos-Cordoba,Eloy  and Rana,Bhaskar  and Rask,Alan E.  and Rettig,Adam  and Richard,Ryan M.  and Rob,Fazle  and Rossomme,Elliot  and Scheele,Tarek  and Scheurer,Maximilian  and Schneider,Matthias  and Sergueev,Nickolai  and Sharada,Shaama M.  and Skomorowski,Wojciech  and Small,David W.  and Stein,Christopher J.  and Su,Yu-Chuan  and Sundstrom,Eric J.  and Tao,Zhen  and Thirman,Jonathan  and Tornai,G{\'a}bor J.  and Tsuchimochi,Takashi  and Tubman,Norm M.  and Veccham,Srimukh Prasad  and Vydrov,Oleg  and Wenzel,Jan  and Witte,Jon  and Yamada,Atsushi  and Yao,Kun  and Yeganeh,Sina  and Yost,Shane R.  and Zech,Alexander  and Zhang,Igor Ying  and Zhang,Xing  and Zhang,Yu  and Zuev,Dmitry  and Aspuru-Guzik,Alán  and Bell,Alexis T.  and Besley,Nicholas A.  and Bravaya,Ksenia B.  and Brooks,Bernard R.  and Casanova,David  and Chai,Jeng-Da  and Coriani,Sonia  and Cramer,Christopher J.  and Cserey,Gy{\"o}rgy  and DePrince,A. Eugene and DiStasio,Robert A.  and Dreuw,Andreas and Dunietz,Barry D.  and Furlani,Thomas R. and Goddard,William A. and Hammes-Schiffer,Sharon  and Head-Gordon,Teresa  and Hehre,Warren J.  and Hsu,Chao-Ping  and Jagau,Thomas-C.  and Jung,Yousung  and Klamt,Andreas  and Kong,Jing  and Lambrecht,Daniel S.  and Liang,WanZhen and Mayhall,Nicholas J.  and McCurdy,C. William and Neaton,Jeffrey B.  and Ochsenfeld,Christian and Parkhill,John A.  and Peverati,Roberto and Rassolov,Vitaly A.  and Shao,Yihan  and Slipchenko,Lyudmila V.  and Stauch,Tim  and Steele,Ryan P.  and Subotnik,Joseph E.  and Thom,Alex J. W.  and Tkatchenko,Alexandre  and Truhlar,Donald G. and Van Voorhis,Troy and Wesolowski,Tomasz A.  and Whaley,K. Birgitta  and Woodcock,H. Lee  and Zimmerman,Paul M.  and Faraji,Shirin  and Gill,Peter M. W.  and Head-Gordon,Martin  and Herbert,John M.  and Krylov,Anna I. },
title = {Software for the frontiers of quantum chemistry: {An} overview of developments in the {Q-Chem} 5 package},
journal = {J. Chem. Phys.},
volume = {155},
number = {8},
pages = {084801},
year = {2021},
doi = {10.1063/5.0055522},
URL = { 
        https://doi.org/10.1063/5.0055522
},
}

@article{Vidal2019fcCVS,
author = "Marta L. Vidal and Xintian Feng and Evgeny Epifanovsky and Anna I. Krylov and Sonia Coriani",
title = "{A New and Efficient Equation-of-Motion Coupled-Cluster Framework for Core-Excited and Core-Ionized States}",
year = "2019",
journal="J. Chem. Theory Comput.",
volume={15}, 
pages={3117--3133},
doi = {10.1021/acs.jctc.9b00039},
}

@article{botschwina2010calculated,
  title={Calculated photoelectron spectra of isotopomers of the propargyl radical (\ce{H2C3H}): {A}n explicitly correlated coupled cluster study},
  author={Botschwina, Peter and Oswald, Rainer},
  journal={Chem. Phys.},
  volume={378},
  number={1-3},
  pages={4--10},
  year={2010},
  publisher={Elsevier},
  doi = {https://doi.org/10.1016/j.chemphys.2010.07.030}
}

@article{krylov2020orbitals,
    author = {Krylov, Anna I.},
    title = "{From orbitals to observables and back}",
    journal = {J. Chem. Phys.},
    volume = {153},
    number = {8},
    pages = {080901},
    year = {2020},
    month = {08},
    issn = {0021-9606},
    doi = {10.1063/5.0018597},
    OPTurl = {https://doi.org/10.1063/5.0018597},
    OPTeprint = {https://pubs.aip.org/aip/jcp/article-pdf/doi/10.1063/5.0018597/15577789/080901\_1\_online.pdf},
}

@misc{DTU-DCC,
    author    = {{DTU Computing Center}},
    title     = {{DTU Computing Center resources}},
    year      = {2024},
    publisher = {Technical University of Denmark},
    note      = {https://doi.org/10.48714/DTU.HPC.0001},
    doi       = {10.48714/DTU.HPC.0001},
    OPTurl       = {https://doi.org/10.48714/DTU.HPC.0001},
}

@article{peterson2008a,
    author = {Peterson, Kirk A. and Adler, Thomas B. and Werner, Hans-Joachim},
    title = {{Systematically convergent basis sets for explicitly correlated wavefunctions: The atoms H, He, B-Ne, and Al-Ar}},
    journal = {J. Chem. Phys.},
    volume = {128},
    pages = {084102},
    year = {2008},
    doi = {10.1063/1.2831537}
}

@article{adler2007simple,
  title={A simple and efficient {CCSD(T)-F12} approximation},
  author={Adler, Thomas B and Knizia, Gerald and Werner, Hans-Joachim},
  journal={J. Chem. Phys.},
  volume={127},
  number={22},
  year={2007},
  pages={221106},
  doi={10.1063/1.2817618},
  publisher={AIP Publishing}
}

@article{knizia2009simplified,
  title={Simplified {CCSD(T)-F12} methods: {T}heory and benchmarks},
  author={Knizia, Gerald and Adler, Thomas B and Werner, Hans-Joachim},
  journal={J. Chem. Phys.},
  volume={130},
  number={5},
  year={2009},
  publisher={AIP Publishing},
pages={054104}, 
doi={10.1063/1.3054300},
}

@article{marchetti2008accurate,
  title={Accurate calculations of intermolecular interaction energies using explicitly correlated wave functions},
  author={Marchetti, Oliver and Werner, Hans-Joachim},
  journal={Phys. Chem. Chem. Phys.},
  volume={10},
  number={23},
  pages={3400--3409},
  year={2008},
  doi={10.1021/jp9059467},
  publisher={Royal Society of Chemistry}
}

@article{Alagia_Allyl,
   author = {Alagia, M. and Bodo, E. and Decleva, P. and Falcinelli, S. and Ponzi, A. and Richter, R. and Stranges, S.},
   title = {The soft {X}-ray absorption spectrum of the allyl free radical},
   journal = {Phys. Chem. Chem. Phys.},
   volume = {15},
   number = {4},
   pages = {1310-1318},
   ISSN = {1463-9076},
   DOI = {10.1039/C2CP43466K},
   url = {http://dx.doi.org/10.1039/C2CP43466K},
   year = {2013},
   type = {Journal Article}
}

@article{Schaffner_2024,
   author = {Schaffner, Dorothee and Juncker von Buchwald, Theo and Karaev, Emil and Alagia, Michele and Richter, Robert and Stranges, Stefano and Coriani, Sonia and Fischer, Ingo},
   title = {The x-ray absorption spectrum of the tert-butyl radical: {A}n experimental and computational investigation},
   journal = {J. Chem. Phys.},
   volume = {161},
   number = {3},
   pages={034309},
   ISSN = {0021-9606},
   DOI = {10.1063/5.0216364},
   url = {https://doi.org/10.1063/5.0216364},
   year = {2024},
   type = {Journal Article}
}

@article{de_Simone_2022,
   author = {{de Simone}, M. and Coreno, M. and Alagia, M. and Richter, R. and Prince, K. C.},
   title = {Inner shell excitation spectroscopy of the tetrahedral molecules {CX}$_4$ ({X = H, F, Cl})},
   journal = {J. Phys. B: At. Mol. Opt. Phys.},
   volume = {35},
   number = {1},
   pages = {61},
   ISSN = {0953-4075},
   DOI = {10.1088/0953-4075/35/1/305},
   url = {https://dx.doi.org/10.1088/0953-4075/35/1/305},
   year = {2002},
   type = {Journal Article}
}

@article{Holmes_1979,
   author = {Holmes, John L. and Lossing, F. P.},
   title = {The reactivity of [\ce{C3H3+}] ions; a thermochemical study},
   journal = {Can. J.  Chem.},
   volume = {57},
   number = {2},
   pages = {249-252},
   OPTnote = {doi: 10.1139/v79-041},
   ISSN = {0008-4042},
   DOI = {10.1139/v79-041},
   url = {https://doi.org/10.1139/v79-041},
   year = {1979},
   type = {Journal Article}
}

@article{Schuessler_2005,
   author = {Sch{\"u}{\ss}ler, T. and Roth, W. and Gerber, T. and Alcaraz, C. and Fischer, I.},
   title = {The 
{VUV} photochemistry of radicals: \ce{C3H3} and \ce{C2H5}},
   journal = {Phys. Chem. Chem. Phys.},
   volume = {7},
   number = {5},
   pages = {819-825},
   ISSN = {1463-9076},
   DOI = {10.1039/B414163F},
   url = {http://dx.doi.org/10.1039/B414163F},
   year = {2005},
   type = {Journal Article}
}

@article{Blyth_1999,
   author = {Blyth, R. R. and Delaunay, R. and Zitnik, M. and Krempasky, J. and Krempaska, R. and Slezak, J. and Prince, K. C. and Richter, R. and Vondracek, M. and Camilloni, R. and Avaldi, L. and Coreno, M. and Stefani, G. and Furlani, C. and de Simone, M. and Stranges, S. and Adam, M. Y.},
   title = {The high resolution {G}as {P}hase {P}hotoemission beamline, {E}lettra},
   journal = {J. Electron Spectrosc. Relat. Phenom.},
   volume = {101-103},
   pages = {959-964},
   ISSN = {0368-2048},
   doi = {10.1016/S0368-2048(98)00381-8},
   url = {https://www.sciencedirect.com/science/article/pii/S0368204898003818},
   year = {1999},
   type = {Journal Article}
}

@article{Alagia_2003,
   author = {Alagia, M. and Avaldi, L. and Coreno, M. and Camilloni, R. and Furlani, C. and Prince, K. C. and Richter, R. and de Simone, M. and Stefani, G. and Stranges, S.},
   title = {The gas phase photoemission beamline at {E}lettra},
   journal = {Synchrotron Radiat. News},
   volume = {16},
   number = {2},
   pages = {19-27},
   ISSN = {0894-0886},
   doi = {10.1080/08940880308603010},
   url = {https://doi.org/10.1080/08940880308603010},
   year = {2003},
   type = {Journal Article}
}

@article{Alagia_Methyl,
   author = {Alagia, M. and Lavollée, M. and Richter, R. and Ekström, U. and Carravetta, V. and Stranges, D. and Brunetti, B. and Stranges, S.},
   title = {Probing the potential energy surface by high-resolution x-ray absorption spectroscopy: The umbrella motion of the core-excited \ce{CH3} free radical},
   journal = {Phys. Rev. A},
   volume = {76},
   number = {2},
   pages = {022509},
   OPTnote = {PRA},
   DOI = {10.1103/PhysRevA.76.022509},
   url = {https://link.aps.org/doi/10.1103/PhysRevA.76.022509},
   year = {2007},
   type = {Journal Article}
}

@article{Ekstrom_CD3_CH3,
   author = {Ekström, U. and Carravetta, V. and Alagia, M. and Lavollée, M. and Richter, R. and Bolcato, C. and Stranges, S.},
   title = {The umbrella motion of core-excited \ce{CH3} and \ce{CD3} methyl radicals},
   journal = {J. Chem. Phys.},
   volume = {128},
   number = {4},
   pages={044302},
   ISSN = {0021-9606},
   DOI = {10.1063/1.2822246},
   url = {https://doi.org/10.1063/1.2822246},
   year = {2008},
   type = {Journal Article}
}

@article{Agundez2021,
	author = {{Agúndez, M.} and {Cabezas, C.} and {Tercero, B.} and {Marcelino, N.} and {Gallego, J. D.} and {de Vicente, P.} and {Cernicharo, J.}},
	title = {{Discovery of the propargyl radical (CH$_2$CCH) in {TMC-1}: One of the most abundant radicals ever found and a key species for cyclization to benzene in cold dark clouds}},
	DOI= "10.1051/0004-6361/202140553",
	url= "https://doi.org/10.1051/0004-6361/202140553",
	journal = {Astron. Astrophys.},
	year = 2021,
	volume = 647,
	pages = "L10",
}

@article{Silva_2023,
   author = {Silva, W. G. D. P. and Cernicharo, J. and Schlemmer, S. and Marcelino, N. and Loison, J.-C. and Agúndez, M. and Gupta, D. and Wakelam, V. and Thorwirth, S. and Cabezas, C. and Tercero, B. and Doménech, J. L. and Fuentetaja, R. and Kim, W.-J. and de Vicente, P. and Asvany, O.},
   title = {{Discovery of H$_2$CCCH$^+$ in TMC-1}},
   journal = {Astron. Astrophys.},
   volume = {676},
   pages = {L1},
   url = {https://doi.org/10.1051/0004-6361/202347174},
   year = {2023},
   type = {Journal Article}
}

@article{Deyerl1999,
   author = {Deyerl, H.-J. and Fischer, I. and Chen, P.},
   title = {Photodissociation dynamics of the propargyl radical},
   journal = {J. Chem. Phys.},
   volume = {111},
   number = {8},
   pages = {3441-3448},
   year = {1999},
   type = {Journal Article},
    issn = {0021-9606},
    doi = {10.1063/1.479629},
    url = {https://doi.org/10.1063/1.479629},
}

@article{Zheng2009,
   author = {Zheng, X. and Song, Y. and Zhang, J.},
   title = {Ultraviolet photodissociation dynamics of the propargyl radical},
   journal = {J. Phys. Chem. A},
   volume = {113},
   number = {16},
   pages = {4604-12},
   ISSN = {1089-5639},
   DOI = {10.1021/jp8113336},
   year = {2009},
   type = {Journal Article}
}

@article{Broderick2018,
   author = {Broderick, Bernadette M. and Suas-David, Nicolas and Dias, Nureshan and Suits, Arthur G.},
   title = {Isomer-specific detection in the UV photodissociation of the propargyl radical by chirped-pulse mm-wave spectroscopy in a pulsed quasi-uniform flow},
   journal = {Phys. Chem. Chem. Phys.},
   volume = {20},
   number = {8},
   pages = {5517-5529},
   ISSN = {1463-9076},
   DOI = {10.1039/C7CP06211G},
   url = {http://dx.doi.org/10.1039/C7CP06211G},
   year = {2018},
   type = {Journal Article}
}

@article{Alkemade1989,
   author = {Alkemade, U. and Homann, K.-H.},
   title = {Formation of \ce{C6H6} {I}somers by {R}ecombination of {P}ropynyl in the {S}ystem {S}odium {V}apour/{P}ropynylhalide},
   journal = {Z. Phys. Chem., Neue Folge},
   volume = {161},
   pages = {19-34},
   year = {1989},
   type = {Journal Article},
   doi={10.1524/zpch.1989.161.Part_1_2.019},
}

@article{Stein_1991,
   author = {Stein, Stephen E. and Walker, James A. and Suryan, Mahendra M. and Fahr, Askar},
   title = {A new path to benzene in flames},
   journal = {Symp. (Int.) Combust.},
   volume = {23},
   number = {1},
   pages = {85-90},
   ISSN = {0082-0784},
   DOI = {https://doi.org/10.1016/S0082-0784(06)80245-6},
   url = {https://www.sciencedirect.com/science/article/pii/S0082078406802456},
   year = {1991},
   type = {Journal Article}
}

@article{Hebrard2013,
Author = {Hebrard, E. and Dobrijevic, M. and Loison, J. C. and Bergeat, A. and
Hickson, K. M. and Caralp, F.},
Title = {Photochemistry of {C$_3$H$_p$} hydrocarbons in
{T}itan's stratosphere revisited},
Journal = {Astron. Astrophys.},
Year = {2013},
Volume = {552},
Article-Number = {A132},
pages={A132},
ISSN = {0004-6361},
EISSN = {1432-0746},
doi = {10.1051/0004-6361/201220686},
}

@article{Savee2022,
   author = {Savee, John D. and Sztáray, Bálint and Hemberger, Patrick and Zádor, Judit and Bodi, Andras and Osborn, David L.},
   title = {Unimolecular isomerisation of 1,5-hexadiyne observed by threshold photoelectron photoion coincidence spectroscopy},
   journal = {Faraday Discuss.},
   volume = {238},
   number = {0},
   pages = {645-664},
   ISSN = {1359-6640},
   DOI = {10.1039/D2FD00028H},
   url = {http://dx.doi.org/10.1039/D2FD00028H},
   year = {2022},
   type = {Journal Article}
}

@article{Agundez2022,
	author = {{Agúndez, M.} and {Marcelino, N.} and {Cabezas, C.} and {Fuentetaja, R.} and {Tercero, B.} and {de Vicente, P.} and {Cernicharo, J.}},
	title = {Detection of the propargyl radical at $\lambda$ 3 mm},
	DOI= "10.1051/0004-6361/202142678",
	url= "https://doi.org/10.1051/0004-6361/202142678",
	journal = {Astron. Astrophys.},
	year = 2022,
	volume = 657,
	pages = "A96",
}

@article{stauber2005,
  title={X-ray chemistry in the envelopes around young stellar objects},
  author={St{\"a}uber, Pascal and Doty, SD and Van Dishoeck, EF and Benz, AO},
  journal={Astron. Astrophys.},
  volume={440},
  number={3},
  pages={949--966},
  year={2005},
  publisher={EDP Sciences},
  doi={10.1051/0004-6361:20052889},
}

@article{Levey2022,
author = {Levey, Zachariah D. and Laws, Benjamin A. and Sundar, Srivathsan P. and Nauta, Klaas and Kable, Scott H. and da Silva, Gabriel and Stanton, John F. and Schmidt, Timothy W.},
title = {{PAH Growth in Flames and Space: Formation of the Phenalenyl Radical}},
journal = {J. Phys. Chem. A},
volume = {126},
number = {1},
pages = {101-108},
year = {2022},
doi = {10.1021/acs.jpca.1c08310},
OPTnote ={PMID: 34936357},
OPTURL = { https://doi.org/10.1021/acs.jpca.1c08310},
OPTeprint = {https://doi.org/10.1021/acs.jpca.1c08310}
}

@article{tielens2013molecular,
  title={The molecular universe},
  author={Tielens, A. G. G. M.},
  journal={Rev. Mod. Phys.},
  volume={85},
  number={3},
  pages={1021--1081},
  year={2013},
  publisher={APS},
  doi={10.1103/RevModPhys.85.1021},
}

@article{tielens2008interstellar,
  title={Interstellar polycyclic aromatic hydrocarbon molecules},
  author={Tielens, Alexander G. G. M.},
  journal={Annu. Rev. Astron. Astrophys.},
  volume={46},
  number={1},
  pages={289--337},
  year={2008},
  publisher={Annual Reviews},
  doi={10.1146/annurev.astro.46.060407.145211},
}

@article{Hemberger_2011,
   author = {Hemberger, Patrick and Lang, Melanie and Noller, Bastian and Fischer, Ingo and Alcaraz, Christian and Cunha de Miranda, Bárbara K. and Garcia, Gustavo A. and Soldi-Lose, Héloïse},
   title = {{Photoionization of Propargyl and Bromopropargyl Radicals: A Threshold Photoelectron Spectroscopic Study}},
   journal = {J. Phys. Chem. A},
   volume = {115},
   number = {11},
   pages = {2225-2230},
   ISSN = {1089-5639},
   DOI = {10.1021/jp112110j},
   url = {https://doi.org/10.1021/jp112110j},
   year = {2011},
   type = {Journal Article}
}

@article{Schleier_2025,
   author = {Arns, Rahel and McClish, Rory and Hemberger, Patrick and Bodi, Andras and Bouwman, Jordy and Kasper, Tina and Schleier, Domenik},
   title = {{Is Phenylnitrene a Missing Link in the Formation of Polycyclic Aromatic Nitrogen Heterocycles?}},
   journal = {Angew. Chem. Int. Ed.},
   volume = {64},
   number = {31},
   pages = {e202503940},
   ISSN = {1433-7851},
   DOI = {https://doi.org/10.1002/anie.202503940},
   url = {https://onlinelibrary.wiley.com/doi/abs/10.1002/anie.202503940},
   year = {2025},
   type = {Journal Article}
}

@article{Klippenstein_2003,
   author = {Miller, James A. and Klippenstein, Stephen J.},
   title = {{The Recombination of Propargyl Radicals and Other Reactions on a \ce{C6H6}  Potential}},
   journal = {J. Phys. Chem. A},
   volume = {107},
   pages = {7783-7799},
   ISSN = {1089-5639},
   doi = {10.1021/jp030375h},
   url = {https://ui.adsabs.harvard.edu/abs/2003JPCA..107.7783M},
   year = {2003},
   type = {Journal Article}
}

@article{Merkt_2013,
   author = {Jacovella, U. and Gans, B. and Merkt, F.},
   title = {On the adiabatic ionization energy of the propargyl radical},
   journal = {J. Chem. Phys.},
   volume = {139},
   number = {8},
   pages={084308},
   ISSN = {0021-9606},
   doi = {10.1063/1.4818982},
   OPTurl = {https://doi.org/10.1063/1.4818982},
   year = {2013},
   type = {Journal Article}
}

@article{Garcia_2018,
   author = {Garcia, Gustavo A. and Gans, Bérenger and Kr{\"u}ger, Julia and Holzmeier, Fabian and R{\"o}der, Anja and Lopes, Allan and Fittschen, Christa and Alcaraz, Christian and Loison, Jean-Christophe},
   title = {Valence shell threshold photoelectron spectroscopy of {C}$_3${H}$_x$ ($x$ = 0-3)},
   journal = {Phys. Chem. Chem. Phys.},
   volume = {20},
   number = {13},
   pages = {8707-8718},
   ISSN = {1463-9076},
   DOI = {10.1039/C8CP00510A},
   url = {http://dx.doi.org/10.1039/C8CP00510A},
   year = {2018},
   type = {Journal Article}
}

@article{Kaiser_2021,
   author = {Zhao, Long and Lu, Wenchao and Ahmed, Musahid and Zagidullin, Marsel V. and Azyazov, Valeriy N. and Morozov, Alexander N. and Mebel, Alexander M. and Kaiser, Ralf I.},
   title = {Gas-phase synthesis of benzene via the propargyl radical self-reaction},
   journal = {Sci. Adv.},
   volume = {7},
   number = {21},
   pages = {eabf0360},
   DOI = {doi:10.1126/sciadv.abf0360},
   url = {https://www.science.org/doi/abs/10.1126/sciadv.abf0360},
   year = {2021},
   type = {Journal Article}
}

@article{Kaiser_2023,
   author = {Zhenghai Yang  and Galiya R. Galimova  and Chao He  and Shane J. Goettl  and Dababrata Paul  and Wenchao Lu  and Musahid Ahmed  and Alexander M. Mebel  and Xiaohu Li  and Ralf I. Kaiser},
   title = {Gas-phase formation of the resonantly stabilized 1-indenyl (\ch{C9H7^{.}}) radical in the interstellar medium},
   journal = {Sci. Adv.},
   volume = {9},
   number = {36},
   pages = {eadi5060},
   DOI = {10.1126/sciadv.adi5060},
   url = {https://www.science.org/doi/abs/10.1126/sciadv.adi5060},
   year = {2023},
   type = {Journal Article}
}

@article{Zhang_2025,
   author = {Zhang, Jinyang and Gao, Jiao and Wang, Hong and Guan, Jiwen and Xu, Guangxian and Xing, Lili and Truhlar, Donald G. and Wang, Zhandong},
   title = {Less-Dominant Resonance Configuration of Propargyl Radical Leads to a Growth Mechanism for Polycyclic Aromatic Hydrocarbons that Preserves the Cyclopenta Ring},
   journal = {J. Am. Chem. Soc.},
   volume = {147},
   number = {11},
   pages = {9283-9293},
   DOI = {10.1021/jacs.4c15155},
   url = {https://doi.org/10.1021/jacs.4c15155},
   year = {2025},
   type = {Journal Article}
}

@article{Hrodmarsson_2024,
   author = {Hrodmarsson, Helgi Rafn and Garcia, Gustavo A. and Bourehil, Lyna and Nahon, Laurent and Gans, Bérenger and Boyé-Péronne, Séverine and Guillemin, Jean-Claude and Loison, Jean-Christophe},
   title = {{The isomer distribution of C$_6$H$_6$ products from the propargyl radical gas-phase recombination investigated by threshold-photoelectron spectroscopy}},
   journal = {Commun. Chem.},
   volume = {7},
   number = {1},
   pages = {156},
   ISSN = {2399-3669},
   DOI = {10.1038/s42004-024-01239-7},
   url = {https://doi.org/10.1038/s42004-024-01239-7},
   year = {2024},
   type = {Journal Article}
}

@article{Constantinidis_2017,
   author = {Constantinidis, P. and Hirsch, F. and Fischer, I. and Dey, A. and Rijs, A. M.},
   title = {{Products of the Propargyl Self-Reaction at High Temperatures Investigated by IR/UV Ion Dip Spectroscopy}},
   journal = {J.  Phys. Chem. A},
   volume = {121},
   number = {1},
   pages = {181-191},
   ISSN = {1089-5639},
   DOI = {10.1021/acs.jpca.6b08750},
   url = {https://doi.org/10.1021/acs.jpca.6b08750},
   year = {2017},
   type = {Journal Article}
}

@article{Osborn_2023,
   author = {Selby, T. M. and Goulay, F. and Soorkia, S. and Ray, A. and Jasper, A. W. and Klippenstein, S. J. and Morozov, A. N. and Mebel, A. M. and Savee, J. D. and Taatjes, C. A. and Osborn, D. L.},
   title = {{Radical-Radical Reactions in Molecular Weight Growth: The Phenyl + Propargyl Reaction}},
   journal = {J. Phys. Chem. A},
   volume = {127},
   number = {11},
   pages = {2577-2590},
   ISSN = {1089-5639},
   DOI = {10.1021/acs.jpca.2c08121},
   year = {2023},
   type = {Journal Article}
}

@article{Savee_2015,
   author = {Savee, J. D. and Selby, T. M. and Welz, O. and Taatjes, C. A. and Osborn, D. L.},
   title = {Time- and Isomer-Resolved Measurements of Sequential Addition of Acetylene to the Propargyl Radical},
   journal = {J. Phys. Chem. Lett.},
   volume = {6},
   number = {20},
   pages = {4153-8},
   OPTnote = {1948-7185},
   ISSN = {1948-7185},
   DOI = {10.1021/acs.jpclett.5b01896},
   year = {2015},
   type = {Journal Article}
}

@article{Lindstedt_2002,
   author = {Lindstedt, Peter and Maurice, Lourdes and Meyer, Michael},
   title = {Thermodynamic and kinetic issues in the formation and oxidation of aromatic species},
   journal = {Faraday Discuss.},
   volume = {119},
   number = {0},
   pages = {409-432},
   ISSN = {1359-6640},
   DOI = {10.1039/B104056C},
   url = {http://dx.doi.org/10.1039/B104056C},
   year = {2002},
   type = {Journal Article}
}

@article{Mebel_2020,
   author = {Morozov, Alexander N. and Mebel, Alexander M.},
   title = {Theoretical study of the reaction mechanism and kinetics of the phenyl + propargyl association},
   journal = {Phys. Chem. Chem. Phys.},
   volume = {22},
   number = {13},
   pages = {6868-6880},
   ISSN = {1463-9076},
   DOI = {10.1039/D0CP00306A},
   url = {http://dx.doi.org/10.1039/D0CP00306A},
   year = {2020},
   type = {Journal Article}
}

@article{Matsugi_2012,
   author = {Matsugi, Akira and Miyoshi, Akira},
   title = {Computational study on the recombination reaction between benzyl and propargyl radicals},
   journal = {Int. J. Chem. Kinet.},
   volume = {44},
   number = {3},
   pages = {206-218},
   ISSN = {0538-8066},
   DOI = {https://doi.org/10.1002/kin.20625},
   url = {https://onlinelibrary.wiley.com/doi/abs/10.1002/kin.20625},
   year = {2012},
   type = {Journal Article}
}

@article{Jacox_1974,
   author = {Jacox, Marilyn E. and Milligan, Dolphus E.},
   title = {Matrix isolation study of the vacuum ultraviolet photolysis of allene and methylacetylene. Vibrational and electronic spectra of the species \ce{C3}, \ce{C3H}, \ce{C3H2}, and \ce{C3H3}},
   journal = {Chem. Phys.},
   volume = {4},
   number = {1},
   pages = {45-61},
   ISSN = {0301-0104},
   DOI = {https://doi.org/10.1016/0301-0104(74)80047-9},
   url = {https://www.sciencedirect.com/science/article/pii/0301010474800479},
   year = {1974},
   type = {Journal Article}
}

@article{Jochnowitz_2005,
   author = {Jochnowitz, E. B. and Zhang, X. and Nimlos, M. R. and Varner, M. E. and Stanton, J. F. and Ellison, G. B.},
   title = {Propargyl radical: ab initio anharmonic modes and the polarized infrared absorption spectra of matrix-isolated \ce{HCCCH2}},
   journal = {J. Phys. Chem. A},
   volume = {109},
   number = {17},
   pages = {3812-21},
   ISSN = {1089-5639 (Print) 1089-5639},
   DOI = {10.1021/jp040719j},
   year = {2005},
   type = {Journal Article}
}

@article{Ramsay_1966,
   author = {Ramsay, D. A. and Thistlethwaite, P.},
   title = {The Electronic Absorption Spectrum of the Propargyl Radical},
   journal = {Can. J. Phys.},
   volume = {44},
   number = {7},
   pages = {1381-1386},
   DOI = {10.1139/p66-116},
   url = {https://cdnsciencepub.com/doi/abs/10.1139/p66-116},
   year = {1966},
   type = {Journal Article}
}

@article{Fahr_2005,
   author = {Fahr, A. and Laufer, A. H.},
   title = {{UV-absorption spectra of the radical transients generated from the 193-nm photolysis of allene, propyne, and 2-butyne}},
   journal = {J. Phys. Chem. A},
   volume = {109},
   number = {11},
   pages = {2534-9},
   OPTnote = {Fahr, Askar},
   ISSN = {1089-5639 (Print)
1089-5639},
   DOI = {10.1021/jp0406058},
   year = {2005},
   type = {Journal Article}
}

@article{Savic_2005,
   author = {Savić, I. and Gerlich, D.},
   title = {{Temperature variable ion trap studies of C$_3$H$_n^+$ with H$_2$ and HD}},
   journal = {Phys. Chem. Chem. Phys.},
   volume = {7},
   number = {5},
   pages = {1026-1035},
   ISSN = {1463-9076},
   DOI = {10.1039/B417965J},
   url = {http://dx.doi.org/10.1039/B417965J},
   year = {2005},
   type = {Journal Article}
}

@article{Gerlich_2005,
   author = {Savić, I. and Schlemmer, S. and Gerlich, D.},
   title = {{Low-Temperature Experiments on the Formation of Deuterated C$_3$H$_3^+$}},
   journal1 = {Astrophys. J.},
   journal = {ApJ},
   volume = {621},
   number = {2},
   pages = {1163},
   ISSN = {0004-637X},
   DOI = {10.1086/427648},
   url = {https://dx.doi.org/10.1086/427648},
   year = {2005},
   type = {Journal Article}
}

@article{Reinwardt_2024,
   author = {Reinwardt, Simon and Cieslik, Patrick and Buhr, Ticia and Perry-Sassmannshausen, Alexander and Schippers, Stefan and Müller, Alfred and Trinter, Florian and Martins, Michael},
   title = {Isomer-specific photofragmentation of \ce{C3H3+} at the carbon {K}-edge},
   journal = {Phys. Chem. Chem. Phys.},
   volume = {26},
   number = {21},
   pages = {15519-15529},
   ISSN = {1463-9076},
   DOI = {10.1039/D4CP00370E},
   url = {http://dx.doi.org/10.1039/D4CP00370E},
   year = {2024},
   type = {Journal Article}
}

@article{Tanaka1997,
   author = {Tanaka, K. and Sumiyoshi, Y. and Ohshima, Y. and Endo, Y. and Kawaguchi, K.},
   title = {Pulsed discharge nozzle {Fourier} transform microwave spectroscopy of the propargyl radical ({H$_2$CCCH})},
   journal = {J. Chem. Phys.},
   volume = {107},
   pages = {2728--2733},
   year = {1997},
   type = {Journal Article},
   doi={10.1063/1.474631},
}

@article{herbst1989gas,
  title={Gas-phase production of complex hydrocarbons, cyanopolyynes, and related compounds in dense interstellar clouds},
  author={Herbst, Eric and Leung, Chun Ming},
  journal={Astrophys. J. Suppl. Ser.},
  volume={69},
  pages={271--300},
  year={1989},
  doi = {10.1086/191314},
}

@article{Richter2000,
title = {Formation of polycyclic aromatic hydrocarbons and their growth to soot—a review of chemical reaction pathways},
journal = {Prog. Energy Combust. Sci.},
volume = {26},
number = {4},
pages = {565-608},
year = {2000},
issn = {0360-1285},
doi = {https://doi.org/10.1016/S0360-1285(00)00009-5},
url = {https://www.sciencedirect.com/science/article/pii/S0360128500000095},
author = {H Richter and J.B Howard}
}

@article{Kress2010,
title = {The ‘soot line’: Destruction of presolar polycyclic aromatic hydrocarbons in the terrestrial planet-forming region of disks},
journal = {Adv. Space Res.},
volume = {46},
number = {1},
pages = {44-49},
year = {2010},
issn = {0273-1177},
doi = {https://doi.org/10.1016/j.asr.2010.02.004},
url = {https://www.sciencedirect.com/science/article/pii/S0273117710001092},
author = {Monika E. Kress and Alexander G.G.M. Tielens and Michael Frenklach}
}

@article{Changala_2024,
   author = {Changala, P. Bryan and Franke, Peter R. and Stanton, John F. and Ellison, G. Barney and McCarthy, Michael C.},
   title = {{Direct Probes of $\pi$-Delocalization in Prototypical Resonance-Stabilized Radicals: Hyperfine-Resolved Microwave Spectroscopy of Isotopic Propargyl and Cyanomethyl}},
   journal = {J. Am. Chem. Soc.},
   volume = {146},
   number = {2},
   pages = {1512-1521},
   ISSN = {0002-7863},
   DOI = {10.1021/jacs.3c11220},
   url = {https://doi.org/10.1021/jacs.3c11220},
   year = {2024},
   type = {Journal Article}
}

@article{Byrne_2023,
   author = {Byrne, Alex N. and Xue, Ci and Cooke, Ilsa R. and McCarthy, Michael C. and McGuire, Brett A.},
   title = {Astrochemical Modeling of Propargyl Radical Chemistry in {TMC-1}},
   journal1 = {Astrophys. J.},
   journal = {ApJ},
   volume = {957},
   number = {2},
   pages = {88},
   ISSN = {0004-637X},
   DOI = {10.3847/1538-4357/acf863},
   url = {https://doi.org/10.3847/1538-4357/acf863},
   year = {2023},
   type = {Journal Article}
}

\end{document}


\preprint{SI/propargyl}

\title{Supplementary Information.\\
The X-ray absorption spectrum of the propargyl radical, \ch{C3H3^{.}}\,. }

\author{Dorothee Schaffner}
\affiliation{Institute of Physical and Theoretical Chemistry, University of W{\"u}rzburg, 
 D-97074 W{\"u}rzburg, Germany.}
\author{Theo Juncker von Buchwald}%
\affiliation{DTU Chemistry, Technical University of Denmark, DK-2800 Kgs. Lyngby, Denmark.}

\author{Jacob Pedersen}%
\affiliation{DTU Chemistry, Technical University of Denmark, DK-2800 Kgs. Lyngby, Denmark.}
\affiliation{Department of Chemistry, Norwegian University of Science and Technology, N-7491 Trondheim, Norway.}

\author{Andreas Rasp}%
\affiliation{Institute of Physical and Theoretical Chemistry, University of W{\"u}rzburg, 
 D-97074 W{\"u}rzburg, Germany.}
 
\author{Emil Karaev}%
\affiliation{Institute of Physical and Theoretical Chemistry, University of W{\"u}rzburg, 
 D-97074 W{\"u}rzburg, Germany.}
 
 \author{Valentin von Laffert}%
\affiliation{Institute of Physical and Theoretical Chemistry, University of W{\"u}rzburg, 
 D-97074 W{\"u}rzburg, Germany.}

\author{Alessio Bruno}%
\affiliation{Dipartimento di Chimica e Tecnologia dei Farmaci, Universit{\`a} degli Studi di Roma  ``La Sapienza'', piazzale Aldo Moro 5, I-00185 Rome, Italy.}

\author{Michele Alagia}
\affiliation{CNR - Istituto Officina dei Materiali (IOM), Laboratorio TASC, 34149 Trieste, Italy.}

\author{Stefano Stranges}
\affiliation{CNR - Istituto Officina dei Materiali (IOM), Laboratorio TASC, 34149 Trieste, Italy.}
\affiliation{Dipartimento di Chimica e Tecnologia dei Farmaci, Universit{\`a} degli Studi di Roma  ``La Sapienza'', piazzale Aldo Moro 5, I-00185 Rome, Italy.}

\author{Ingo Fischer}
\homepage{E-mail: ingo.fischer@uni-wuerzburg.de (experiment)}
\affiliation{Institute of Physical and Theoretical Chemistry, University of W{\"u}rzburg, 
 D-97074 W{\"u}rzburg, Germany.}
 
\author{Sonia Coriani}
\homepage{E-mail: soco@kemi.dtu.dk (theory)}
\affiliation{DTU Chemistry, Technical University of Denmark, DK-2800 Kgs. Lyngby, Denmark.}

\date{\today}

\maketitle

\section*{Content}
This SI contains:
\begin{itemize}
\item[\ref{sec:exp_SI}.] 
TOF mass spectra at different excitation energies
\item[\ref{sec:theory_SI}.] Computational information
\begin{itemize}
\item[\ref{sec:comp_details_SI}] Extended Computational Details
\item[\ref{sec:XAS-calculated}] X-ray absorption spectra from fc-CVS-EOM-CCSD and CVS-ADC(2)-x
\item[\ref{sec:CoreIEs}] Core Ionization Energies
\item[\ref{sec:exciene_and_f_SI}] Excitation Energies and Oscillator Strengths
\item[\ref{sec:NTOs}] Natural Transition Orbitals (NTOs) 
\item[\ref{sec:theory_SI_vib}] Vibrationally Resolved Spectra
\item[\ref{sec:OptimizedStructures}] Optimized structures
\item[\ref{sec:fcclasses:input}] \texttt{FCclasses3} input file example
\end{itemize}
\end{itemize}

\section{TOF mass spectra at different excitation energies} \label{sec:exp_SI}

\begin{figure}[hbpt!]
    \centering
\includegraphics[scale=1.0]{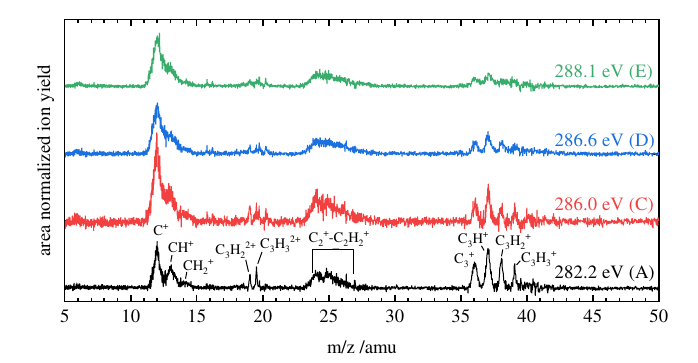}
    \caption{TOF mass spectra of single ion events upon excitation of the propargyl radical at a photon energy of 282.2\,eV (band A), 286.0\,eV (band C), 286.6\,eV (band D) and 288.1\,eV (band E).}
    \label{fig:TOF_Singles_all_peaks}
\end{figure}

Fig.~\ref{fig:TOF_Singles_all_peaks} shows a comparison of the single ion mass spectra after excitation at 282.2\,eV (band A), 286.0\,eV (band C), 286.6\,eV (band D) and 288.1\,eV (band E). The spectra were obtained after subtraction of the contribution from off-resonant ionization which was recorded at 280.2\,eV.
The precursor shows resonant absorptions in the regions of 286.0\,eV (band C), 286.6\,eV (band D) and 288.1\,eV (band E). To obtain a spectrum of purely the propargyl radical for these excitation energies, the spectrum of the precursor was measured at the same photon energy at 0\,W and (after its correction for off-resonant contributions) subtracted. This additional correction procedure results in a lower signal to noise ratio, especially at 286.0\,eV (band C).
All four spectra show the formation of the same fragment ions, whose branching ratios are summarized in Table~\ref{tab:br_singles}. Small, sharp signals observed at \textit{m/z} 40.5/39.5 (Br\textsuperscript{2+}), 27/26.34 (Br\textsuperscript{3+}), 20.25/19.75 (Br\textsuperscript{4+}), 16.2/15.8 (Br\textsuperscript{5+}) result from the bromine radical that is not completely subtracted by the correction for the off-resonant contribution. 
For higher excitations the relative intensity of the \ce{C3} fragments decreases significantly from more than 30\% to below 20\%, whereas the branching ratios for \ce{C2} and \ce{C1} fragments increase correspondingly. In addition, the ratio between the fragments \ce{C1} and \ce{C2} increases with the excitation energy. This shows that the fragmentation tendency increases with the excitation of higher core excited states as was observed for the \textit{tert}-butyl radical.\cite{Schaffner_2024}

\begin{table}
\caption{\label{tab:br_singles}Branching ratios of single ion events after excitation of the propargyl radical at different resonances. Relative intensities were determined from the mass spectrum (Fig.~\ref{fig:TOF_Singles_all_peaks}) by integration. Integrals of overlapping masses were derived from fitting with Gauss functions. The branching ratios were normalized so that all channels sum up to 100\%. Very low branching ratios are not specified and are summarized in ``others''. The error of each branching ratio is estimated to be around $\pm$20\% of the respective branching ratio.}
\begin{ruledtabular}
\begin{tabular}{llcccc} 
\textit{m/z}&species&\mbox{282.2\,eV /\%}&\mbox{286.0\,eV /\%}&\mbox{286.6\,eV /\%}&\mbox{288.1\,eV /\%}\\
\hline
39&\ce{C3H3+}   &4&2&2&2\\
38&\ce{C3H2+}   &8&2&5&4\\
37&\ce{C3H+}    &13&6&8&7\\
36&\ce{C3+}     &10&4&5&4\\
26&\ce{C2H2+}   &6&8&9&7\\
25&\ce{C2H+}    &12&9&8&7\\
24&\ce{C2+}     &9&13&11&11\\
19.5&\ce{C3H3}\textsuperscript{2+}&2&--&--&--\\
19&\ce{C3H2}\textsuperscript{2+}&--&--&--&--\\
14&\ce{CH2+}    &4&3&3&4\\
13&\ce{CH+}     &13&18&17&15\\
12&\ce{C+}      &17&30&29&35\\
6&C\textsuperscript{2+}&--&--&--&2\\
others& &2&2&3&3\\
\end{tabular}
\end{ruledtabular}
\end{table}

\clearpage
\newpage

\section{Computational Information} \label{sec:theory_SI}

\subsection{Extended Computational Details}
\label{sec:comp_details_SI}

The X-ray absorption spectrum of the propargyl radical is calculated using the frozen-core core-valence-separated equation-of-motion coupled cluster singles and doubles (fc-CVS-EOM-CCSD) method~\cite{Vidal2019fcCVS} in \texttt{Q-Chem 6.2}~\cite{Qchem541} using both restricted open-shell (ROHF) and unrestricted (UHF) Hartree-Fock reference orbitals. We used the optimized ground state structure by \citeauthor{botschwina2010calculated}~\cite{botschwina2010calculated} obtained at the CCSD(T*)-F12a~\cite{adler2007simple,knizia2009simplified}/cc-pVTZ-F12~\cite{peterson2008a} level of theory. The X-ray absorption spectrum is calculated using the 6-311++G** basis set,~\cite{clark1983a,krishnan1980a,marchetti2008accurate} the cc-pVTZ basis set,~\cite{dunning1989a} and the aug-cc-pVTZ basis set.~\cite{kendall1992a} 
When reporting the results, we use usual C$_{2\text{v}}$ point group symmetry conventions, i.e. the molecule is oriented such that the C$_2$ rotational axis is aligned with the Z-axis and the molecular plane in the YZ-plane. 
We calculated 15 excitation in each irreducible representation (irrep). 
The ground state belongs to the B$_1$ irrep. 
The spectra were found to be of comparable quality regardless of the reference orbitals and basis sets. To test the fc-CVS-EOM-CCSD results we also calculate the X-ray absorption spectrum at the CVS-ADC(2)-x level using UHF reference orbitals and the aug-cc-pVTZ basis set in \texttt{Q-Chem 6.2}. For the assignment of spectral peaks we calculated natural transition orbitals (NTOs)~\cite{Vidal2019fcCVS,krylov2020orbitals} of both fc-CVS-EOM-CCSD and CVS-ADC(2)-x using the aug-cc-pVTZ basis set. 

To investigate the vibrational effects in the X-ray absorption spectra we adopted (i) the nuclear ensemble approach (NEA) and (ii) \textit{a posteriori} treatment based on Franck-Condon (FC) factors and time-dependent spectra computed  within the harmonic approximation with \texttt{FCclasses3}.~\cite{cerezo2023fcclasses3} 
In the NEA the vibrational effects are captured by computing the X-ray absorption spectrum at numerous vibrationally distorted geometries. We sampled 200 structures from a Wigner distribution at a temperature of 300 K.
The FC vibrational treatment requires optimized core-excited state structures, and we used the Vertical Gradient (VG) and Adiabatic Shift (AS) methods.~\cite{macak2000simulations,biczysko2012time,biczysko2012time} We note that the VG approach requires the ground state gradient and Hessian as well as the transition dipole strength. 
We optimized the first three bright transitions (the first, second, and fifth electronic excitations) at the CVS-ADC(2)-x/aug-cc-pVDZ level of theory in \texttt{cvs-adc-grad}~\cite{cvs-adc-grad} in conjunction with \texttt{ADCC}~\cite{herbst2020adcc} and \texttt{PySCF}.~\cite{pyscf,pyscf_2} 
The electronic transition dipole strengths used in both vibrational treatments were obtained at the fc-CVS-EOM-CCSD/aug-cc-pVTZ and CVS-ADC(2)-x/aug-cc-pVTZ levels of theory using \texttt{Q-Chem 6.2}.

All calculations were carried out on the DTU HPC resources.~\cite{DTU-DCC}

\subsection{X-ray absorption spectra from fc-CVS-EOM-CCSD and CVS-ADC(2)-x}
\label{sec:XAS-calculated}

\begin{figure}[H]
    \centering
\includegraphics[width=.49\linewidth]{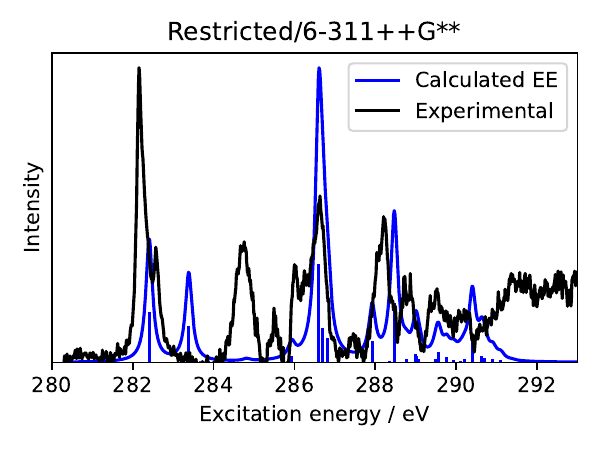}
\includegraphics[width=.49\linewidth]{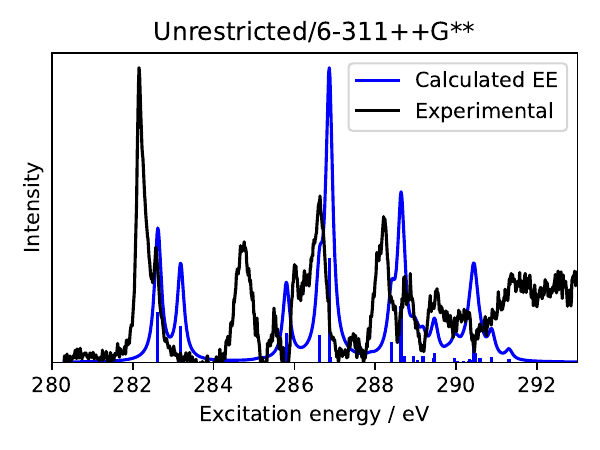} \\
\includegraphics[width=.49\linewidth]{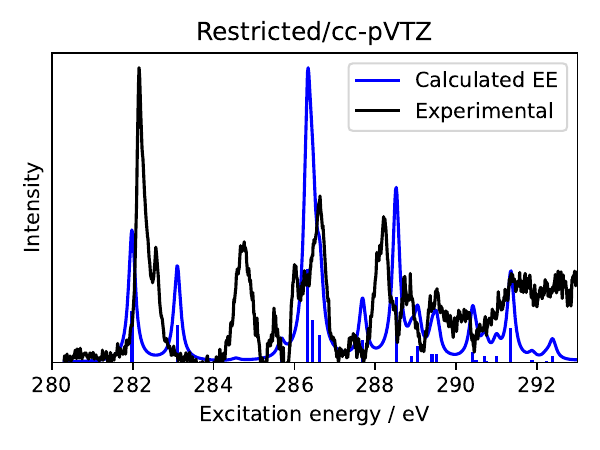}
\includegraphics[width=.49\linewidth]{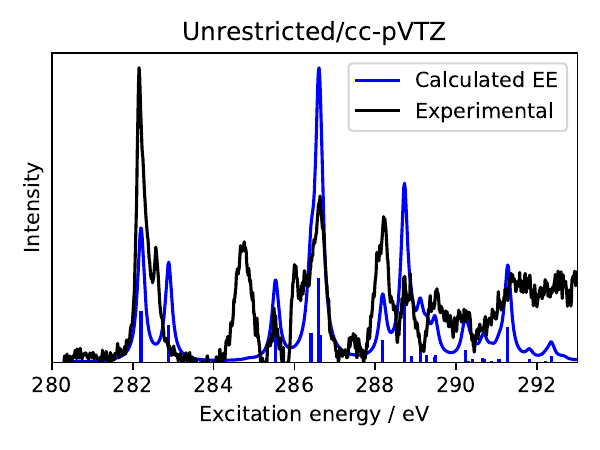} \\
\includegraphics[width=.49\linewidth]{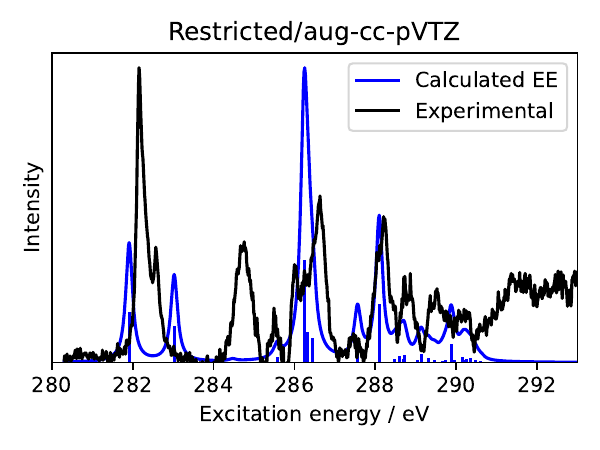}
\includegraphics[width=.49\linewidth]{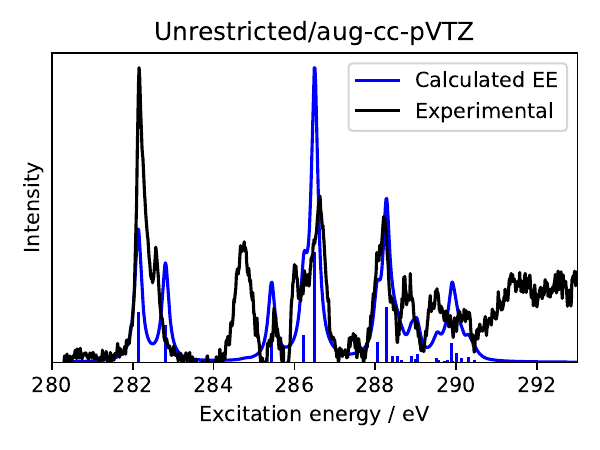}
\caption{Comparison between the experimental spectrum (black line) and the simulated fc-CVS-EOM-CCSD X-ray absorption spectra of the propargyl radical with (top left) ROHF/6-311++G**, (top right) UHF/6-311++G**, (middle left) ROHF/cc-pVTZ, (middle right) UHF/cc-pVTZ, (bottom left) ROHF/aug-cc-pVTZ, and (bottom right) UHF/aug-cc-pVTZ levels of theory in blue. The sticks were convoluted using a Lorentzian lineshape with a full width at half maximum (FWHM) of 200\,meV. 
}
\label{fig:CCSD:XAS}
\end{figure}

\begin{figure}[H]
    \centering
\includegraphics[width=0.85\linewidth]{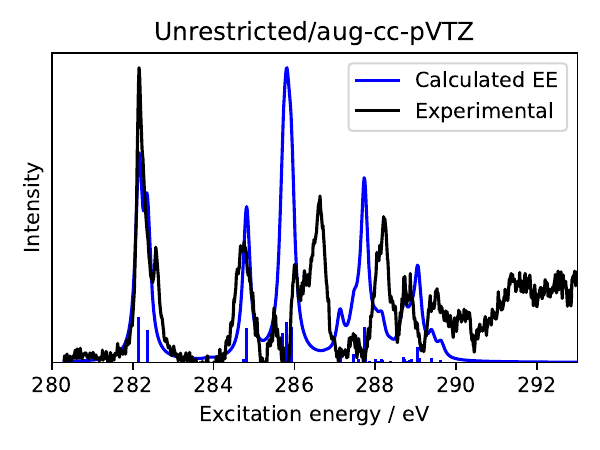}
\caption{Comparison between the experimental spectrum (black line) and the simulated CVS-ADC(2)-x/aug-cc-pVTZ X-ray absorption spectra (blue line) of the propargyl radical. The sticks were convoluted using a Lorentzian lineshape with a FWHM of 200\,meV.}
\label{fig:ADC:XAS}
\end{figure}


\subsection{Core Ionization Energies}
\label{sec:CoreIEs}

\begin{table}[hbpt!]
\centering
\caption{C1s core ionization energies (in eV) of the propargyl radical computed at the fc-CVS-EOM-IP-CCSD/aug-cc-pVTZ level of theory.
The respective carbon atom is highlighted in the last column. The multiplicity stated in the table refers to the cationic final state reached by the ionization.
For the 2 B$_1$ ionized state [(2a$_1$)$^{-1}$], we note that the triplet has a higher ionization energy than the singlet.
}
\label{tab:IPs}
\begin{tabular}{c|c|cc|cc|c}
        \hline
        \hline
        IP state & Ionisation &  \multicolumn{2}{c|}{ROHF} & \multicolumn{2}{c|}{UHF} & Carbon
        \\
        \hline
        & & singlet & triplet & singlet & triplet
        \\\hline
        1~B$_1$ & (3a$_1$)$^{-1}$ & 291.4878 & 291.2337 & 291.4734 & 291.1673 & H$_2$CC{\bf\underline{C}}H \\
        2~B$_1$ & (2a$_1$)$^{-1}$ & 291.6181 & 291.7696 & 291.7233 & 291.8698 & H$_2$C{\bf\underline{C}}CH \\
        3~B$_1$ & (1a$_1$)$^{-1}$ & 292.6061 & 291.9766 & 292.5449 & 291.9304 & H$_2${\bf\underline{C}}CCH 
        \\
        \hline\hline
\end{tabular}
\end{table}

\clearpage

\subsection{Excitation Energies and Oscillator Strengths}
\label{sec:exciene_and_f_SI}

\begingroup
\LTcapwidth=\textwidth
\begin{longtable}{c|cc|cc||cc}
    \caption{fc-CVS-EOM-CCSD and CVS-ADC(2)-x results for the excitation energies (in eV) and oscillator strengths (f) of the propargyl radical, using either UHF or ROHF (only fc-CVS-EOM-CCSD) reference and the aug-cc-pVTZ basis sets. Transitions that are included in Tables~\ref{tab:NTOs:EOM} and~\ref{tab:NTOs:ADC} are shown in gray. The ground state is $\tilde X$\textsuperscript{ 2}B$_1$.
    \label{tab:aug-cc-pvtz}}
    \\
    \hline
     & \multicolumn{4}{|c||}{fc-CVS-EOM-CCSD}
     & \multicolumn{2}{c}{CVS-ADC(2)-x} \\
     \hline
     & \multicolumn{2}{c|}{ROHF reference}
     & \multicolumn{2}{c||}{UHF reference}
     & \multicolumn{2}{c}{UHF reference}
     \\
    Final state 
    & Energy (eV) & f  & Energy (eV) & f & Energy (eV) & f \\\hline
    \rowcolor{lightgray} 
1~${}^{2}\!$A$_1$ & 
281.9118 & 0.0345 & 282.1359 & 0.0345 & 282.1539 & 0.0306 \\
\rowcolor{lightgray} 
2~${}^{2\!}$A$_1$ &
283.0258 & 0.0251 & 282.8106 & 0.0253 & 282.3613 & 0.0221 \\
3~${}^{2}\!$A$_1$ &
284.4676 & 0.0006 & 284.7728 & 0.0002 & 283.8704 & 0.0002 \\
\rowcolor{lightgray} 
4~${}^{2}\!$A$_1$ &
286.4429 & 0.0170 & 286.2346 & 0.0189 & 285.7153 & 0.0201 \\
\rowcolor{lightgray} 
5~${}^{2}\!$A$_1$ &
287.5605 & 0.0146 & 288.0442 & 0.0139 & \cellcolor{white}287.1269 & \cellcolor{white}0.0067 \\
\rowcolor{lightgray} 
6~${}^{2}\!$A$_1$ &
288.0958 & 0.0399 & 288.2749 & 0.0378 & \cellcolor{white}287.4665 & \cellcolor{white}0.0057 \\
\rowcolor{lightgray} 
7~${}^{2}\!$A$_1$ &
\cellcolor{white}288.7771 & \cellcolor{white}0.0000 & \cellcolor{white}288.6950 & \cellcolor{white}0.0000 & 
287.7263 & 0.0244 \\
8~${}^{2}\!$A$_1$ &
289.4533 & 0.0019 & 289.5929 & 0.0004 & 288.6501 & 0.0000 \\
9~${}^{2}\!$A$_1$ &
289.8061 & 0.0001 & 289.6945 & 0.0001 & 288.6995 & 0.0040 \\
10~${}^{2}\!$A$_1$ &
289.8734 & 0.0128 & 289.7803 & 0.0018 & 288.8622 & 0.0000 \\
\rowcolor{lightgray} 
11~${}^{2}\!$A$_1$ &
289.9228 & 0.0004 & 289.8926 & 0.0134 & 289.0442 & 0.0109 \\
12~${}^{2}\!$A$_1$ &
290.1483 & 0.0036 & 289.9998 & 0.0062 & 289.0809 & 0.0029 \\
13~${}^{2}\!$A$_1$ &
290.3517 & 0.0034 & 290.3069 & 0.0040 & 289.4635 & 0.0000 \\
14~${}^{2}\!$A$_1$ &
290.4686 & 0.0020 & 290.5788 & 0.0000 & 289.4754 & 0.0001 \\
15~${}^{2}\!$A$_1$ &
290.6090 & 0.0008 & 290.5813 & 0.0002 & 289.6362 & 0.0003 
\\\hline
1~${}^{2}\!$A$_2$ &
285.4798 & 0.0001 & 285.3688 & 0.0003 & 284.7513 & 0.0025 \\
\rowcolor{lightgray} 
2~${}^{2}\!$A$_2$ &
285.5746 & 0.0038 & 285.4364 & 0.0198 & 284.8203 & 0.0234 \\
\rowcolor{lightgray} 
3~${}^{2}\!$A$_2$ &
286.2398 & 0.0696 & 286.4998 & 0.0748 & 285.8099 & 0.0277 \\
\rowcolor{lightgray}
4~${}^{2}\!$A$_2$ &
\cellcolor{white}286.3313 & \cellcolor{white}0.0210 & \cellcolor{white}286.5326 & \cellcolor{white}0.0001 & 
285.9240 & 0.0240 
\\
5~${}^{2}\!$A$_2$ &
288.5880 & 0.0045 & 288.5411 & 0.0047 & 287.7485 & 0.0034 
\\
6~${}^{2}\!$A$_2$ &
289.0470 & 0.0021 & 288.9873 & 0.0027 & 288.0003 & 0.0025 
\\
7~${}^{2}\!$A$_2$ &
289.1364 & 0.0061 & 289.0403 & 0.0057 & 288.1750 & 0.0018 
\\
8~${}^{2}\!$A$_2$ &
289.3071 & 0.0033 & 289.3753 & 0.0003 & 288.2939 & 0.0000 
\\
9~${}^{2}\!$A$_2$ &
289.5280 & 0.0002 & 289.5154 & 0.0029 & 288.7077 & 0.0022 
\\
10~${}^{2}\!$A$_2$ &
289.7009 & 0.0002 & 289.7118 & 0.0013 & 288.8349 & 0.0014 
\\
11~${}^{2}\!$A$_2$ &
289.7363 & 0.0019 & 289.8679 & 0.0004 & 288.9027 & 0.0026 \\
12~${}^{2}\!$A$_2$ 
& 289.9674 & 0.0014 & 289.9122 & 0.0014 & 289.0198 & 0.0001 
\\
13~${}^{2}\!$A$_2$ &
290.2199 & 0.0014 & 290.1307 & 0.0029 & 289.1723 & 0.0000 \\
14~${}^{2}\!$A$_2$ & 
290.2539 & 0.0027 & 290.2902 & 0.0000 & 289.3913 & 0.0034 \\
15~${}^{2}\!$A$_2$ & 
290.4324 & 0.0003 & 290.4423 & 0.0018 & 289.6223 & 0.0021 
\\\hline
1~${}^{2}$B$_1$ &
287.6665 & 0.0002 & 287.6212 & 0.0004 & 286.7873 & 0.0002 \\
2~${}^{2}$B$_1$ &287.9761 & 0.0004 & 287.8897 & 0.0004 & 287.1690 & 0.0000 \\
3~${}^{2}$B$_1$ & 288.1053 & 0.0010 & 288.1946 & 0.0002 & 287.2745 & 0.0004 \\
4~${}^{2}$B$_1$  & 288.3737 & 0.0002 & 288.2723 & 0.0002 & 287.3794 & 0.0006 \\
5~${}^{2}$B$_1$ &288.4731 & 0.0027 & 288.4283 & 0.0045 & 287.5827 & 0.0022 \\
6~${}^{2}$B$_1$ &
288.6016 & 0.0000 & 288.6555 & 0.0020 & 287.8454 & 0.0011 \\
7~${}^{2}$B$_1$ &
288.6979 & 0.0022 & 288.7501 & 0.0000 & 287.9145 & 0.0001 \\
8~${}^{2}$B$_1$ &
288.7168 & 0.0055 & 288.8905 & 0.0046 & 288.0841 & 0.0011 \\
9~${}^{2}$B$_1$ &
289.1461 & 0.0002 & 289.0523 & 0.0000 & 288.1594 & 0.0027 \\
10~${}^{2}$B$_1$ &
289.3444 & 0.0004 & 289.4157 & 0.0004 & 288.3746 & 0.0008 \\
11~${}^{2}$B$_1$ &
289.4759 & 0.0005 & 289.4661 & 0.0001 & 288.4109 & 0.0000 \\
12~${}^{2}$B$_1$ &
289.6054 & 0.0004 & 289.5514 & 0.0015 & 288.5825 & 0.0007 \\
13~${}^{2}$B$_1$ &
289.6514 & 0.0007 & 289.5692 & 0.0000 & 288.7603 & 0.0008 \\
14~${}^{2}$B$_1$ &
289.7167 & 0.0003 & 289.9055 & 0.0003 & 289.0554 & 0.0002 \\
15~${}^{2}$B$_1$ &
289.9379 & 0.0001 & 290.0618 & 0.0001 & 289.1824 & 0.0001 \\\hline
\hline
\end{longtable}

\clearpage

\LTcapwidth=\textwidth
\begin{longtable}{c|cc|cc||cc|cc}
\caption{fc-CVS-EOM-CCSD results for the excitation energies (in eV) and oscillator strengths (f) of the propargyl radical, using either UHF or ROHF reference and two different basis sets. 
Electronic states are sorted per irreducible representation. The ground state is $\tilde X$\textsuperscript{ 2}B$_1$.
\label{tab:pople-cc-pvtz}}
\\\hline
& \multicolumn{4}{|c||}{6-311++G**}
& \multicolumn{4}{c}{cc-pVTZ} \\
\hline
& \multicolumn{2}{c|}{UHF reference}
& \multicolumn{2}{c||}{ROHF reference}
& \multicolumn{2}{c|}{UHF reference}
& \multicolumn{2}{c}{ROHF reference}
\\
Final state
& Energy (eV) & f  
& Energy (eV) & f 
& Energy (eV) & f  
& Energy (eV) & f  \\
\hline
1~${}^{2}$A$_1$
& 282.4086 & 0.0345 
& 282.6229 & 0.0345 
& 282.2056 & 0.0347 
& 281.9812 & 0.0347 
\\
2~${}^{2}$A$_1$
& 283.3861 & 0.0249 
& 283.1869 & 0.0250 
& 282.8897 & 0.0254 
& 283.1022 & 0.0252 
\\
3~${}^{2}$A$_1$
& 284.8117 & 0.0006 
& 285.1108 & 0.0003 
& 284.8573 & 0.0003 
& 284.5501 & 0.0006 
\\
4~${}^{2}$A$_1$
& 286.8146 & 0.0168 & 286.6201 & 0.0186 & 286.4086 & 0.0203 & 286.6293 & 0.0188 
\\
5~${}^{2}$A$_1$
& 287.9220 & 0.0144 & 288.3984 & 0.0139 & 288.1798 & 0.0151 & 287.6841 & 0.0155 
\\
6~${}^{2}$A$_1$& 288.4702 & 0.0403 & 288.6357 & 0.0384 & 288.7187 & 0.0437 & 288.5134 & 0.0446 
\\
7~${}^{2}$A$_1$ &
289.1469 & 0.0001 & 289.0631 & 0.0000 & 289.4690 & 0.0036 & 289.5083 & 0.0040 
\\
8~${}^{2}$A$_1$ & 
289.9334 & 0.0019 & 290.0664 & 0.0004 & 290.2215 & 0.0087 & 290.4089 & 0.0071 
\\
9~${}^{2}$A$_1$ & 290.3775 & 0.0000 & 290.2587 & 0.0000 & 291.0596 & 0.0022 & 290.9964 & 0.0046 
\\
10~${}^{2}$A$_1$ & 
290.3999 & 0.0164 & 290.3331 & 0.0027 & 294.1167 & 0.0001 & 294.0840 & 0.0007 
\\
11~${}^{2}$A$_1$ & 
290.4845 & 0.0001 & 290.4282 & 0.0159 & 294.5142 & 0.0013 & 294.5371 & 0.0017 
\\
12~${}^{2}$A$_1$ & 
290.6223 & 0.0044 & 290.4862 & 0.0066 & 294.7643 & 0.0337 & 294.7324 & 0.0328 
\\
13~${}^{2}$A$_1$ & 290.9049 & 0.0025 & 290.8742 & 0.0038 & 295.0770 & 0.0041 & 295.2067 & 0.0025 
\\
14~${}^{2}$A$_1$ & 
291.1043 & 0.0017 & 291.3012 & 0.0001 & 295.2898 & 0.0407 & 295.3303 & 0.0410 
\\
15~${}^{2}$A$_1$ & 
291.3588 & 0.0002 & 291.3073 & 0.0025 & 295.8039 & 0.0049 & 295.6744 & 0.0055 
\\\hline
1~${}^{2}$A$_2$ & 285.8427 & 0.0001 & 285.7346 & 0.0002 & 285.4664 & 0.0007 & 285.5833 & 0.0000 \\
2~${}^{2}$A$_2$
& 285.9338 & 0.0042 & 285.8002 & 0.0199 & 285.5344 & 0.0201 & 285.6714 & 0.0042 \\
3~${}^{2}$A$_2$
& 286.6000 & 0.0666 & 286.8594 & 0.0707 & 286.5977 & 0.0574 & 286.3260 & 0.0636 
\\
4~${}^{2}$A$_2$
& 286.6972 & 0.0234 & 286.8934 & 0.0040 & 286.6514 & 0.0187 & 286.4492 & 0.0286 \\
5~${}^{2}$A$_2$
& 289.0048 & 0.0046 & 288.9527 & 0.0048 & 288.8919 & 0.0047 & 288.8958 & 0.0045 
\\
6~${}^{2}$A$_2$
& 289.5016 & 0.0021 & 289.4464 & 0.0029 & 289.2728 & 0.0052 & 289.4003 & 0.0055 
\\
7~${}^{2}$A$_2$
& 289.5570 & 0.0069 & 289.4699 & 0.0065 & 290.6580 & 0.0029 & 290.6990 & 0.0042 
\\
8~${}^{2}$A$_2$
& 289.7613 & 0.0038 & 289.8033 & 0.0002 & 291.2767 & 0.0244 & 291.3516 & 0.0233 
\\
9~${}^{2}$A$_2$
& 289.9594 & 0.0003 
& 289.9671 & 0.0033 
& 292.0554 & 0.0001 
& 292.1860 & 0.0002 
\\
10~${}^{2}$A$_2$
& 290.1500 & 0.0000 
& 290.1907 & 0.0013 
& 292.2138 & 0.0008 
& 292.2264 & 0.0007 
\\
11~${}^{2}$A$_2$
& 290.2074 & 0.0022 & 290.3088 & 0.0011 & 292.5841 & 0.0001 & 292.4240 & 0.0006 
\\
12~${}^{2}$A$_2$
& 290.3968 & 0.0021 
& 290.3454 & 0.0019 
& 292.6530 & 0.0007 
& 292.4578 & 0.0000 
\\
13~${}^{2}$A$_2$
& 290.6573 & 0.0030 
& 290.5882 & 0.0032 
& 293.9671 & 0.0031 
& 293.9882 & 0.0013 
\\
14~${}^{2}$A$_2$ & 290.7105 & 0.0028 & 290.7855 & 0.0003 & 294.5844 & 0.0220 & 294.5188 & 0.0249 
\\
15~${}^{2}$A$_2$
& 290.9425 & 0.0005 & 290.8826 & 0.0028 & 294.8267 & 0.0069 & 294.9005 & 0.0025
\\\hline
1~${}^{2}$B$_1$  
& 287.9224 & 0.0002 
& 287.8728 & 0.0005 
& 288.2358 & 0.0007 
& 288.2672 & 0.0002 
\\
2~${}^{2}$B$_1$
& 288.3442 & 0.0012 & 288.2767 & 0.0005 & 288.4986 & 0.0004 & 288.5707 & 0.0003 \\
3~${}^{2}$B$_1$
& 288.3638 & 0.0004 & 288.4443 & 0.0004 & 289.1105 & 0.0107 & 289.0494 & 0.0110 \\
4~${}^{2}$B$_1$
& 288.6269 & 0.0002 & 288.5248 & 0.0004 & 289.4800 & 0.0054 & 289.5130 & 0.0056 \\
5~${}^{2}$B$_1$
& 288.7638 & 0.0025 & 288.7102 & 0.0042 & 289.7680 & 0.0001 & 289.8586 & 0.0001 \\
6~${}^{2}$B$_1$
& 288.8680 & 0.0000 & 289.0172 & 0.0003 & 290.2500 & 0.0010 & 290.1775 & 0.0001 \\
7~${}^{2}$B$_1$
& 289.0029 & 0.0058 & 289.0342 & 0.0018 & 290.3139 & 0.0012 & 290.3320 & 0.0000 \\
8~${}^{2}$B$_1$
& 289.0697 & 0.0021 & 289.1799 & 0.0048 & 290.4071 & 0.0022 & 290.4035 & 0.0055 \\
9~${}^{2}$B$_1$
& 289.3930 & 0.0002 & 289.3006 & -0.0000 & 290.6992 & 0.0026 & 290.4894 & 0.0016 \\
10~${}^{2}$B$_1$
& 289.6057 & 0.0005 & 289.7522 & 0.0002 & 290.8658 & 0.0013 & 290.7364 & 0.0012 \\
11~${}^{2}$B$_1$
& 289.8530 & 0.0006 & 289.7904 & 0.0007 & 291.8116 & 0.0021 & 291.8761 & 0.0017 \\
12~${}^{2}$B$_1$
& 289.9621 & 0.0004 & 289.8920 & 0.0005 & 291.8803 & 0.0000 & 291.9192 & 0.0000 \\
13~${}^{2}$B$_1$
& 290.0814 & 0.0003 & 290.0213 & 0.0010 & 292.2258 & 0.0002 & 292.3016 & 0.0003 \\
14~${}^{2}$B$_1$
& 290.1147 & 0.0011 & 290.2737 & 0.0004 & 292.3517 & 0.0044 & 292.3812 & 0.0047 \\
15~${}^{2}$B$_1$
& 290.3320 & 0.0002 & 290.4565 & 0.0001 & 292.5758 & 0.0001 & 292.5004 & 0.0000 \\\hline
\hline
\end{longtable}

\endgroup

\clearpage

\subsection{Natural Transition Orbitals (NTOs)}
\label{sec:NTOs}
\vspace{-0.2cm}
\begin{table}[hbpt!]
\caption{NTOs from (unrestricted) fc-CVS-EOM-CCSD/aug-cc-pVTZ calculations for selected core excitations (MO isovalue = 0.05). The ground state is $\tilde{X}~{}^2$B$_1$.\label{tab:NTOs:EOM}}
\begin{tabular}{l|cc|cc}
  \hline
& Hole & Particle & Hole & Particle  \\\hline
\textbf{A}
& 
\multirow{4}{0.2\linewidth}{\includegraphics[width=\linewidth]{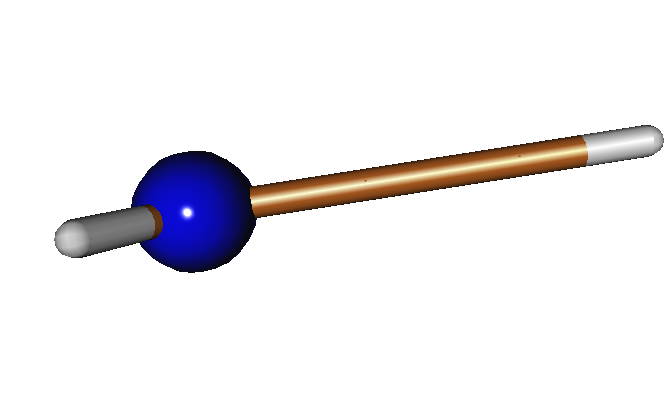}} & 
\multirow{4}{0.2\linewidth}{\includegraphics[width=\linewidth]{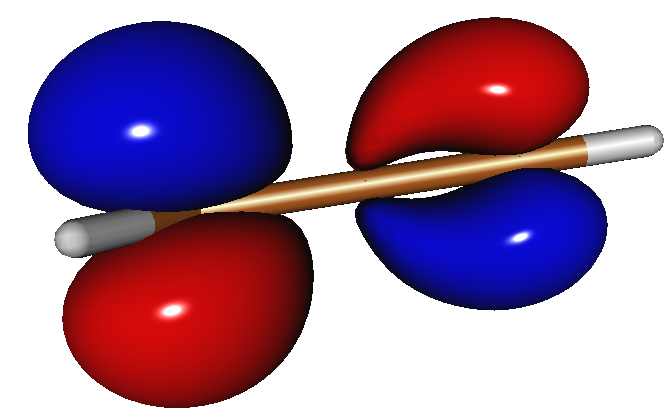}} & 
& 
\\[-.25cm]
E = 282.136 eV & & & & \\[-.25cm]
f = 0.0345 & & & &  \\[-.2cm]
1~$^2\!$A$_1$ & & & & 
\\\hline
\textbf{A} 
& 
\multirow{4}{0.2\linewidth}{\includegraphics[width=\linewidth]{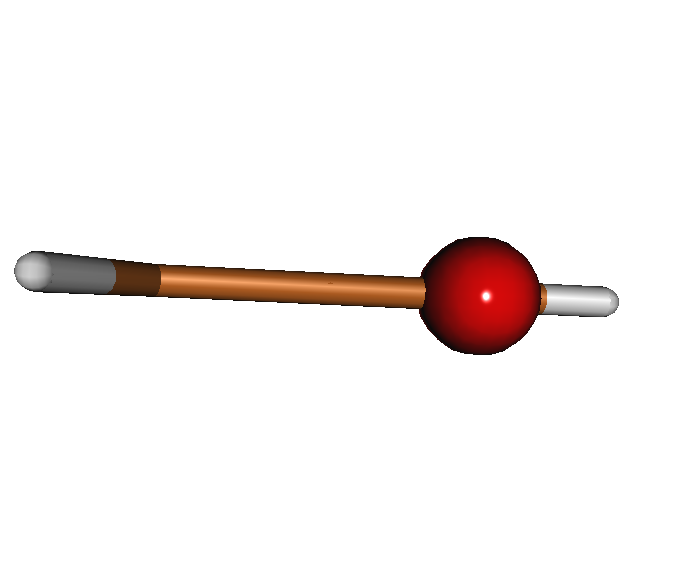}} & 
\multirow{4}{0.2\linewidth}{\includegraphics[width=\linewidth]{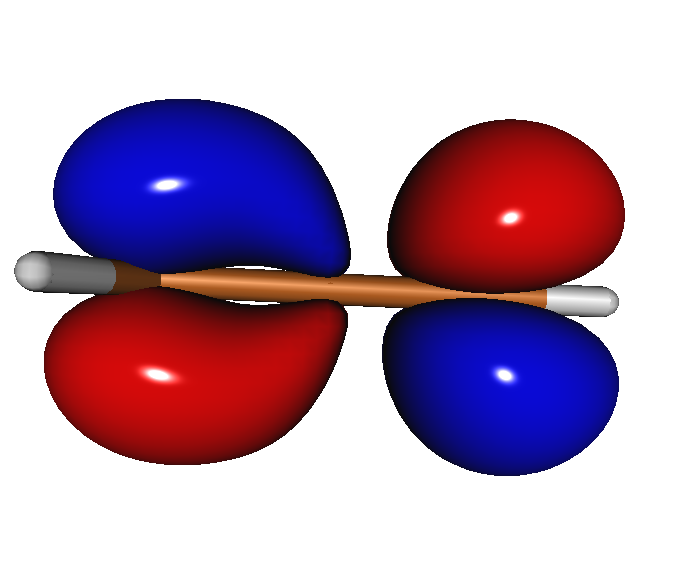}} & & 
\\[-.25cm]
E = 282.811 eV & & & & \\[-.25cm]
f = 0.0253 & & & & \\[-.25cm]
2~$^2\!$A$_1$ & & & & \\
\hline
\textbf{B} 
& 
\multirow{4}{0.2\linewidth}{\includegraphics[width=\linewidth]{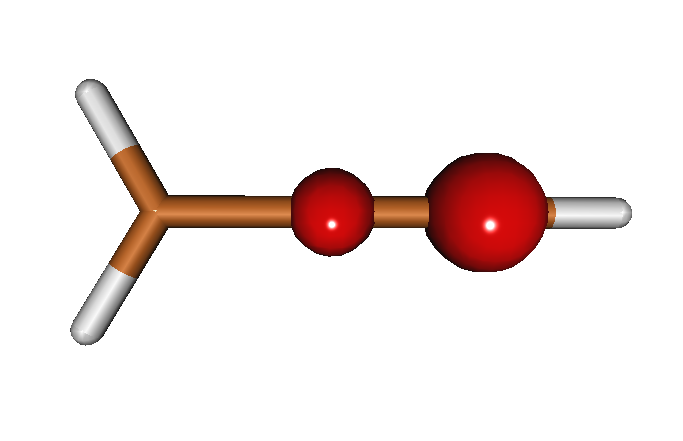}} & 
\multirow{4}{0.2\linewidth}{\includegraphics[width=\linewidth]{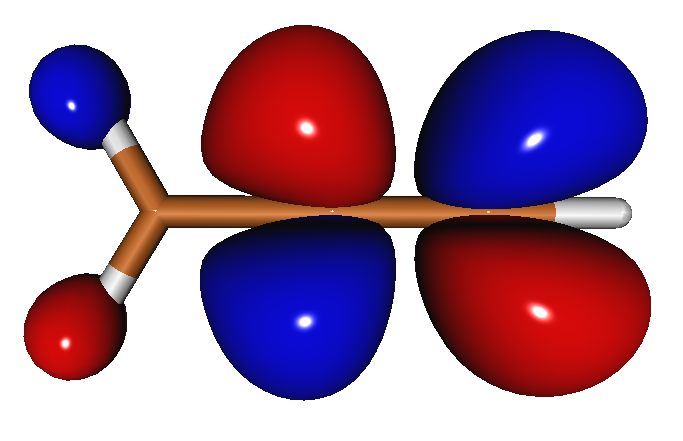}} & 
\multirow{4}{0.2\linewidth}{\includegraphics[width=\linewidth]{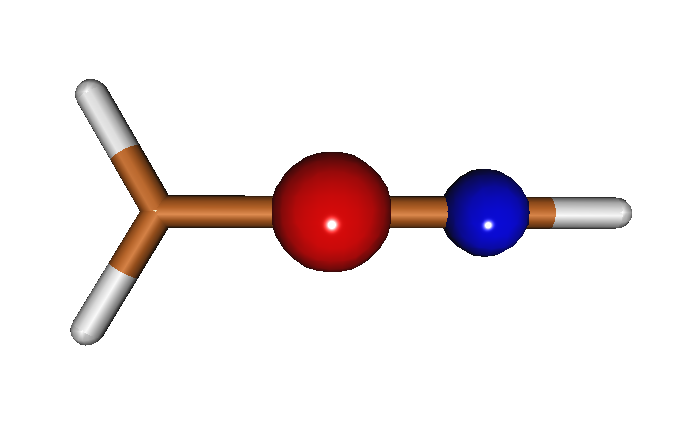}} & 
\multirow{4}{0.2\linewidth}{\includegraphics[width=\linewidth]{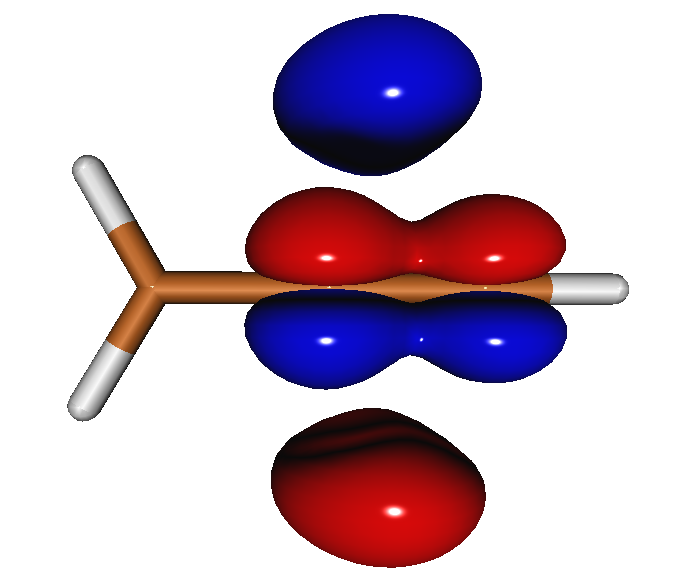}} 
\\[-.25cm]
E = 285.436 eV & & & & \\[-.25cm]
f = 0.0198 & & & & \\[-.25cm]
2~$^2\!$A$_2$ & & & & \\[.3cm]
\hline
\textbf{C} 
& 
\multirow{4}{0.2\linewidth}{\includegraphics[width=\linewidth]{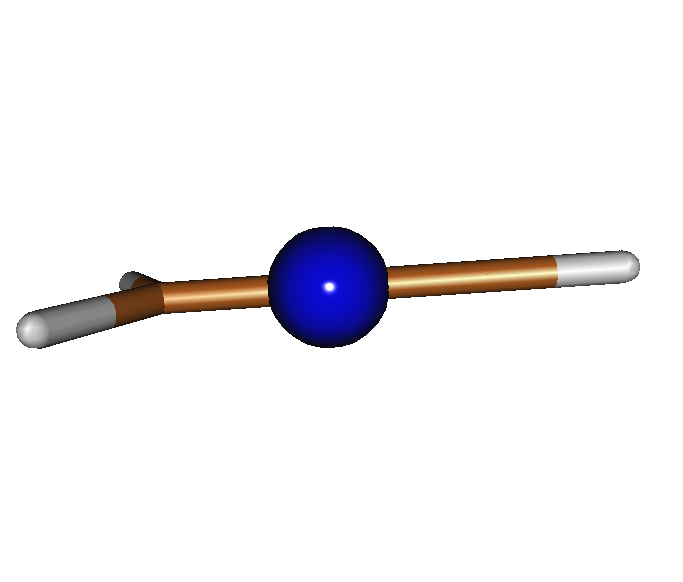}} & 
\multirow{4}{0.2\linewidth}{\includegraphics[width=\linewidth]{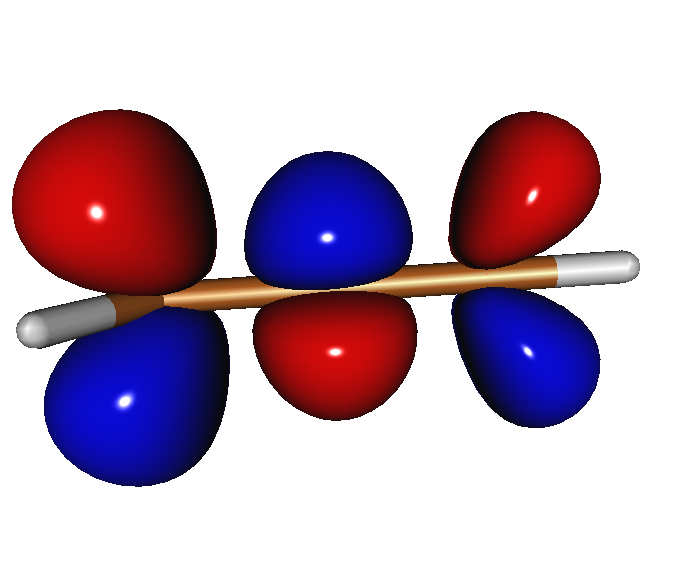}} & & \\[-.2cm]
E = 286.235 eV & & & & \\[-.2cm]
f = 0.0189 & & & & \\[-.2cm]
4~$^2\!$A$_1$ & & & & \\
\hline
\textbf{D} 
& 
\multirow{4}{0.2\linewidth}{\includegraphics[width=\linewidth]{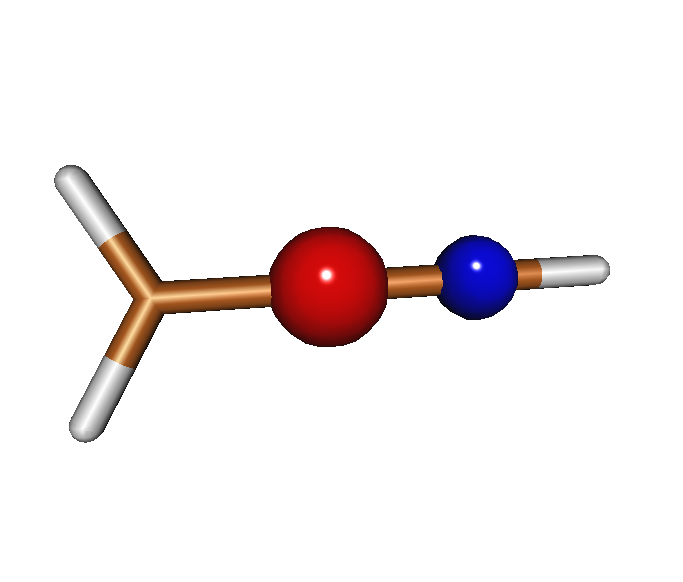}} & 
\multirow{4}{0.2\linewidth}{\includegraphics[width=\linewidth]{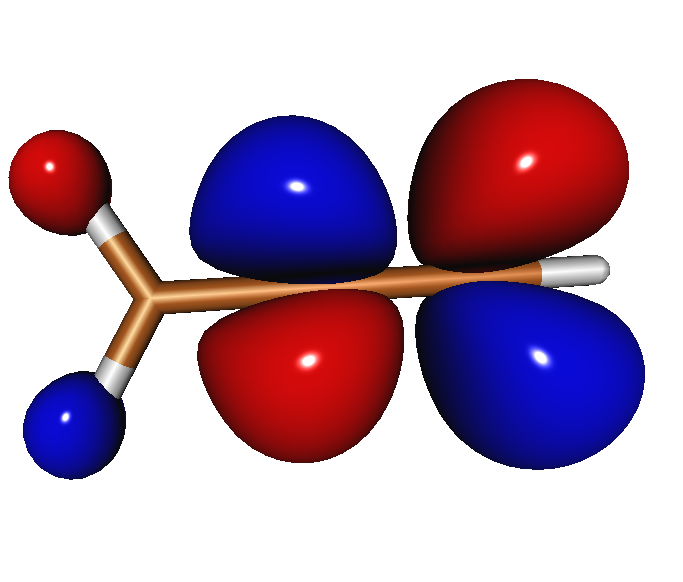}} & & \\[-.25cm]
E = 286.500 eV & & & & \\[-.2cm]
f = 0.0747 & & & & \\[-.2cm]
3~$^2\!$A$_2$ & & & & \\
\hline
\textbf{E} 
& 
\multirow{4}{0.2\linewidth}{\includegraphics[width=\linewidth]{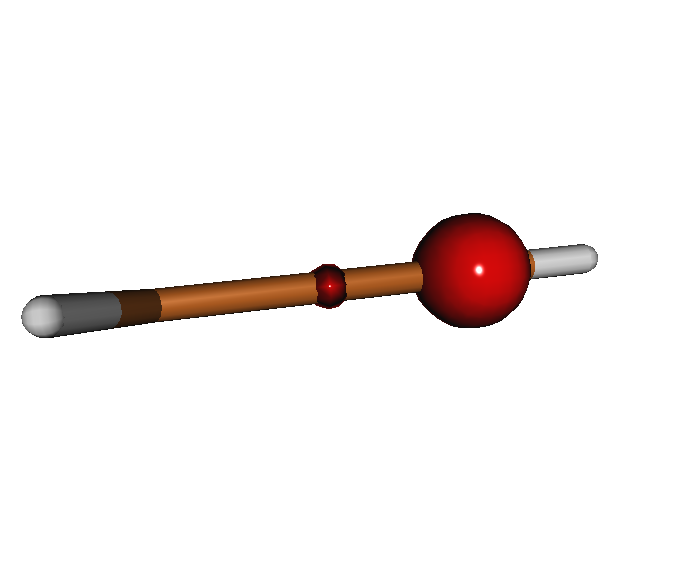}} & 
\multirow{4}{0.2\linewidth}{\includegraphics[width=\linewidth]{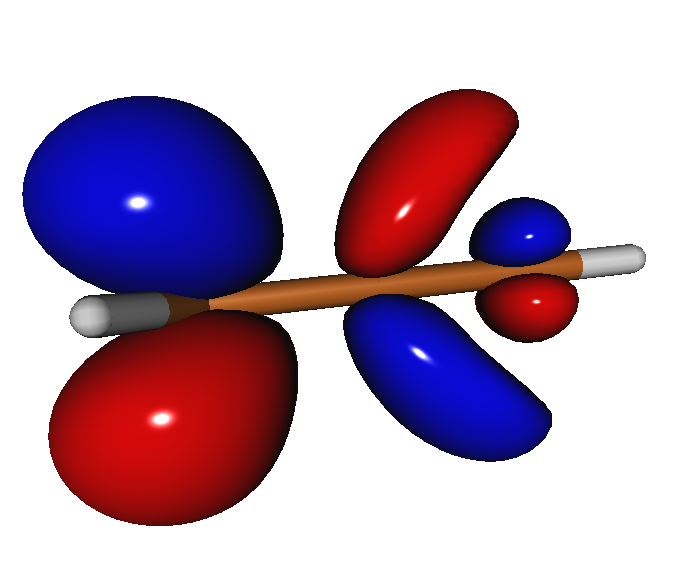}} & & \\[-.2cm]
E = 288.044 eV & & & & \\[-.2cm]
f = 0.0139 & & & & \\[-.2cm]
5~$^2\!$A$_1$ & & & & \\[.2cm]
\hline
\textbf{E} 
& 
\multirow{4}{0.2\linewidth}{\includegraphics[width=\linewidth]{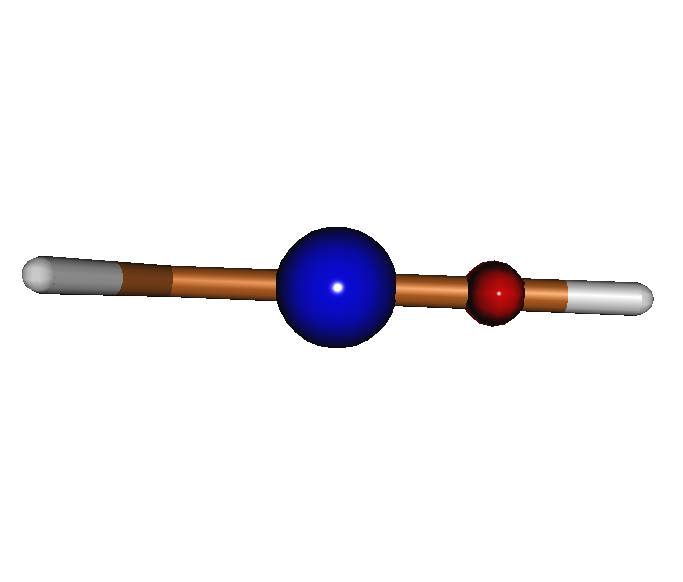}} & 
\multirow{4}{0.2\linewidth}{\includegraphics[width=\linewidth]{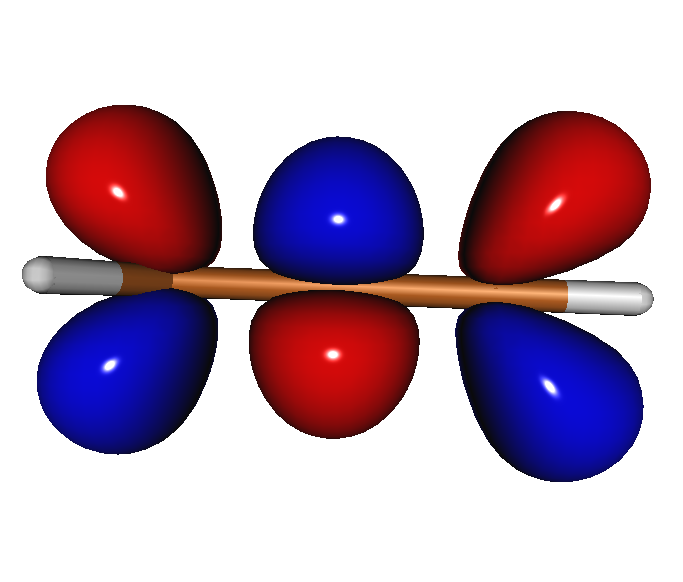}} & & \\[-.2cm]
E = 288.275 eV & & & & \\[-.22cm]
f = 0.037774 & & & & \\[-.22cm]
6~$^2\!$A$_1$ & & & & 
\\\hline
& \multirow{4}{0.2\linewidth}{\includegraphics[width=\linewidth]{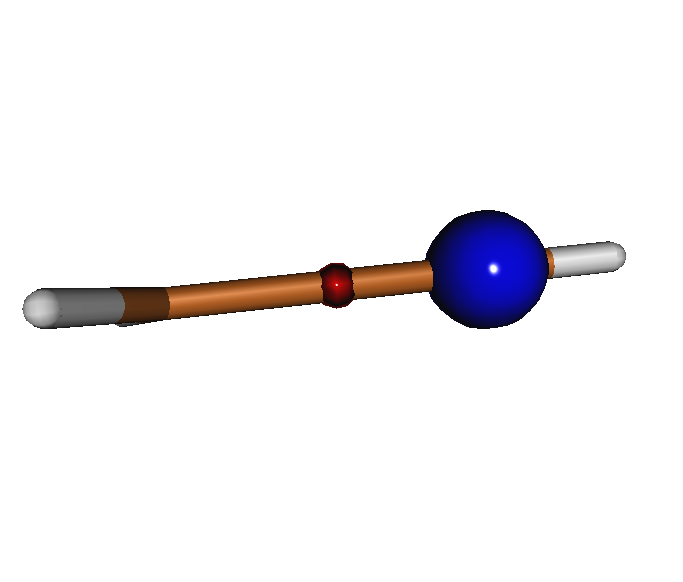}} 
& \multirow{4}{0.2\linewidth}{\includegraphics[width=\linewidth]{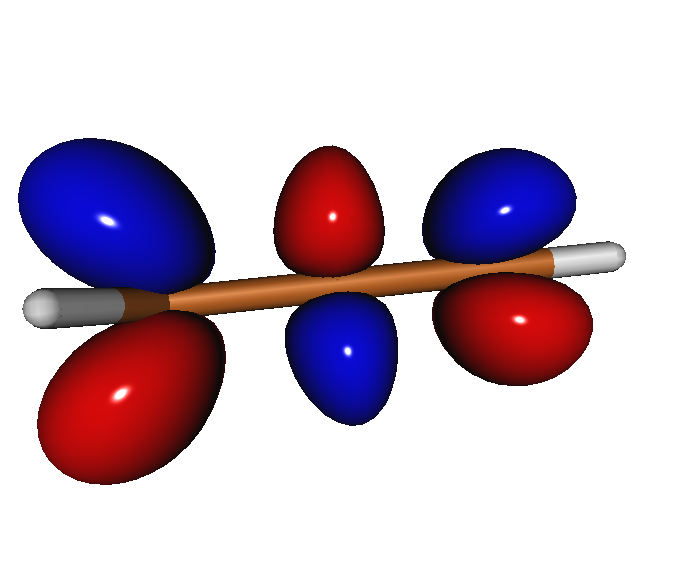}} & & \\[-.22cm]
E = 289.893 eV & & & & \\[-.22cm]
f = 0.0133 & & & & \\[-.2cm]
11~$^2\!$A$_1$ & & & & 
\\\hline
\end{tabular}
\end{table}

\clearpage

\begin{table}[hbpt!]
\caption{NTOs from (unrestricted) CVS-ADC(2)-x/aug-cc-pVTZ for selected core excitations (MO isovalue = 0.05). The ground state is $\tilde X$\textsuperscript{ 2}B$_1$. 
\label{tab:NTOs:ADC}}
\begin{tabular}{l|cc|cc}
\hline 
& Hole & Particle 
& Hole & Particle \\
\hline
\textbf{A} 
& \multirow{4}{0.2\linewidth}{\includegraphics[width=\linewidth]{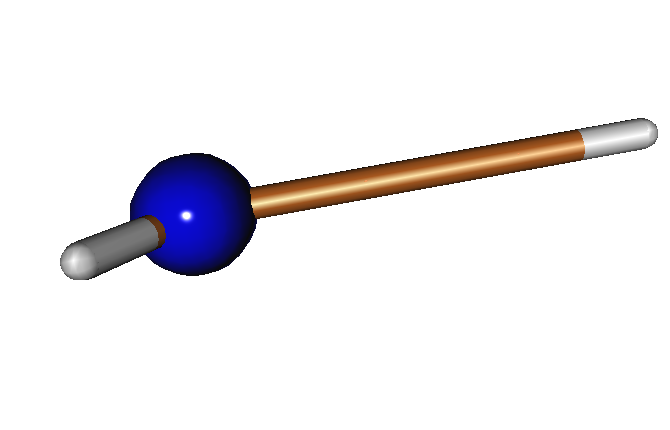}} 
& \multirow{4}{0.2\linewidth}{\includegraphics[width=\linewidth]{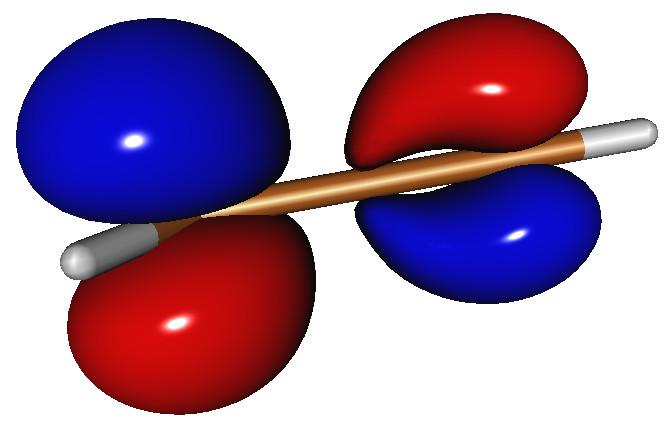}} & & \\[-.2cm]
E = 282.154 eV & & & & \\[-.2cm]
f = 0.0306 & & & & \\[-.2cm]
1~$^2\!$A$_1$ & & & & \\
\hline
\textbf{A} 
& \multirow{4}{0.2\linewidth}{\includegraphics[width=\linewidth]{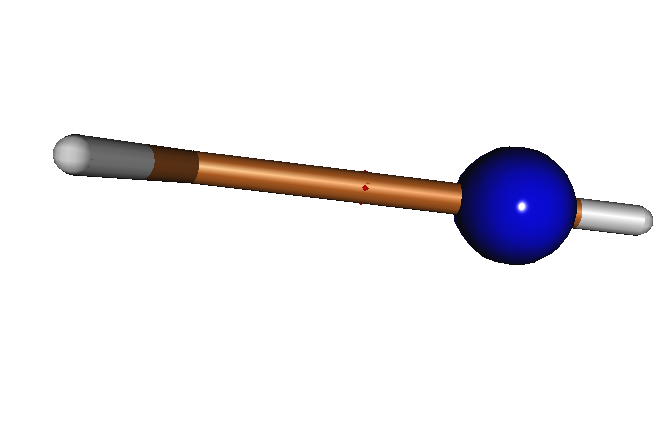}} 
& \multirow{4}{0.2\linewidth}{\includegraphics[width=\linewidth]{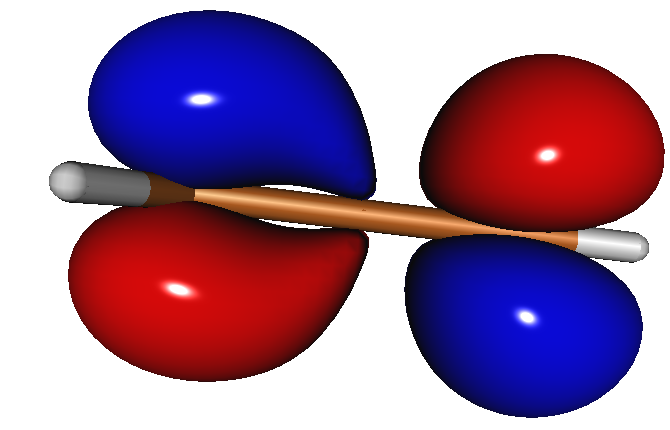}} & & \\[-.2cm]
E = 282.361  eV & & & & \\[-.2cm]
f = 0.0221 & & & & \\[-.2cm]
2~$^2\!$A$_1$ & & & & \\
\hline
\textbf{B} 
& 
\multirow{4}{0.2\linewidth}{\includegraphics[width=\linewidth]{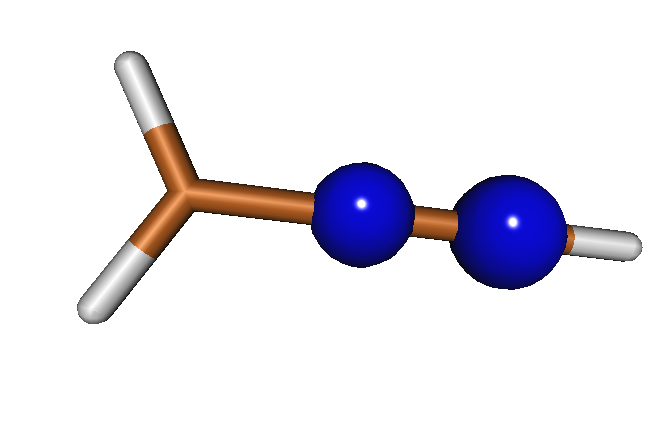}} 
& \multirow{4}{0.2\linewidth}{\includegraphics[width=\linewidth]{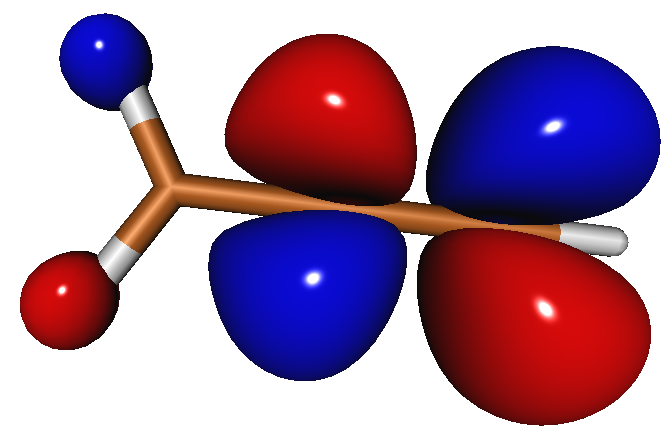}} & 
\multirow{4}{0.2\linewidth}{\includegraphics[width=\linewidth]{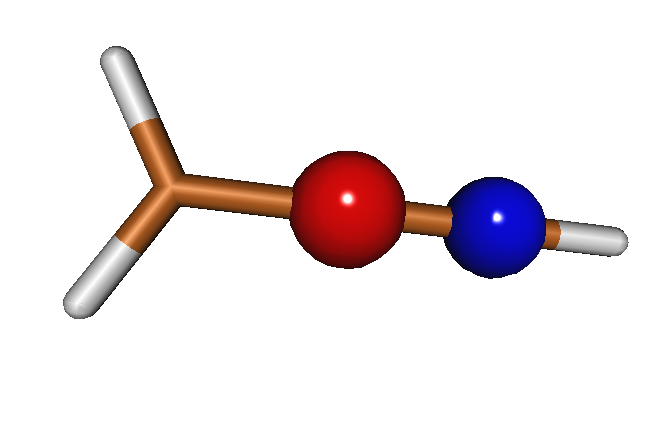}} 
& \multirow{4}{0.2\linewidth}{\includegraphics[width=\linewidth]{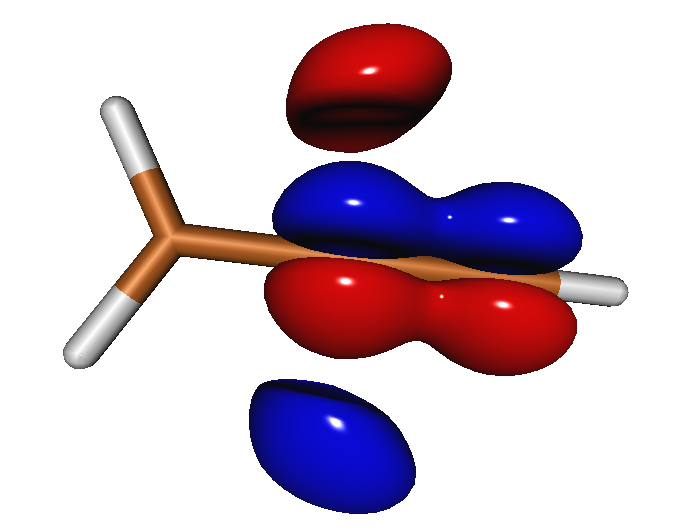}} \\[-.2cm]
E = 284.820 eV & & & & \\[-.2cm]
f = 0.0234 & & & & \\[-.2cm]
2~$^2\!$A$_2$ & & & & \\
\hline
\textbf{C-D} & 
\multirow{4}{0.2\linewidth}{\includegraphics[width=\linewidth]{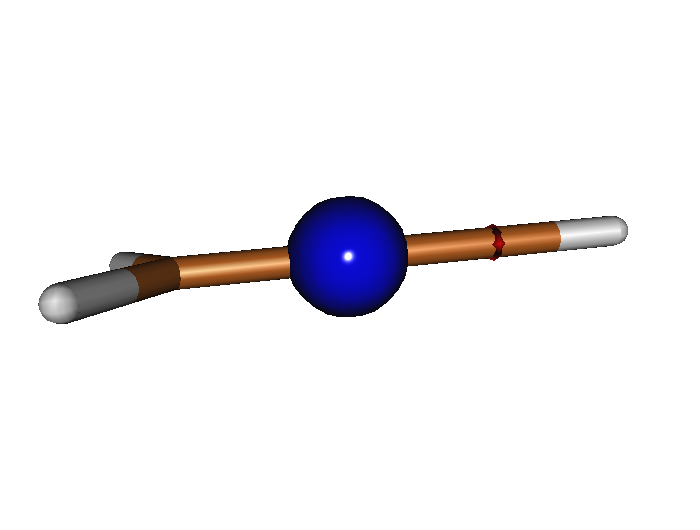}} & 
\multirow{4}{0.2\linewidth}{\includegraphics[width=\linewidth]{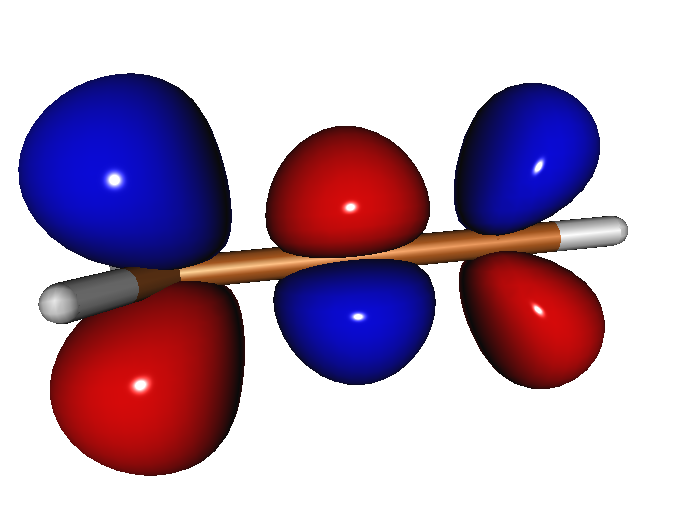}} & & \\[-.2cm]
E = 285.715 eV & & & & \\[-.2cm]
f = 0.0201 & & & & \\[-.2cm]
4~$^2\!$A$_1$ & & & & \\
\hline
\textbf{C-D} & 
\multirow{4}{0.2\linewidth}{\includegraphics[width=\linewidth]{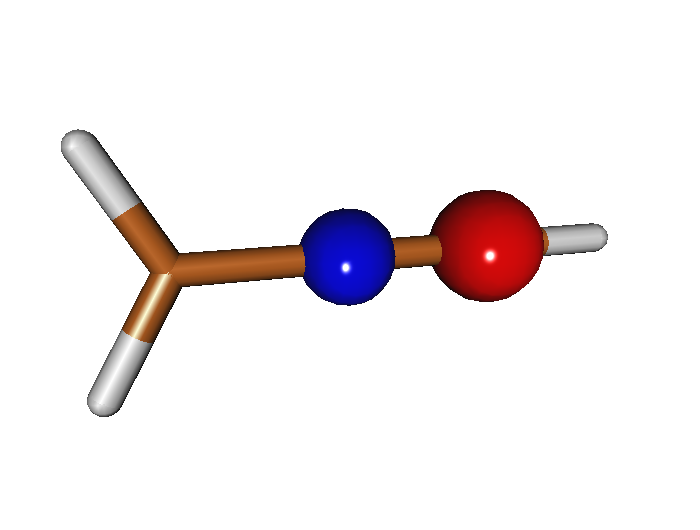}} & \multirow{4}{0.2\linewidth}{\includegraphics[width=\linewidth]{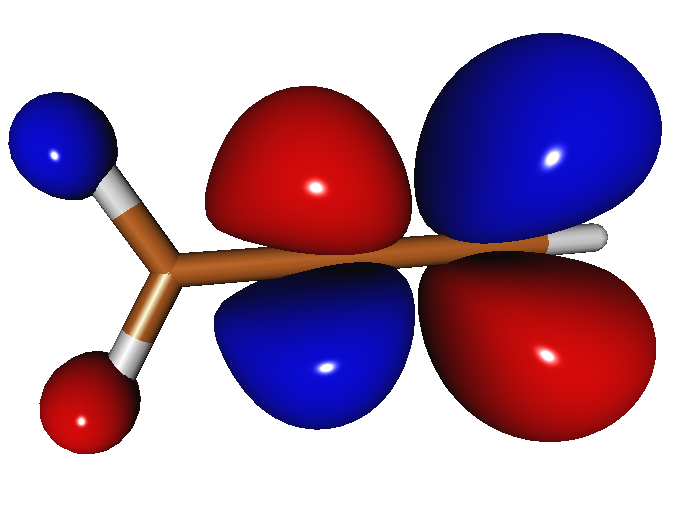}} & & \\[-.2cm]
E = 285.810 eV & & & & \\[-.2cm]
f = 0.0277 & & & & \\[-.2cm]
3~$^2\!$A$_2$ & & & & \\
\hline
\textbf{C-D} & 
\multirow{4}{0.2\linewidth}{\includegraphics[width=\linewidth]{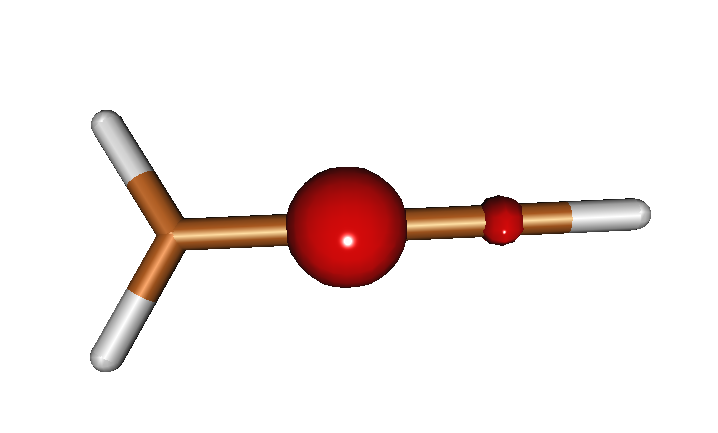}} & \multirow{4}{0.2\linewidth}{\includegraphics[width=\linewidth]{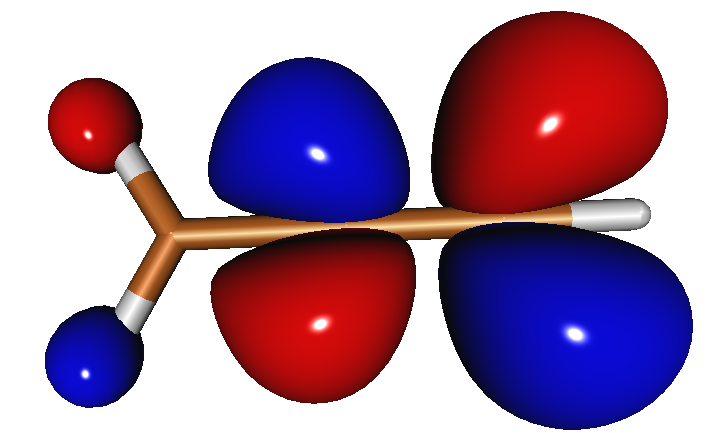}} & & \\[-.2cm]
E = 285.924 eV & & & & \\[-.2cm]
f = 0.024001 & & & & \\[-.2cm]
4~$^2\!$A$_2$ & & & & 
\\
\hline
        %
\textbf{E} & 
\multirow{4}{0.2\linewidth}{\includegraphics[width=\linewidth]{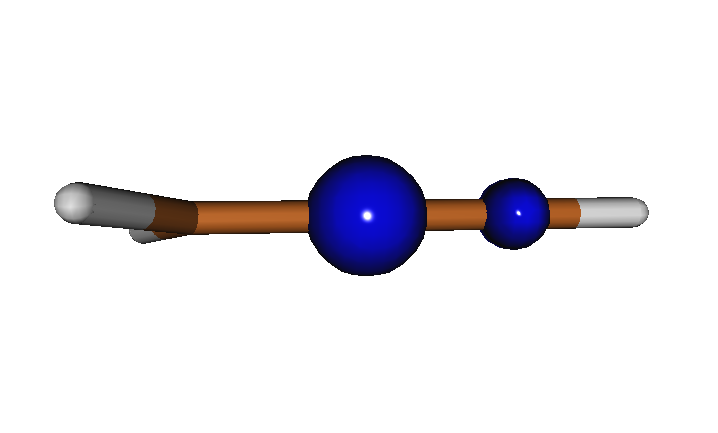}} & \multirow{4}{0.2\linewidth}{\includegraphics[width=\linewidth]{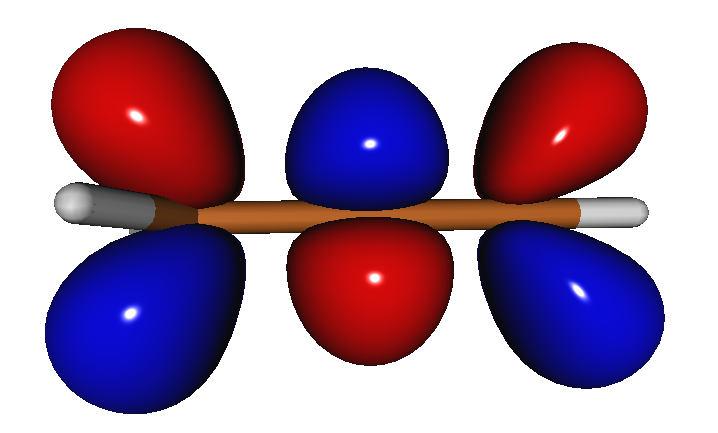}} & & \\[-.2cm]
E = 287.726 eV & & & & \\[-.2cm]
f = 0.0244 & & & & \\[-.2cm]
7~$^2\!$A$_1$ & & & & \\
\hline
 & 
 \multirow{4}{0.2\linewidth}{\includegraphics[width=\linewidth]{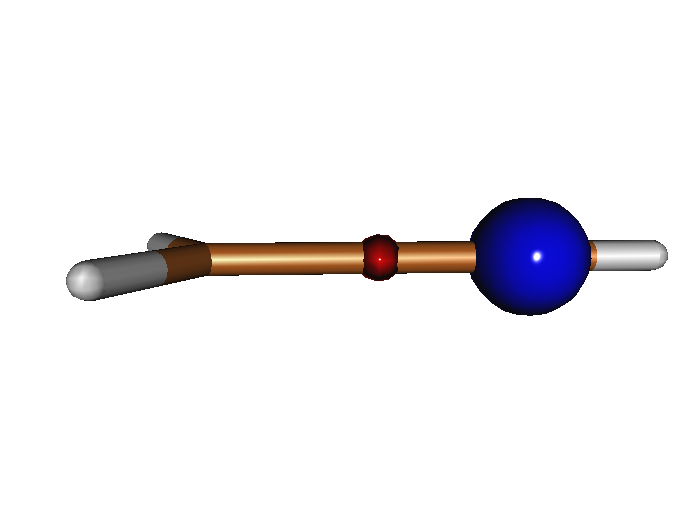}} & \multirow{4}{0.2\linewidth}{\includegraphics[width=\linewidth]{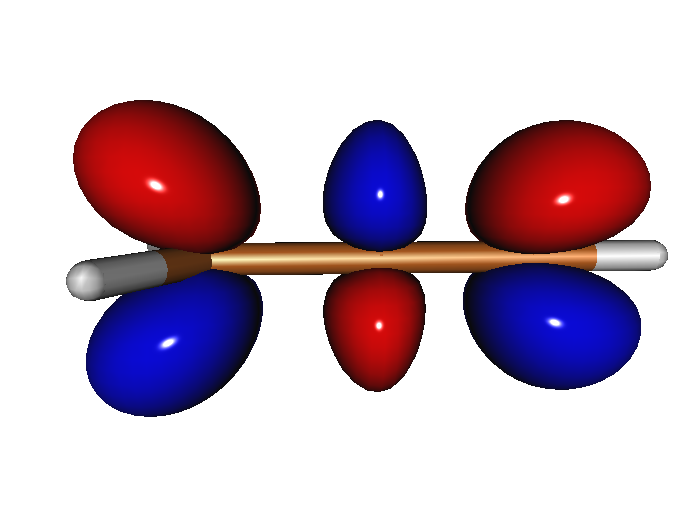}} & & \\[-.2cm]
E = 289.044 eV & & & & \\[-.2cm]
f = 0.0109 & & & & \\[-.2cm]
11~$^2\!$A$_1$ & & & & \\\hline
\end{tabular}
\end{table}

\clearpage

\subsection{Vibrationally Resolved Spectra} 
\label{sec:theory_SI_vib}



\noindent



\begin{figure}[H]
    \centering
    \includegraphics[width=0.49\linewidth]{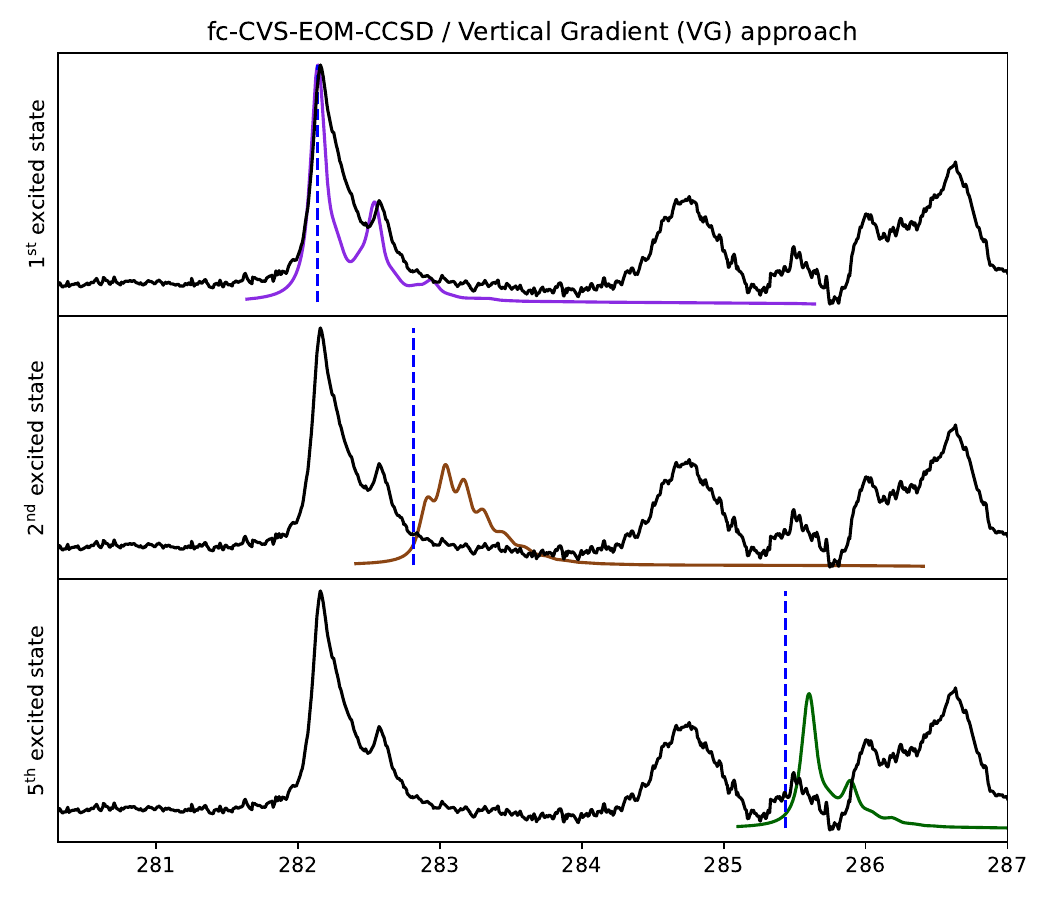}
    \includegraphics[width=0.49\linewidth]{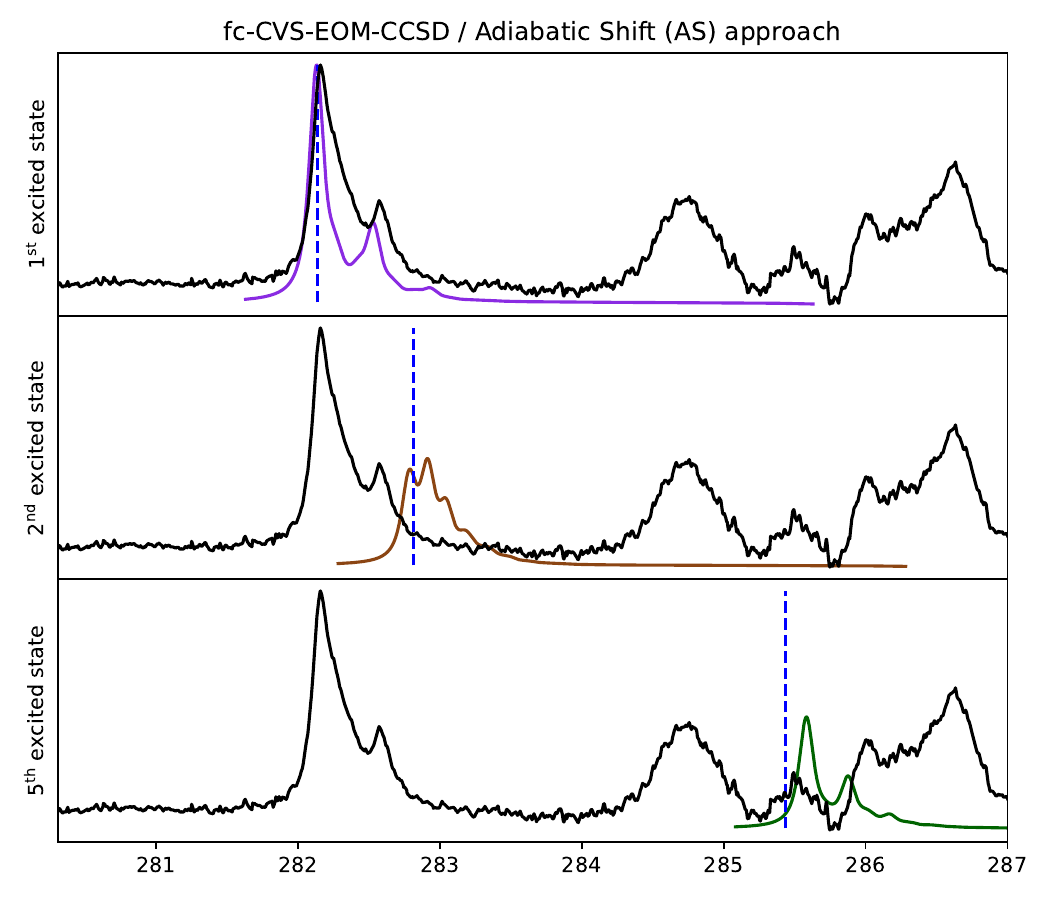}
    \includegraphics[width=0.49\linewidth]{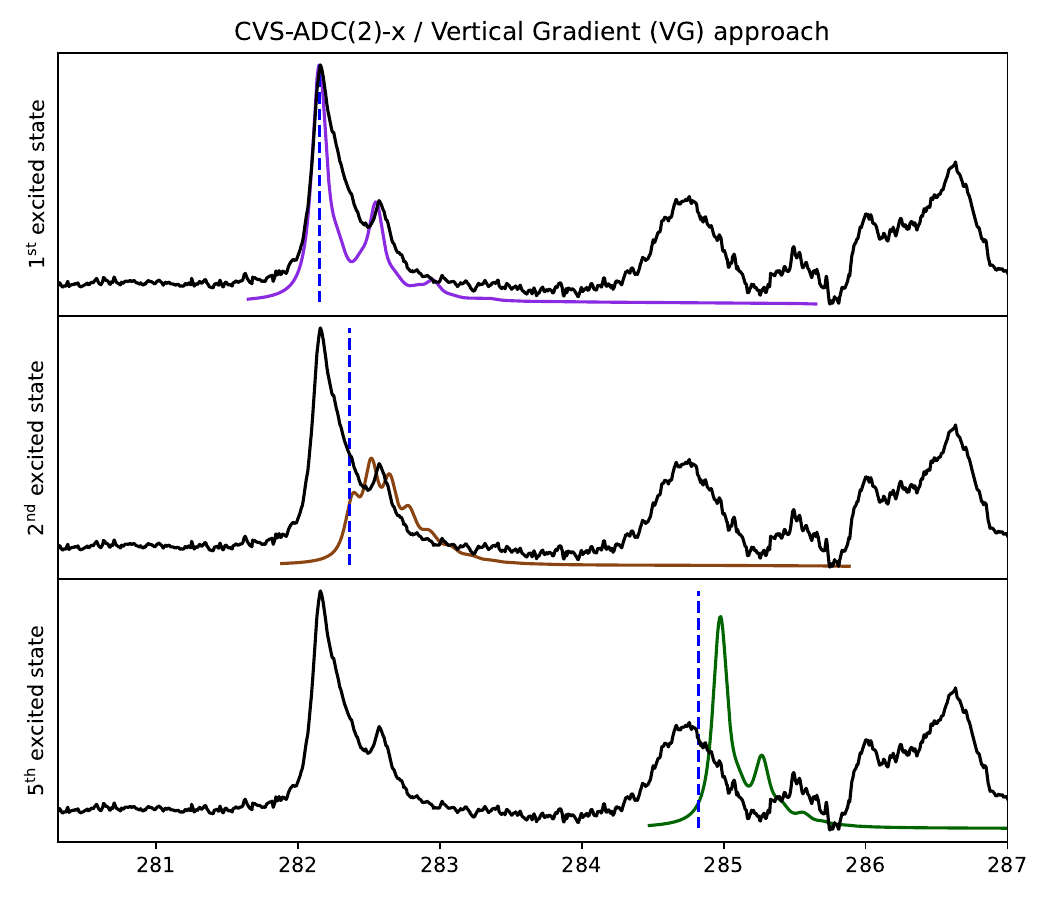}
    \includegraphics[width=0.49\linewidth]{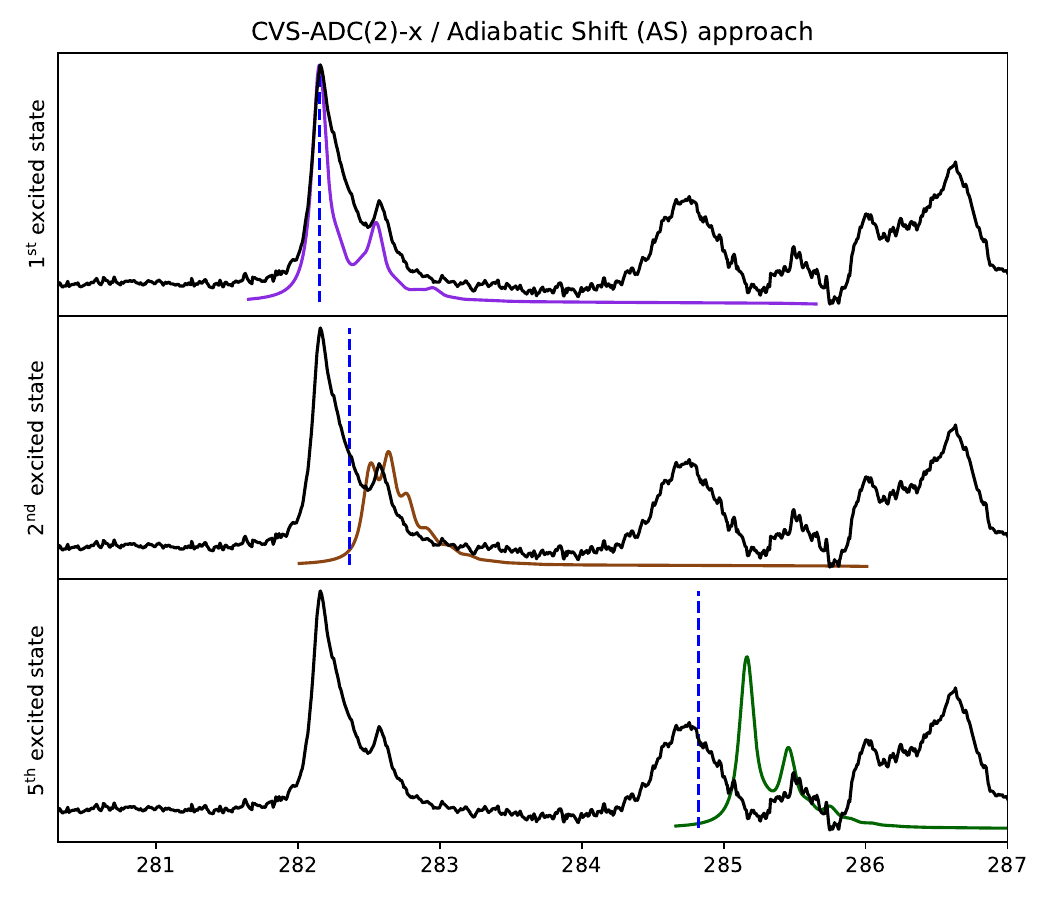}
    \caption{Vibrationally resolved spectral bands from a time-dependent approach based on (left) the Vertical Gradient approximation for the potential energy surfaces (PES), and (right) the Adiabatic Shift approximation. 
    Top: Excitation energies and electronic transition densities from fc-CVS-EOM-CCSD. Bottom: excitation energies and electronic transition densities from CVS-ADC(2)-x. The spectral bands were broadened by a Voigt function with a Gaussian broadening of 100\,meV and Lorentzian broadening of 50\,meV. The fc-CVS-EOM-CCSD VG spectra are shifted by +200\,meV and the fc-CVS-EOM-CCSD AS spectra by +250\,meV. The CVS-ADC(2)-x VG results are shifted by +175\,meV and the CVS-ADC(2)-x AS results were shifted by +318\,meV. Vertical lines correspond to the excitations energies of Figs.~\ref{fig:CCSD:XAS} and~\ref{fig:ADC:XAS}, {and of Fig.~4 of the main paper}.
    }
    \label{fig:FC}
\end{figure}

\begin{figure}[H]
    \centering
    \includegraphics[width=0.49\linewidth]{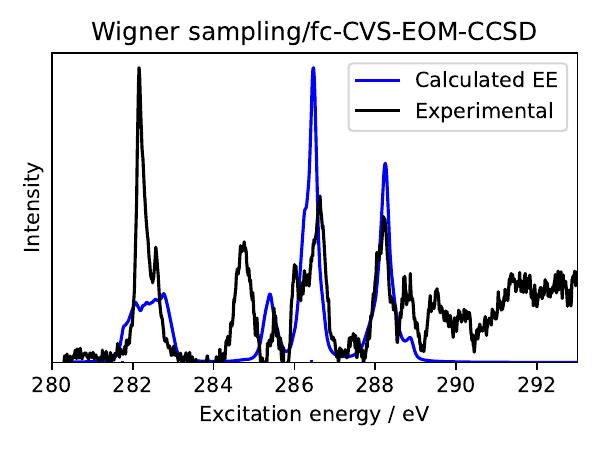}
    \includegraphics[width=0.49\linewidth]{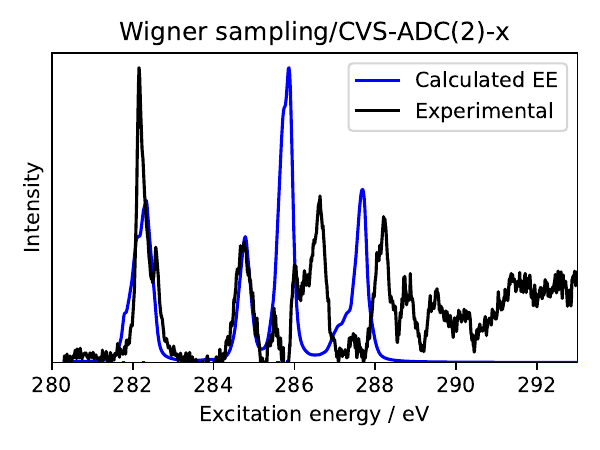}
    \caption{Comparison between experimental NEXAFS spectrum (black line) and the Wigner sampling at 300\,K consisting of 200 geometries in blue. Calculations were performed at the (left) fc-CVS-EOM-CCSD/aug-cc-pVTZ and (right) CVS-ADC(2)-x/aug-cc-pVTZ level of theory.
    }
    \label{fig:Wigner}
\end{figure}

\begin{figure}[H]
    \centering
    \includegraphics[width=\linewidth]{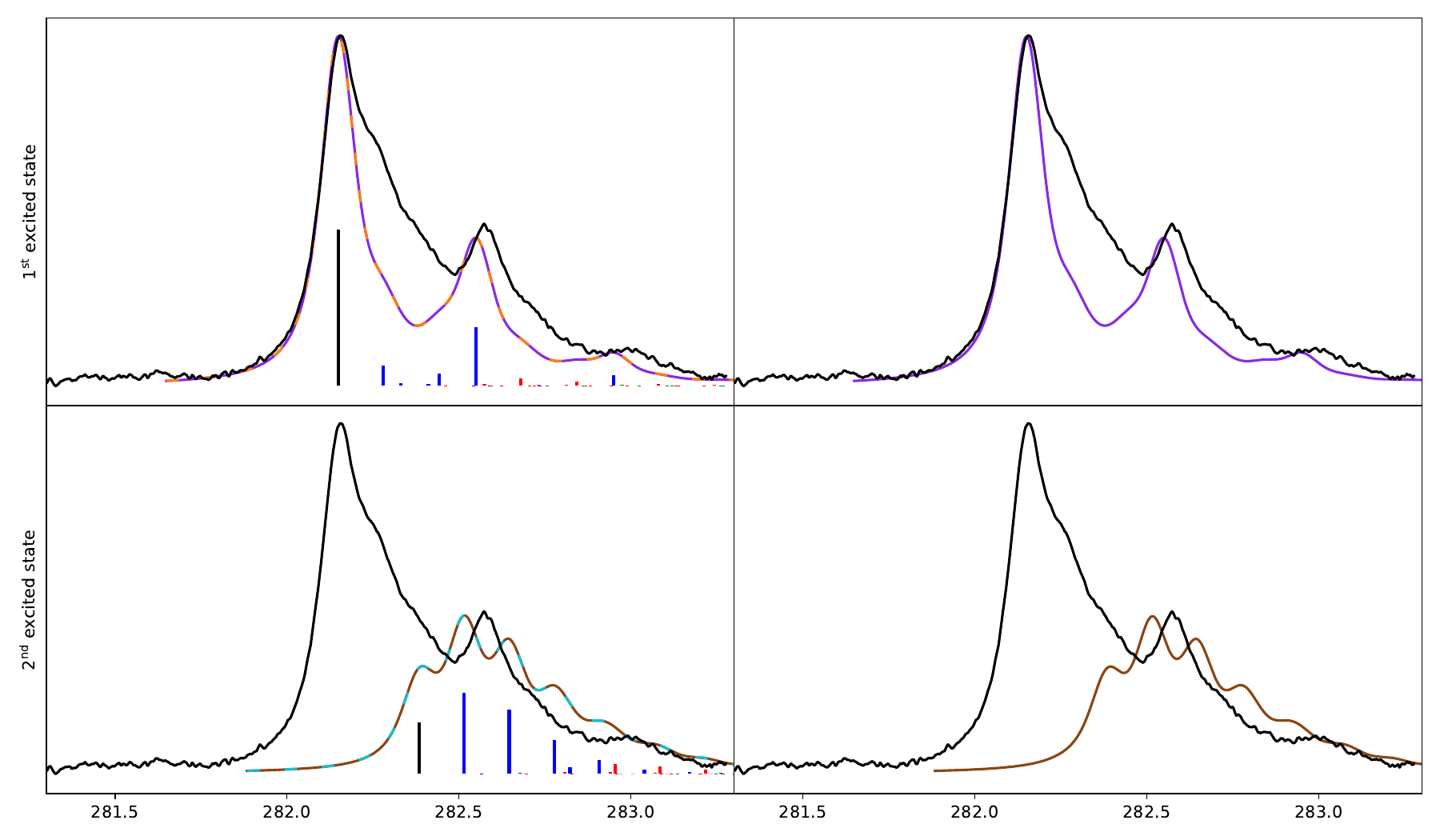}
    \caption{Comparison between the high-resolution experimental NEXAFS spectrum (black line) and vibrational structure of the first and second CVS-ADC(2)-x electronic transitions at temperatures of 0\,K (left) and 300\,K (right). For 0\,K the results were obtained by the TI (dashed orange and turquoise line) and TD approach (violet and brown line) while only the TD approach was employed at 300\,K. The PES were obtained from the Vertical Gradient approximation.
    The colors of the sticks correspond to the number of different vibrational modes involved in a transition; blue sticks consists of one vibrational mode, red sticks consists of two vibrational modes, and green sticks consists of three vibrational modes.
    The spectra have been shifted by +175\,meV to align with the first experimental peak. The spectral bands were broadened by a Voigt function with a Gaussian FWHM of 50\,meV and a Lorentzian FWHM of 100\,meV.
    }
    \label{fig:temperature:VG}
\end{figure}

\begin{figure}[H]
    \centering
    \includegraphics[width=\linewidth]{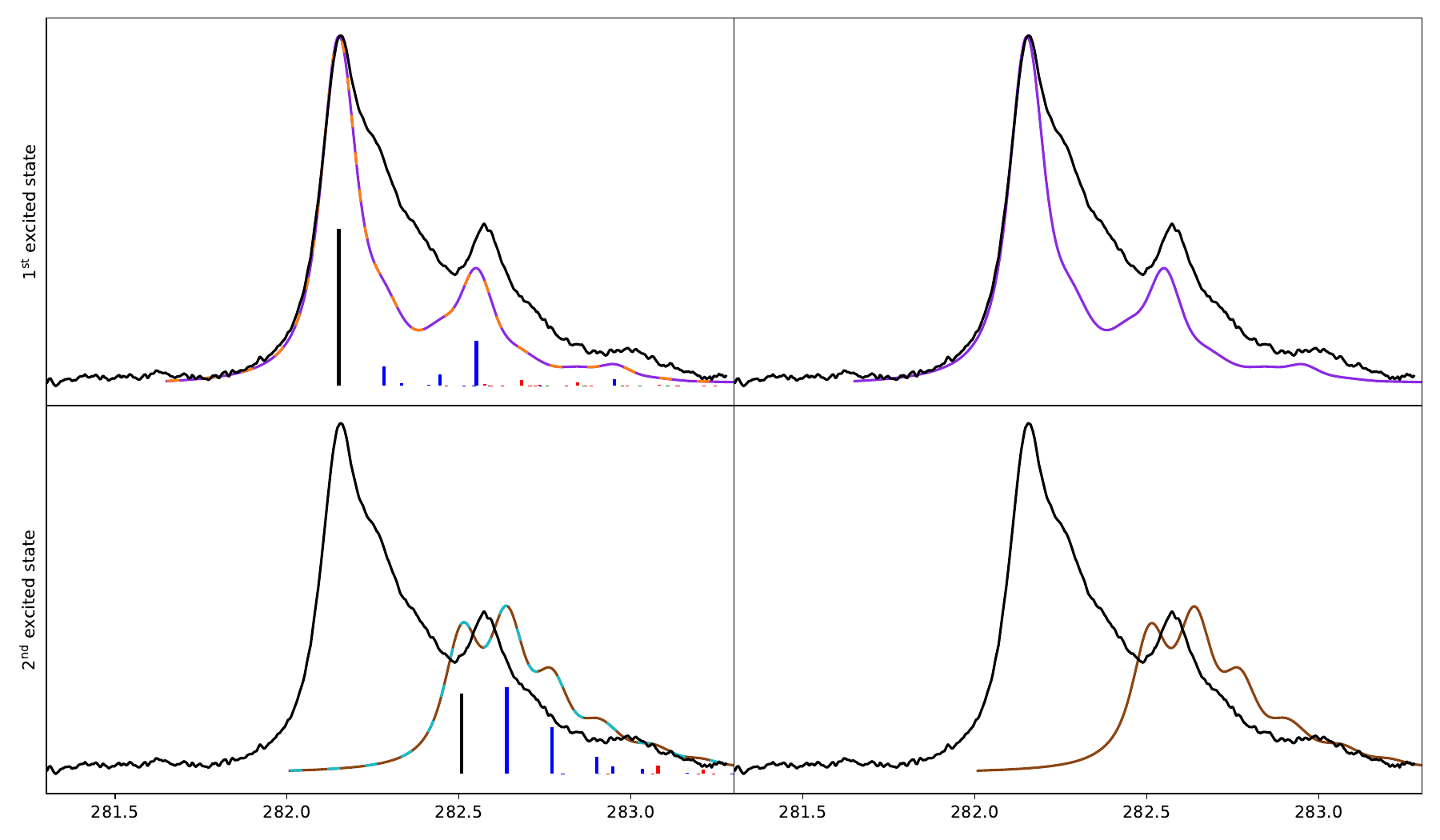}
    \caption{
    Comparison between the high-resolution experimental NEXAFS spectrum (black line) and vibrational structure of the first and second CVS-ADC(2)-x electronic transitions at temperatures of 0\,K (left) and 300\,K (right). For 0\,K the results were obtained by the TI (dashed orange and turquoise line) and TD approach (violet and brown line) while only the TD approach was employed at 300\,K. The PES were obtained from the Adiabatic Shift approximation.
    The colors of the sticks correspond to the number of different vibrational modes involved in a transition; blue sticks consists of one vibrational mode, red sticks consists of two vibrational modes, and green sticks consists of three vibrational modes.
    The spectra have been shifted by +318\,meV to align with the first experimental peak. The spectral bands were broadened by a Voigt function with a Gaussian FWHM of 50\,meV and a Lorentzian FWHM of 100\,meV. 
    }
    \label{fig:temperature:AS}
\end{figure}

\clearpage

\begin{table}
\centering
\begin{tabular}{l|c|c|c}
\caption{Vibrational modes of the propargyl radical and their frequencies in cm$^{-1}$ as obtained on the MP2/aug-cc-pVDZ level of theory. An illustration of these modes can be found in Fig.~\ref{fig:modes}.}
\label{tab:modes}\\
    \hline
    Vibration & Symmetry & Description & 
    Energy (cm$^{-1}$)
    \\\hline
    mode 12   & A$_1$  & C3-H stretching & 3542.59   \\
    mode 11   & B$_1$  & asymm. \ce{-CH2} stretching & 3349.32  \\
    mode 10   & A$_1$  & symm. \ce{-CH2} stretching & 3227.90  \\
    mode 9    & A$_1$  & C3-C2 stretching & 2369.24  \\
    mode 8    & A$_1$  &  \ce{-CH2} scissoring  & 1465.92  \\
    mode 7    & A$_1$  &  C1-C2 stretching & 1057.24   \\
    mode 6    & B$_1$  & in-plane rocking @C1 & 1048.32   \\
    mode 5    & B$_1$  & in-plane rocking @C3 & 768.244  \\
    mode 4    & B$_2$  & oop ($\perp$) wagging @C3 & 681.703  \\
    mode 3    & B$_2$  & oop ($\perp$) wagging \ce{-CH2} & 623.043  \\
    mode 2    & B$_2$  & oop bending @C2 (wagging)& 399.720  \\
    mode 1 & B$_1$  & in-plane rocking @C2 & 258.101  \\
    \hline
    \end{tabular}
\end{table}

\clearpage
\begin{figure}
\centering
\includegraphics[width=1\linewidth]{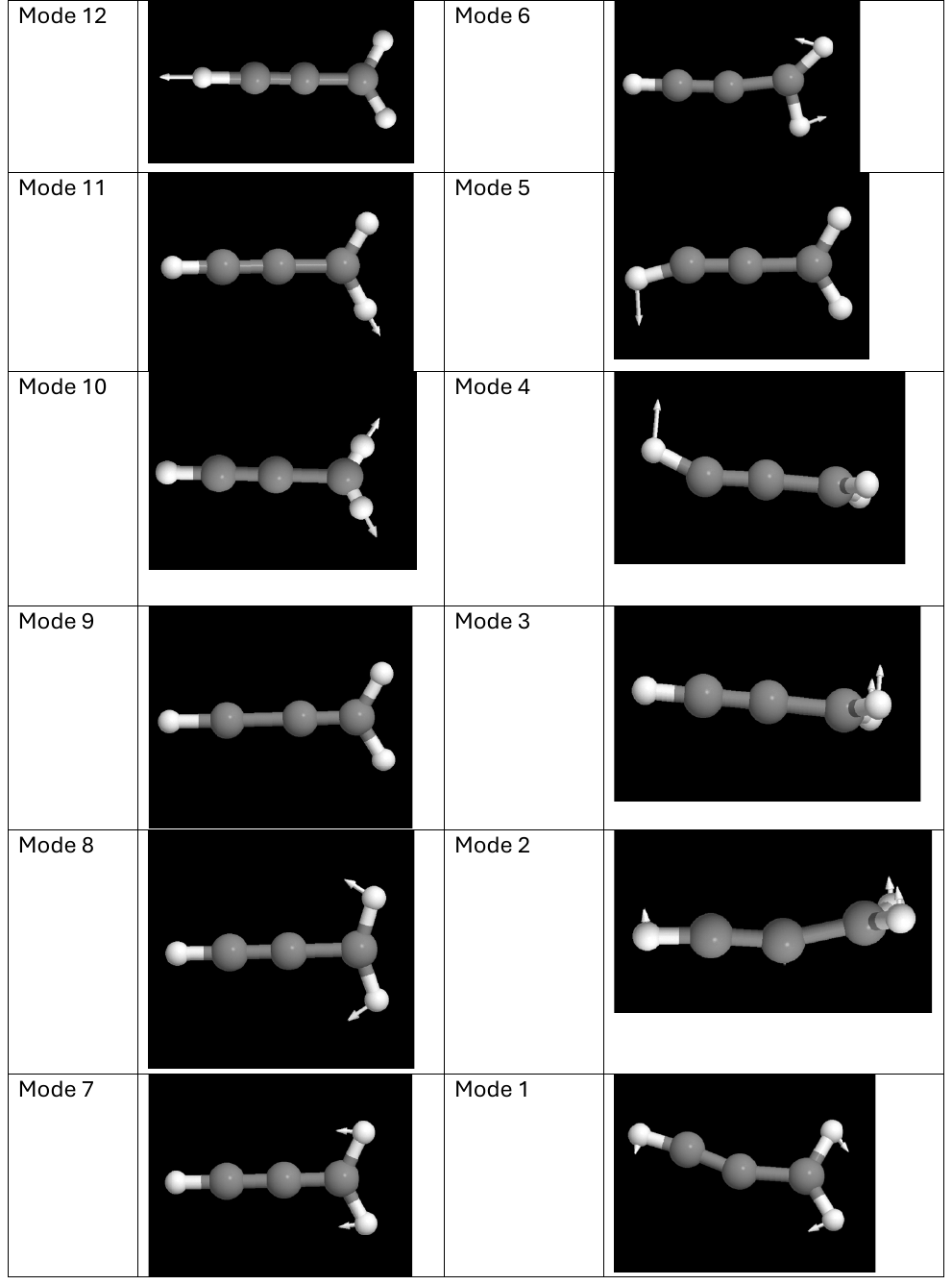}
\caption{Illustration of the fundamental modes of the propargyl radical as specified in Table~\ref{tab:modes}.}
    \label{fig:modes}
\end{figure}

\clearpage

\subsection{Optimized structures}
\label{sec:OptimizedStructures}

\begin{table}[hbpt!]
\centering
\caption{Bond lengths (in \r{A}) for the ground state [optimized both at CCSD(T*)-F12a/cc-pVQC-F12 (from Ref.~\citenum{botschwina2010calculated}) and fc-MP2/aug-cc-pVDZ (this work) levels] and the first three bright core excited states (optimized using CVS-ADC(2)-x/aug-cc-pVDZ, this work) of the propargyl radical. The semi-experimental bond-lengths for the ground state from Ref.~\citenum{Changala_2024} are also reported for comparison.}
\label{tab:bond_length}
\begin{tabular}{l*{3}{|l}|l}
\hline\hline
{State/method/basis} & C1--H1/2 & C1--C2 & C2--C3 & C3--H3 \\\hline
        GS/CCSD(T*)-F12a/cc-pVQC-F12~\cite{botschwina2010calculated} & 1.07997 & 1.37722 & 1.22507 & 1.06302 \\\hline
    GS/semi-experimental~\cite{Changala_2024}
        & 1.0789(1) & 1.3749(4) & 1.2220(4) & 1.0610(2)\\\hline
        GS/fc-MP2/aug-cc-pVDZ & 1.08723 & 1.40144 & 1.20988 & 1.07130 \\\hline
    
        %
        %
        1$^\text{st}$/CVS-ADC(2)-x/aug-cc-pVDZ & 1.03232 & 1.38785 & 1.19007 & 1.06943 \\
        2$^\text{nd}$/CVS-ADC(2)-x/aug-cc-pVDZ & 1.08436 & 1.31830 & 1.21496 & 1.02749 \\
        5$^\text{th}$/CVS-ADC(2)-x/aug-cc-pVDZ & 1.09665 & 1.35510 & 1.25539 & 1.05839 \\
        \hline\hline
    \end{tabular}
\end{table}

\begin{samepage}
\noindent \textbf{Ground state ($\tilde X$\textsuperscript{ 2}B$_1$)/fc-MP2/aug-cc-pVDZ}
\begin{verbatim}
Ground state energy:      -115.7801414328 Ha.
1st excited state energy: -105.4107274234 Ha. (E_exc: 282.166 eV) 
2nd excited state energy: -105.3995097829 Ha. (E_exc: 282.472 eV)
5th excited state energy: -105.3096408480 Ha. (E_exc: 284.917 eV)
C1      0.000000000000     0.000000000000     1.268908999959
C2      0.000000000000     0.000000000000    -0.134232999996
C3      0.000000000000     0.000000000000    -1.344778999957
H1     -0.940295999970     0.000000000000     1.816694999942
H2      0.940295999970     0.000000000000     1.816694999942
H3      0.000000000000     0.000000000000    -2.416981999922
\end{verbatim}
\end{samepage}

\begin{samepage}
\noindent \textbf{First excited state (1~$^2\!$A$_1$)/CVS-ADC(2)-x/aug-cc-pVDZ}
\begin{verbatim}
Energy: -105.4229786608 Ha. (E_adia: 281.833 eV)
C1    -0.000109386715      0.000000000000     1.260963322953
C2    -0.000131754750     -0.000000000000    -0.126882435583
C3     0.000040587964     -0.000000000000    -1.316950220391
H1    -0.887534699379      0.000000000000     1.788009294302
H2     0.887588569164     -0.000000000000     1.787545919345
H3     0.000146683717     -0.000000000000    -2.386380880667
\end{verbatim}
\end{samepage}

\begin{samepage}
\noindent \textbf{Second excited state (2~$^2\!$A$_1$)/CVS-ADC(2)-x/aug-cc-pVDZ}
\begin{verbatim}
Energy: -105.4098358009 Ha. (E_adia: 282.190 eV)
C1    -0.000018395481      0.000000000000     1.226614758055
C2    -0.000012583657     -0.000000000000    -0.091682480706
C3    -0.000075431347     -0.000000000000    -1.306642411737
H1    -0.946311572364      0.000000000000     1.756118582801
H2     0.946332709390      0.000000000000     1.756027123128
H3     0.000085273461     -0.000000000000    -2.334130571527
\end{verbatim}
\end{samepage}
\newpage
\begin{samepage}
\noindent \textbf{Fifth excited state (third bright) (2~$^2\!$A$_2$)/CVS-ADC(2)-x/aug-cc-pVDZ}
\begin{verbatim}
Energy: -105.3123007185 (E_adia: 284.845 eV)
C1    -0.000000000000      0.000000000000     1.251064629150
C2     0.000000000002      0.000000000000    -0.104036636939
C3    -0.000000000003     -0.000000000000    -1.359426986977
H1    -0.938578669184      0.000000000000     1.818259546634
H2     0.938578669181     -0.000000000000     1.818259546637
H3     0.000000000002     -0.000000000000    -2.417815098541
\end{verbatim}
\end{samepage}

\noindent





\noindent \textbf{Ground state ($\tilde X$\textsuperscript{ 2}B$_1$) gradient}
\begin{verbatim}
 0.00000000E+00  0.00000000E+00 -1.76300000E-04  
 0.00000000E+00  0.00000000E+00 -2.76190000E-04
 0.00000000E+00  0.00000000E+00  1.20780000E-04
 6.29100000E-05  0.00000000E+00  1.31660000E-04
-6.29100000E-05  0.00000000E+00  1.31660000E-04
 0.00000000E+00  0.00000000E+00  6.83900000E-05
\end{verbatim}

\begin{table}[H]
\centering
\begin{tabular}{lccccr}
\textbf{First excited state/VG gradient (1~$^2\!$A$_1$)$\quad$} 
&
&
$\quad\quad$
&
& 
\textbf{$\quad$ First excited state/AS gradient (1~$^2\!$A$_1$)}
\end{tabular}
\end{table}

\vspace{-1cm}
\begin{verbatim}
 0.         0.        -0.0279717           -0.0000275  0.         0.0000821
-0.         0.         0.0434325            0.0000149 -0.         0.0001105
 0.        -0.        -0.0582662            0.0000035  0.        -0.0001316
-0.0405327 -0.         0.0224284            0.0000617 -0.        -0.0000189
 0.0405327  0.         0.0224284           -0.0000526  0.        -0.0000464
-0.         0.        -0.0020514           -0.0000002  0.         0.0000043
\end{verbatim}

\clearpage

\begin{table}[H]
\centering
\begin{tabular}{lccccr}
\textbf{Second excited state/VG gradient (2~$^2\!$A$_1$) $\quad$} 
&
&
$\quad\quad$
&
& 
\textbf{$\quad$ Second excited state/AS gradient (2~$^2\!$A$_1$)}
\end{tabular}
\end{table}

\vspace{-1cm}
\begin{verbatim}
-0.         0.         0.064691            -0.0000093 -0.         0.0000169
 0.        -0.        -0.0734743            0.0000104  0.        -0.000017
-0.         0.         0.0357863           -0.0000088 -0.         0.0000254
-0.0000707  0.         0.0060424           -0.0000003 -0.        -0.0000041
 0.0000707 -0.         0.0060424            0.0000042  0.        -0.0000073
 0.        -0.        -0.0390877            0.0000038 -0.        -0.0000139
\end{verbatim}

\begin{table}[H]
\centering
\begin{tabular}{lccccr}
\textbf{Fifth excited state/VG gradient (2~$^2\!$A$_2$)$\quad$} 
&
&
$\quad\quad$
&
& 
\textbf{$\quad$ Fifth excited state/AS gradient (2~$^2\!$A$_2$)}
\end{tabular}
\end{table}

\vspace{-1cm}
\begin{verbatim}
 0.         0.         0.0291388            0.        -0.         0.0000031
-0.        -0.        -0.1115036           -0.         0.         0.0000032
 0.         0.         0.0938297            0.        -0.         0.0000081
-0.0032512 -0.        -0.0014204           -0.0000002  0.         0.0000003
 0.0032512  0.        -0.0014204            0.0000002 -0.         0.0000003
-0.        -0.        -0.008624            -0.        -0.        -0.000015
\end{verbatim}

\clearpage
\textbf{Ground state ($\tilde X$\textsuperscript{ 2}B$_1$) Hessian (used as excited state Hessian for VG and AS)}
\begin{verbatim}
  6.52783110E-01  0.00000000E+00  5.14137200E-02  0.00000000E+00  0.00000000E+00
  6.14648070E-01 -8.54368300E-02  0.00000000E+00  0.00000000E+00  1.06219140E-01
  0.00000000E+00 -3.26353300E-02  0.00000000E+00  0.00000000E+00  5.78502400E-02
 -2.50000000E-07  0.00000000E+00 -2.43717860E-01  1.40000000E-07  0.00000000E+00
  1.52805322E+00  6.89528000E-03  0.00000000E+00  0.00000000E+00 -3.52666100E-02
  0.00000000E+00 -3.00000000E-08  5.70601600E-02  0.00000000E+00  4.61399000E-03
  0.00000000E+00  0.00000000E+00 -3.18150700E-02  0.00000000E+00  0.00000000E+00
  4.73997900E-02  2.50000000E-07  0.00000000E+00 -1.14237770E-01 -1.50000000E-07
  0.00000000E+00 -1.27217487E+00  3.00000000E-08  0.00000000E+00  1.83390845E+00
 -2.91589690E-01  0.00000000E+00  1.34164210E-01  7.00990000E-03  0.00000000E+00
 -1.54037000E-03 -5.12740000E-04  0.00000000E+00 -2.47472000E-03  2.97539300E-01
  0.00000000E+00 -1.41775700E-02  0.00000000E+00  0.00000000E+00  2.62153000E-03
  0.00000000E+00  0.00000000E+00  5.74010000E-04  0.00000000E+00  0.00000000E+00
  5.38284000E-03  1.33176130E-01  0.00000000E+00 -1.29106600E-01  2.69865100E-02
  0.00000000E+00 -7.41189000E-03 -8.96110000E-04  0.00000000E+00 -9.01005000E-03
 -1.44068000E-01  0.00000000E+00  1.35099830E-01 -2.91589690E-01  0.00000000E+00
 -1.34164200E-01  7.00990000E-03  0.00000000E+00  1.54037000E-03 -5.12740000E-04
  0.00000000E+00  2.47472000E-03 -1.22295000E-02  0.00000000E+00 -1.38866100E-02
  2.97539300E-01  0.00000000E+00 -1.41778100E-02  0.00000000E+00  0.00000000E+00
  2.62182000E-03  0.00000000E+00  0.00000000E+00  5.74090000E-04  0.00000000E+00
  0.00000000E+00  5.41794000E-03  0.00000000E+00  0.00000000E+00  5.38281000E-03
 -1.33176150E-01  0.00000000E+00 -1.29106600E-01 -2.69864900E-02  0.00000000E+00
 -7.41189000E-03  8.96100000E-04  0.00000000E+00 -9.01005000E-03  1.38866100E-02
  0.00000000E+00  1.04793800E-02  1.44068000E-01  0.00000000E+00  1.35099830E-01
  8.93781000E-03  0.00000000E+00  0.00000000E+00  4.64510000E-04  0.00000000E+00
  1.40000000E-07 -2.76633500E-02  0.00000000E+00 -1.40000000E-07 -2.17270000E-04
  0.00000000E+00 -1.31192000E-03 -2.17270000E-04  0.00000000E+00  1.31193000E-03
  1.86955700E-02  0.00000000E+00  4.96299000E-03  0.00000000E+00  0.00000000E+00
  1.35682000E-03  0.00000000E+00  0.00000000E+00 -2.13468100E-02  0.00000000E+00
  0.00000000E+00  1.81250000E-04  0.00000000E+00  0.00000000E+00  1.81160000E-04
  0.00000000E+00  0.00000000E+00  1.46645800E-02  2.00000000E-08  0.00000000E+00
  1.52076000E-03 -1.00000000E-08  0.00000000E+00  2.66329000E-03  0.00000000E+00
  0.00000000E+00 -4.29475710E-01  3.22700000E-05  0.00000000E+00 -5.06700000E-05
 -3.22700000E-05  0.00000000E+00 -5.06700000E-05  0.00000000E+00  0.00000000E+00
  4.25393020E-01
\end{verbatim}


\clearpage
\subsection{\texttt{FCclasses3} input file example}
\label{sec:fcclasses:input}

As an example of the construction of the input files for \texttt{fcclasses3}, we here provide the general structure of the input for the Adiabatic Shift of the first excited state

\subsubsection{fcc.inp}
\begin{verbatim}
$$$
PROPERTY     =   OPA
MODEL        =   AS
DIPOLE       =   FC
TEMP         =   0.00
BROADFUN     =   VOI
HWHM         =   0.025
HWHM2        =   0.05
METHOD       =   TD
NORMALMODES  =   COMPUTE
COORDS       =   CARTESIAN
STATE1_FILE  =   GS.fcc
STATE2_FILE  =   ES.fcc
ELDIP_FILE   =   eldip
\end{verbatim}

\subsubsection{ES.fcc}
\begin{verbatim}
GEOM      UNITS=ANGS
     6

First excited state/CVS-ADC(2)-x/aug-cc-pVDZ

ENER      UNITS=AU
-105.4229786608

GRAD      UNITS=AU
First excited state/AS gradient
\end{verbatim}

\subsubsection{GS.fcc}
\begin{verbatim}
GEOM      UNITS=ANGS
     6

Ground state/fc-MP2/aug-cc-pVDZ

ENER      UNITS=AU
-115.7801414328

GRAD      UNITS=AU
Ground state gradient

HESS      UNITS=AU
Ground state Hessian
\end{verbatim}

\subsubsection{eldip}
\begin{verbatim}
0.000000 1.000000 0.000000
0.000000 1.000000 0.000000
\end{verbatim}


%